\documentclass[twocolumn,times]{aastex63}

\shorttitle{kinematic models of narrow line region outflows}
\shortauthors{Polack et al.}

\usepackage{amsmath}
\usepackage{natbib}
\defcitealias{Polack2023}{Paper I}
\usepackage{float}
\usepackage{color}
\usepackage{comment}
\usepackage{graphics}
\usepackage{epstopdf}
\usepackage{microtype}
\usepackage{subfigure}
\usepackage[normalem]{ulem}
\usepackage{natbib}
\usepackage{longtable}
\graphicspath{{./}{Plots/}}
\newcommand{\lbol}{\ensuremath{L_{bol}}}
\newcommand{\mbh}{\ensuremath{M_{BH}}}
\newcommand{\ledd}{\ensuremath{\lbol/L_{Edd}}}

\newcommand{\halpha}{H$\alpha$~}

\definecolor{malachite}{rgb}{0.04, 0.85, 0.32}

\usepackage{soul,xcolor}
\setstcolor{red}

\begin{document}

\title{Determining the Extents, Geometries, and Kinematics of Narrow-Line Region Outflows in Nearby Seyfert Galaxies}

\author[0000-0001-5862-2150]{Garrett E. Polack}
\affiliation{Department of Physics and Astronomy, Georgia State University, 25 Park Place, Suite 605, Atlanta, GA 30303, USA}

\author[0000-0002-4917-7873]{Mitchell Revalski}
\affiliation{Space Telescope Science Institute, 3700 San Martin Drive, Baltimore, MD 21218, USA}

\author[0000-0002-6465-3639]{D. Michael Crenshaw}
\affiliation{Department of Physics and Astronomy, Georgia State University, 25 Park Place, Suite 605, Atlanta, GA 30303, USA}

\author[0000-0002-3365-8875]{Travis C. Fischer}
\affiliation{AURA for ESA, Space Telescope Science Institute, 3700 San Martin Drive, Baltimore, MD 21218, USA}

\author[0000-0003-2450-3246]{Henrique R. Schmitt}
\affiliation{Naval Research Laboratory, Washington, DC 20375, USA}

\author[0000-0002-6928-9848]{Steven B. Kraemer} 
\affiliation{Institute for Astrophysics and Computational Sciences, Department of Physics, The Catholic University of America, Washington, DC 20064, USA}

\author[0000-0001-8658-2723]{Beena Meena}
\affiliation{Space Telescope Science Institute, 3700 San Martin Drive, Baltimore, MD 21218, USA}

\author[0000-0002-9946-4731]{Marc Rafelski}
\affiliation{Space Telescope Science Institute, 3700 San Martin Drive, Baltimore, MD 21218, USA}
\affiliation{Department of Physics and Astronomy, Johns Hopkins University, Baltimore, MD 21218, USA}

\begin{abstract}
Outflowing gas from supermassive~black~holes in the centers of active galaxies has been postulated as a major contributor to galactic evolution. To explore the interaction between narrow-line region (NLR) outflows and their host galaxies, we use Hubble~Space~Telescope~(HST) Space~Telescope~Imaging~Spectrograph~(STIS) spectra and Wide~Field~Camera~3~(WFC3) images of 15 nearby (z~$<$~0.02) active~galactic~nuclei (AGN) to determine the extents and geometries of their NLRs. We combine new HST~WFC3 continuum and [\ion{O}{3}] $\lambda$5007~\AA~images of 11 AGN with 4 archival AGN to match existing spectra from HST~STIS. For the 6 AGN with suitable long-slit coverage of their NLRs, we use isophotal fitting of ground-based images, continuum-subtracted [\ion{O}{3}] images, and the STIS spectra, to resolve, measure, and de-project the gas kinematics to the plane of the host galaxy disk and distinguish NLR outflows from galaxy rotation and/or kinematically disturbed gas. We find an average [\ion{O}{3}] extent of $\sim$680~pc with a correlation between gas extent and [\ion{O}{3}] luminosity of R$_\mathrm{[O~III]}$~$\propto$~L$_{\text{[O~III]}}^{0.39}$. The measured extents depend strongly on the depth of the [\ion{O}{3}] images, highlighting the importance of adopting uniform thresholds when analyzing scaling relationships. The outflows reach from 39--88\% of the full NLR extents, and we find that all 6 of the AGN with STIS coverage of their entire NLRs show strong kinematic evidence for outflows, despite previous uncertainty for these AGN. This suggests that NLR outflows are ubiquitous in moderate luminosity AGN and that standard criteria for kinematic modeling are essential for identifying outflows.
\end{abstract}


\section{Introduction} \label{sec:intro}
\subsection{Active Galaxy Feedback} \label{sec:Science}

Observations of the nearby universe reveal the existence of supermassive black holes (SMBHs) residing at the centers of most massive galaxies \citep{Magorrian_1998, Marconi2004}. A few percent of these black holes are accreting their surrounding material at high rates, funneling nearby stars and gas into highly energetic accretion disks. The accretion process releases incredible amounts of radiation from the galaxy nucleus in the form of both relativistic jets and ionized winds \citep{Kormendy2013, Harrison2024}. The accreting SMBH, it's energetic emanation, and the resulting interaction with the host galaxy comprise complex systems, termed active galactic nuclei (AGN), and are important for understanding the co-evolution of supermassive black holes and their host galaxies \citep{Kormendy1995,Ferrarese2000,Gebhardt2000}.

The geometry and observed properties of these AGN are described by the generally accepted ``unified model" \citep{unified}. A key feature of this model is that the SMBH and its accretion disk are confined within an optically thick, ring-like structure of molecular material -- the torus -- that channels the outflowing radiation into two conical shapes, facing opposite directions and with vertices intersecting at the nucleus \citep{Almeida2017}. These biconical outflows are theorized to be powerful enough to regulate SMBH accretion and galaxy star-formation rates (SFR) by displacing stellar material at bulge-radius scales \citep{fabian_cooling_1994, Harrison2024}. However, the impact of this displacement is yet to be fully understood. The redistribution of stellar material back into the galaxy has the potential to positively impact a galaxy's SFR by either compressing the surrounding interstellar medium or triggering the generation of stars within the outflows themselves \citep{Ferland1998, Riffel2024, Zubovas2017}. On the other hand, both simulated and observational studies have detailed the negative impact of outflows on stellar formation via the heating or complete evacuation of the star-forming gas \citep{Sim1, Harrison2018}. Several recent simulated studies have shown AGN as an effective solution for galaxy cooling and SFR quenching \citep{Harrison2024,Sim2,Sim3}. Verifying these possibilities observationally is not easy, as the role that these outflows play in galaxy evolution is heavily influenced by the characteristics of each individual AGN, such as mass, luminosity and galaxy environment \citep{Ciotti2001, Heckman2014}.

Additionally, spatially-resolved observations are generally only available for lower redshift AGN (z less than $\sim$0.11) \citep{Barbosa2009, Riffel2008, Storchi-Bergmann2010, Muller2011}. Thus, the difficulty of obtaining reliable constraints lead many investigations to use spatially averaged outflow characteristics (e.g., assuming constant electron density or outflow energetics as a function of radial distance)\citep{Cicone2018}. \cite{Revalski2021} suggests, however, that generalizing AGN parameters in this way may lead to estimates of outflow properties that are systematically biased by as much as 2--3~dex when compared with spatially resolved data and calculations.

Our goal is to utilize spatially resolved observations of nearby AGN to quantify the effects of outflows on galaxy and AGN evolution, given that several fundamental AGN parameters (masses, bulge sizes, luminosities, etc.) correlate with outflow characteristics. This work, therefore, focuses on the resolvable outflow regions of relatively nearby (z $\leq$ 0.1) Seyfert galaxies. Seyferts are a moderate luminosity (L$\mathrm{_{bol}}$ $\approx 10^{43} - 10^{45}$ erg s$^{-1}$) subgroup of AGN, typically comprised of spirals and divided into types based on the angle of the obscuring torus to our line of sight \citep{Osterbrock_Ferland2006, Harrison2024}. Type 1 Seyferts allow for the observation of the inner torus and the accretion disk, whereas type 2s have their accretion disk mostly obscured by the torus and their biconical radiation funneled in a more perpendicular direction to the line of sight. Our sample mainly consists of type 2 Seyferts, as the angle of the bicone across the line of sight allows for a more extensive radial measurement of the outflow.

The narrow-line region is comprised of low density (n$_{H}$ $\approx$ 10$^{2}$ -- 10$^{6}$ cm$^{-3}$ \citealp{Revalski2022}), high ionization clouds, ejected at velocities of $\sim$100 to 2000 km~s$^{-1}$, and can be found anywhere from sub-pc to kpc scales \citep{Fischer2013,Crenshaw2015,Meena2021,ubler2024}. The term ``narrow-line" originates from the optical spectra of this region that is comprised of narrow emission lines, possessing a full width at half maximum (FWHM) of $<$ 2000 km~s$^{-1}$. We focus on these outflows because their interactions with the surrounding environment occur on scales that can be spatially resolved in nearby galaxies, thereby allowing us to gain insight into the effects of AGN feedback on the structure and dynamics of the host galaxies \citep{Kormendy2013, Osterbrock_Ferland2006}. 

Recent results from previous analyses find positive correlations for both NLR outflow mass ($M$) and peak mass outflow rate ($\dot M$) with AGN luminosity ($\lbol$)\citep{Revalski2018b, Revalski2019, Revalski2021}. Additionally, \cite{Revalski2021} finds the same correlation with $\lbol$ for the radial extent of NLR outflows in multiple targets, reinforcing the connections found in several recent studies \citep{Fischer2018, Storchi-Bergmann2018, Luo_2021}. While these analyses provide strong evidence for their targets, a larger sample is required to better constrain any true correlation between AGN attributes and outflow characteristics. Moreover, most of the targets selected for outflow analysis reside on the higher luminosity end of AGN candidates, excluding high-z quasars. It is necessary to expand these samples and determine if AGN with [\ion{O}{3}] luminosity lower than $L_\mathrm{[O~III]}$$<$ 10$^{41}$ erg~s$^{-1}$ (or $L_\mathrm{bol}$$<$ 10$^{44.5}$ erg~s$^{-1}$, adopting the conversion from \cite{Heckman2004} with $L_\mathrm{bol}$~=~3500$\times$$L_{\mathrm{[O~III]}}$) continue to demonstrate these empirical correlations. This is an important range to examine, because there is a relatively large population of lower luminosity AGN compared to higher luminosity ones that has yet to be investigated in this manner \citep{Osterbrock_Ferland2006,Liu2013, Gatto2024}.

Our previous work with spatially-resolved observations has produced well constrained measurements of outflow parameters. The simplified procedure developed in \cite{Revalski2022} allows us to model more AGN with lower quality spectra and expand to larger samples more efficiently, as the modeling is less intensive. In this work we perform the analyses of the geometry, kinematics, and radial extents for the targets whose spectroscopic observations provide enough data to perform said mass outflow and energetics analyses. We also investigate different ways to identify and analyze outflows, using multiple techniques, to examine the inherent biases that come with empirically developed relationships.

\begin{deluxetable*}{lcccccccc}
\vspace{-0.5em}
\setlength{\tabcolsep}{0.14in}
\def\arraystretch{0.95}
\tablecaption{Physical Properties of the Active Galaxy Sample}
\tablehead{
\colhead{Catalog} & \colhead{Seyfert} & \colhead{Redshift} & \colhead{Distance} &\colhead{Scale} & \colhead{Inclination} & \colhead{$\log$(\lbol)} & \colhead{$\log$(\mbh)$^\star$} & \colhead{\ledd \vspace{-0.7em}}\\
\colhead{Name} & \colhead{(Type)} & \colhead{(unitless)} & \colhead{(Mpc)} &\colhead{(pc/$\arcsec$)} & \colhead{(deg)} & \colhead{(erg s$^{-1}$)} & \colhead{($M_{\odot}$)} & \colhead{(unitless) \vspace{-0.7em}} \\
\colhead{(1)} & \colhead{(2)} & \colhead{(3)} &\colhead{(4)} & \colhead{(5)} & \colhead{(6)} & \colhead{(7)} & \colhead{(8)} & \colhead{(9)}
}
\startdata
NGC 788 & 2   & 0.0136 & 55.8 & 271 & 31.8 & 44.2 & 7.5 & 0.036 \\
NGC 1358 & 2    & 0.0134 & 55.2 & 268 & 46.4 & 43.7 & 8.0 & 0.004 \\
NGC 1667 & 2 & 0.0153 & 62.7 & 304 & 45.6 & 43.6 & 7.8 & 0.005 \\
NGC 2273 & 2    & 0.0060 & 17.9 & 87 & 53.1 & 44.0$^\star$ & 7.3 & 0.040$^\star$ \\
NGC 3393 & 2   & 0.0125 & 51.4 & 249 & 33.9 & 44.8 & 7.7 & 0.107 \\
NGC 5135 & 2   & 0.0137 & 56.2 & 273 & 57.3 & 44.4 & 7.4 & 0.078 \\
NGC 5283 & 2    & 0.0104 & 42.7 & 207 & 30.7 & 43.9 & 7.6 & 0.015 \\
NGC 5347 & 2    & 0.0080 & 39.0 & 189 & 54.6 & 43.8$^\star$ & 6.8 & 0.079$^\star$ \\
NGC 5427 & 2    & 0.0087 & 27.0 & 131 & 34.9 & 42.8 & 7.6 & 0.001 \\
NGC 5643 & 2   & 0.0040 & 11.8 & 57 & 47.9 & 43.6$^\star$ & 6.8 & 0.050$^\star$ \\
NGC 5695 & 2    & 0.0141 & 57.9 & 281 & 46.4 & 43.7 & 8.0 & 0.004 \\
NGC 6300 & 2    & 0.0037 & 13.1 & 64 & 60.0 & 42.2 & 6.8 & 0.002 \\
NGC 7682 & 2 & 0.0171 & 70.4 & 341 & 47.2 & 44.3 & 7.3 & 0.072 \\
IC 3639  & 2   & 0.0109 & 44.9 & 217 & 32.8 & 44.0 & 7.0 & 0.087 \\
UGC 1395 & 1.9    & 0.0172 & 70.6 & 342 & 50.2 & 43.7 & 6.7 & 0.072 \\
\enddata
\tablecomments{Columns are (1) target name, (2) Seyfert classification type (3) 21~cm redshift from the NASA/IPAC Extragalactic Database, (4) Hubble distance and (5) spatial scale assuming H$_0$ = 73 km s$^{-1}$ Mpc$^{-1}$ \citep{Riess_2022}, (6) host galaxy inclination based on our isophotal fitting models, (7) bolometric luminosity estimated from [\ion{O}{3}] imaging (see section \ref{sec:Science}), (8) black hole mass, and (9) Eddington ratio (\ledd) calculated using $L_{Edd} = 1.26 \times 10^{38} \left(M/M_{\odot}\right)$ erg~s$^{-1}$. Values for column 8, as well as those marked with asterisks are adopted from \citealp{Fischer2014} and references therein.}
\label{tab:sample}
\vspace{-2em}
\end{deluxetable*}

\subsection{Sample}

Our initial sample consisted of 53 nearby Seyfert galaxies, collected and detailed by \cite{Fischer2013}. These targets were not only chosen for their proximity and morphological type, but also because they contained at least one HST long-slit spectrum covering the [\ion{O}{3}] $\lambda\lambda$4959, 5007~\AA~ and \halpha $\lambda$6563~\AA~emission lines, which are important for tracing the NLR gas \citep{Crenshaw2000a}. It is important to note that this is not a complete sample. Using these spectra and HST imaging, six of these targets have published results for spatially resolved mass outflow rates and scaling relationships \citep{Revalski2021, Revalski2022}. An additional 12 targets were selected based on quantity of sufficient archival data for the performance of similar analyses, missing only HST [\ion{O}{3}] images. HST Program \href{https://www.stsci.edu/hst/phase2-public/16246.pdf}{16246} (PI: M. Revalski, \citealp{Revalski2020}) focused on these targets (NGC~788, NGC~1358, NGC~1667, NGC~5135, NGC~5283, NGC~3227, NGC~5427, NGC~5695, NGC~6300, NGC~7682, IC~3639, UGC~1395) and obtained data in multiple filters to create images that isolate the [\ion{O}{3}] gas. In addition to providing photometry, these images helped to determine whether the archival spectra from STIS were aligned over the nucleus of each AGN, as well as the amount of ionized emission contained within the extent of the slits given their position angles.

The primary sample of this work (listed in Table~\ref{tab:sample}) includes these recently observed galaxies, along with four archival targets (NGC~2273, NGC~3393, NGC~5347, and NGC~5643) with HST STIS spectra. The sample spans a luminosity range of $\sim$ 3.8 dex, contributing significantly to the range of targets used in luminosity-dependent relationships. Due to the variety of multi-wavelength auxiliary data for the HST program target, NGC~3227, we have set it aside for further analysis in a separate publication, although its [\ion{O}{3}] image is still provided here for the sake of completion.

\section{Observations}\label{sec:obs}

\subsection{HST Imaging}\label{ssec:imaging}

\subsubsection{Observing Strategy}\label{sssec:observing}

We observed each galaxy with WFC3/UVIS, using a medium and narrow-band filter, during HST Cycle 28 through Program ID \href{https://www.stsci.edu/hst/phase2-public/16246.pdf}{16246} (PI: M. Revalski, \citealp{Revalski2020}). These filters sample the continuum and [\ion{O}{3}] emission for each AGN, which allows us to create continuum-subtracted [\ion{O}{3}] images that trace the emission line gas ionized by the AGN and star-formation. We observed each bright galaxy for only a single orbit, except for NGC~5283, which is fainter and compact, and so was observed over two orbits.

\begin{figure}[b]
\centering
\includegraphics[width=\columnwidth]{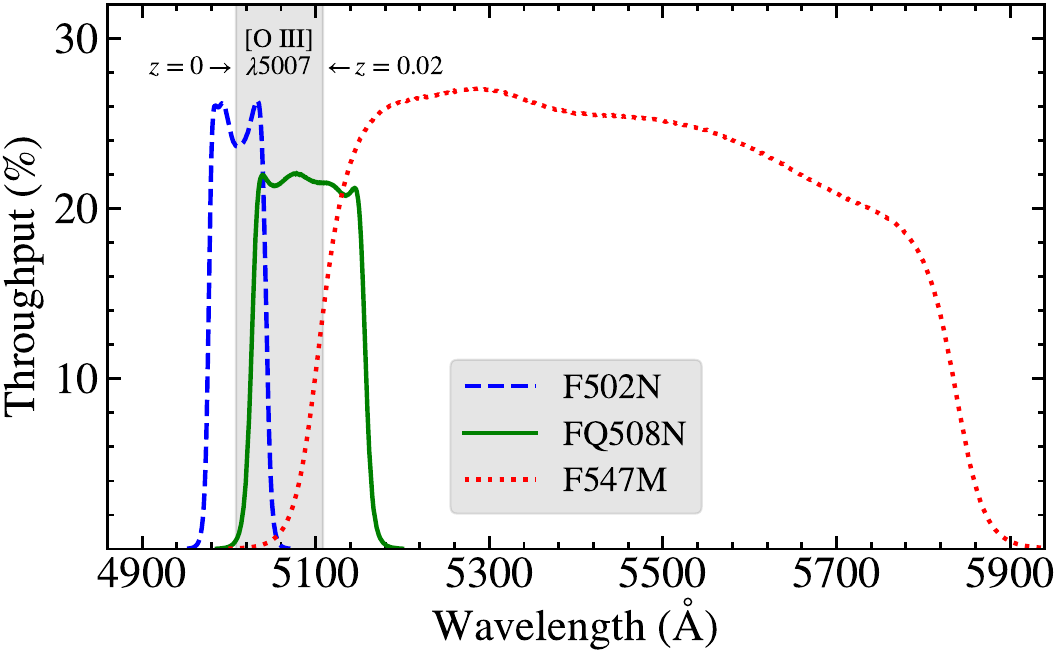}
\caption{The total system throughput for the HST WFC3 direct imaging filters (F502N, FQ508N, and F547M), including the quantum efficiency of the CCD detectors. The gray shaded region shows the centroid of the [\ion{O}{3}] $\lambda$5007~\AA~emission line for $z$ = 0 to 0.02.}
\label{fig:filters}
\end{figure}

We obtained data for 11 of the galaxies that are listed in Table~\ref{tab:sample}, including NGC 3227 (not in table) with observations that were completed between 26~October~2020 and 25~March~2021 (This excludes the archival targets: NGC 2273, NGC 3393, NGC 5347, and NGC 5643). We observed each galaxy with the F547M filter to sample the continuum without strong emission lines, followed by a narrow-band filter to sample the [\ion{O}{3}] emission. We used the FQ508N filter for this purpose for 10 of the galaxies at $z\approx$~0.01. We used the F502N filter for NGC~3227 and NGC~6300, as it most appropriately samples the [\ion{O}{3}] $\lambda$5007~\AA~emission line at z $<$ 0.005. The bandpasses for each filter are shown in Figure~\ref{fig:filters}.

The observations for each galaxy (except NGC~5283) consist of five exposures, with two 350 second exposures in the F547M, and three exposures in the narrow-band that filled each orbit, resulting in exposure times of 1386 -- 1749 seconds per galaxy. Due to its faint, compact nature, NGC~5283 was observed for two orbits. We split the images into two exposures for the continuum and three exposures for the [\ion{O}{3}] images. Three exposures help to remove cosmic rays and bad pixels and improve the image quality by sampling the instrumental Point Spread Function (PSF). Using two F547M exposures is sufficient to correct for cosmic rays with exposure times of $<$ 1000~sec \citep{Dahlen2010} and allowed us to fit each visit within a single orbit. The increased overheads from a third continuum exposure would have required an additional orbit per target due to buffer dump limitations. Finally, we post-flashed the images to reach the recommended background level of 20~electrons per pixel \citep{Biretta2013}, which mitigates charge transfer efficiency losses. We used two and three point dither patterns with three times the default spacing between exposures, which better removes detector artifacts when combining the exposures into a single mosaic. For NGC 3227, a short, 20 second F547M exposure was also included to account for saturation near the bright, unresolved Type I AGN nucleus.

\startlongtable
\begin{deluxetable*}{llclccccc}
\tabletypesize{\footnotesize}
\def\arraystretch{0.94}
\tablecaption{Summary of Observations}
\tablehead{
\colhead{Target} & \colhead{Instrument} & \colhead{Proposal} & \colhead{Observation} & \colhead{Date} & \colhead{Exposure} & \colhead{Grating} & \colhead{Wavelength} & \colhead{Position}\\
\colhead{Name} & \colhead{Name} & \colhead{ID} & \colhead{ID} & \colhead{(yyyy-mm-dd)} & \colhead{Time (s)} & \colhead{or Filter} & \colhead{Range (\AA)} & \colhead{Angle (deg)}
}
\startdata
NGC 788 & WFC3/UVIS & 16246 & iebn02010 (vkq,vlq) & 2020-12-13 & 700 & F547M & 5060-5885 & ... \\
 & WFC3/UVIS & 16246 & iebn02020 (vmq,voq,vqq) & 2020-12-13 & 1389 & FQ508N & 5012-5171 & ... \\ 
 & STIS/CCD & 9143 & o6bu150 (10,20,30) & 2001-09-17 & 840 & G750M & 5450–10140 & -129.56\\
 & STIS/CCD & 9143 & o6bu150 (40,50) & 2001-09-17 & 1080 & G430L & 2900-5700 & -129.56\\ \hline
NGC 1358 & WFC3/UVIS & 16246 & iebn03010 (pyq,pzq) & 2020-12-16 & 700 & F547M & 5060-5885 & ... \\
 & WFC3/UVIS & 16246 & iebn03020 (q0q,q2q,q4q) & 2020-12-16 & 1389 & FQ508N & 5012-5171 & ... \\ 
 & STIS/CCD & 9143 & o6bu030 (10,20,30) & 2002-01-25 & 840 & G750M & 5450–10140 & 23.89\\
 & STIS/CCD & 9143 & o6bu030 (40,50) & 2002-01-25 & 1080 & G430L & 2900-5700 & 23.89\\ \hline
NGC 1667 & WFC3/UVIS & 16246 & iebn04010 (fxq,fyq) & 2021-01-07 & 700 & F547M & 5060-5885 & ... \\
 & WFC3/UVIS & 16246 & iebn04020 (fzq,g1q,g3q) & 2021-01-07 & 1389 & FQ508N & 5012-5171 & ... \\ 
 & STIS/CCD & 9143 & o6bu040 (10,20,30) & 2001-10-14 & 840 & G750M & 5450–10140 & -120.21\\
 & STIS/CCD & 9143 & o6bu040 (40,50) & 2001-10-14 & 1080 & G430L & 2900-5700 & -120.21\\ \hline 
NGC 2273 & STIS/CCD & 9143 & o6bu050 (10,20,30,50) & 2001-10-14 & 900-1226 & G750M & 5450–10140 & -151.61\\
 & STIS/CCD & 9143 & o6bu050 (40) & 2001-11-04 & 840 & G430L & 2900-5700 & -151.61\\ \hline
NGC 3393 & WFC3/UVIS & 12365 & ibly01gwq & 2011-05-17 & 208 & F547M & 5060-5885 & ... \\
 & WFC3/UVIS & 12365 & ibly01010 & 2011-05-16 & 566 & FQ508N & 5012-5171 & ... \\ 
 & STIS/CCD & 8055 & o56c020 (30,40) & 2000-4-22 & 600,865 & G750M & 5450–10140 & 39.98\\
 & STIS/CCD & 8055 & o56c020 (50) & 2000-4-22 & 600 & G430L & 2900-5700 & 39.98\\ \hline
NGC 5135 & WFC3/UVIS & 16246 & iebn06010 (huq,hvq) & 2021-02-26 & 700 & F547M & 5060-5885 & ... \\
 & WFC3/UVIS & 16246 & iebn06020 (hwq,hyq,i0q) & 2021-02-26 & 1407 & FQ508N & 5012-5171 & ... \\ 
 & STIS/CCD & 9143 & o6bu070 (10,20,30) & 2002-01-11 & 840 & G750M & 5450–10140 & -115.81\\
 & STIS/CCD & 9143 & o6bu070 (40,50) & 2002-01-11 & 1080 & G430L & 2900-5700 & -115.81\\ \hline
NGC 5283 & WFC3/UVIS & 16246 & iebn10010 (rsq,rtq,ruq) & 2021-03-24 & 2727 & F547M & 5060-5885 & ... \\
 & WFC3/UVIS & 16246 & iebn10020 (rvq,rzq,s5q) & 2021-03-25 & 2808 & FQ508N & 5012-5171 & ... \\ 
 & STIS/CCD & 9143 & o6bu080 (10,20,30) & 2001-10-11 & 840 & G750M & 5450–10140 & -37.06\\
 & STIS/CCD & 9143 & o6bu080 (40,50) & 2001-10-11 & 1080 & G430L & 2900-5700 & -37.06 \\ \hline 
NGC 5347 & STIS/CCD & 9143 & o6bu090 (10,20,30) & 2001-12-25 & 840-1126 & G750M & 5450–10140 & -102.20\\
& STIS/CCD & 9143 & o6bu090 (40,50) & 2001-12-25 & 840-851 & G430L & 2900-5700 & -102.20\\ \hline
NGC 5427 & WFC3/UVIS & 16246 & iebn07010 (kfq,kgq) & 2020-12-19 & 700 & F547M & 5060-5885 & ... \\
 & WFC3/UVIS & 16246 & iebn07020 (khq,kjq,klq) & 2020-12-19 & 1389 & FQ508N & 5012-5171 & ... \\ 
 & STIS/CCD & 9143 & o6bu100 (10,20,30) & 2002-01-04 & 840 & G750M & 5450–10140 & -113.38\\
 & STIS/CCD & 9143 & o6bu100 (40,50) & 2002-01-04 & 1080 & G430L & 2900-5700 & -113.38\\ \hline 
NGC 5643 & STIS/CCD & 9143 & o52s010 (20) & 2000-03-12 & 2909.7 & G750M & 5450–10140 & -128.04\\
& STIS/CCD & 9143 & o52s010 (10) & 2000-03-12 & 2845 & G430L & 2900-5700 & -128.04\\ \hline
NGC 5695 & WFC3/UVIS & 16246 & iebn08010 (g0q,g1q) & 2020-11-16 & 700 & F547M & 5060-5885 & ... \\
 & WFC3/UVIS & 16246 & iebn08020 (g2q,g4q,g6q) & 2020-11-16 & 1443 & FQ508N & 5012-5171 & ... \\ 
 & STIS/CCD & 9143 & o6bu120 (10,20,30) & 2001-08-11 & 840 & G750M & 5450–10140 & 50.65\\
 & STIS/CCD & 9143 & o6bu120 (40,50) & 2001-08-11 & 1080 & G430L & 2900-5700 & 50.65\\ \hline
NGC 6300 & WFC3/UVIS & 16246 & iebn13010 (aaq,aeq) & 2020-10-26 & 700 & F547M & 5060-5885 & ... \\
 & WFC3/UVIS & 16246 & iebn13020 (agq,ajq,b4q) & 2020-10-26 & 1749 & F502N & 4969-5044 & ... \\
 & STIS/CCD & 9143 & o6bu130 (10,20,30) & 2001-11-08 & 840 & G750M & 5450–10140 & 90.28\\
 & STIS/CCD & 9143 & o6bu130 (40,50) & 2001-11-08 & 900 & G430L & 2900-5700 & 90.28\\ \hline
NGC 7682 & WFC3/UVIS & 16246 & iebn09010 (x4q,x5q) & 2020-11-05 & 700 & F547M & 5060-5885 & ... \\
 & WFC3/UVIS & 16246 & iebn09020 (x6q,x8q,xaq) & 2020-11-05 & 1386 & FQ508N & 5012-5171 & ... \\
 & STIS/CCD & 9143 & o6bu140 (10,20,30) & 2002-07-20 & 840 & G750M & 5450–10140 & 17.85\\
 & STIS/CCD & 9143 & o6bu140 (40,50) & 2002-07-20 & 1080 & G430L & 2900-5700 & 17.85\\ \hline
IC 3639 & WFC3/UVIS & 16246 & iebn05010 (egq,ehq) & 2021-03-05 & 700 & F547M & 5060-5885 & ... \\
 & WFC3/UVIS & 16246 & iebn05020 (eiq,ekq,emq) & 2021-03-05 & 1443 & FQ508N & 5012-5171 & ... \\ 
 & STIS/CCD & 9143 & o6bu010 (10,20,30) & 2002-01-23 & 840 
 & G750M & 5450–10140 & -104.41\\
 & STIS/CCD & 9143 & o6bu010 (40,50) & 2002-01-23 & 1080 & G430L & 2900-5700 & -104.41\\ \hline
UGC 1395 & WFC3/UVIS & 16246 & iebn01010 (hkq,hlq) & 2020-11-14 & 700 & F547M & 5060-5885 & ... \\
 & WFC3/UVIS & 16246 & iebn01020 (hmq,hoq,hqq) & 2020-11-14 & 1389 & FQ508N & 5012-5171 & ... \\ 
 & STIS/CCD & 9143 & o6bu160 (10,20,30) & 2001-12-20 & 840 & G750M & 5450–10140 & 21.33\\
 & STIS/CCD & 9143 & o6bu160 (40,50) & 2001-12-20 & 1080 & G430L & 2900-5700 & 21.33
\enddata
\tablecomments{New observations for this project were obtained under Proposal ID 16246. The HST STIS gratings have spectral dispersions of 0.56 \AA~for the G750M and 2.73 \AA~for the G430L. The pixel scales for WFC3/UVIS and HST/STIS are 0\farcs045 and 0\farcs05078 pix$^{-1}$, respectively.}
\label{tab:obs}
\end{deluxetable*}

\subsubsection{Mosaic Drizzling}\label{ssec:driz}

With the goal of producing high spatial resolution, continuum-subtracted [\ion{O}{3}] emission line images, we must optimally align and combine the exposures to account for undersampling of the UVIS detectors. Undersampling arises because the telescope can provide higher spatial resolution than the detectors are able to capture, which is alleviated with subpixel dithers between exposures. The high spatial resolution is recovered by combining the images into a mosaic using the linear reconstruction process known as ``drizzling" \citep{Fruchter2002}. The Space Telescope Science Institute (STScI) provides several tools for this in the \href{https://www.stsci.edu/scientific-community/software/drizzlepac.html}{\textsc{DrizzlePac}} software package \citep{Gonzaga2012, Hoffmann2021}. We note here that similar methods were applied to the archival target NGC 3393 \citep{Maksym2016, Maksym2017, Maksym2019}, with the resulting [\ion{O}{3}] image shown in Appendix~\ref{fig:3393}. (The remaining three archival targets used in our kinematic analyses: NGC 2273, NGC 5347, and NGC 5643, resulted in images with quality far below that of the rest of the sample and were therefore excluded from our image analyses).

First, we used \textsc{astroquery} on 21 March 2023 to retrieve the most recently calibrated single-visit files (FLCs) for each galaxy. We then ran \textsc{updatewcs} with the \textsc{use\_db} option set to False to reset the problematic world coordinate system (WCS) that can misalign narrow and medium band exposures taken in the same orbit. This process restores the previous WCS solution so that all exposures within an orbit are aligned to within the pointing accuracy of HST. We examined the jitter files for all observations and confirmed that the observatory pointing was stable to within $<$~5 milli-arcseconds for the majority of cases.

We then created drizzles of each object for each filter, and aligned the narrow-band images to their corresponding F547M continuum images using \textsc{tweakreg}. We employed a \textsc{fitgeometry} of \textsc{shift} and required a minimum of two sources for alignment with the \textsc{use\_sharp\_round} set to True. The alignment solutions are based on between two and 48 points depending on the galaxy. We found a consistent offset between the F547M and quadrant filter images of $\delta$x = 6.23 pixels and $\delta$y = --1.04 pixels with typical RMS uncertainties of $\sim$0.1~--~0.2 pixels. This offset is expected between the full-frame and quadrant filters in order to place sources at the optimal center of the quadrant filters, away from regions that are vignetted.  We then used \textsc{tweakback} to apply the alignment solution to the narrow-band FLCs so that all exposures in all filters are inter-aligned. This alignment is critical, as we subtract the continuum images from the narrow-band. There was no measurable offset between the full frame F502N and F547M images for NGC~3227 and NGC~6300 down to the $\sim$0.1 pixel level, as expected for exposures in the same orbit.

We then created the final drizzles using the aligned exposures with pixel scale of 0\farcs04 per pixel. The adopted \textsc{AstroDrizzle} parameters are provided in Table~\ref{tab:driz}, and the central portion of each galaxy in the resulting mosaics are shown in Figure~\ref{fig:images}. Finally, we post-processed the mosaics to replace NaNs and Infs using the \textsc{pixreplace} tool.

\subsubsection{\texorpdfstring{[\ion{O}{3}]}~~Images}\label{ssec:oiii}

We created the continuum-subtracted [\ion{O}{3}] emission-line images by subtracting off a scaled version of the F547M images from their corresponding F502N or FQ508N mosaics. Specifically, we converted the images in counts per second to cgs fluxes by multiplying the data arrays by the PHOTBW and PHOTFLAM parameters in the file headers. In the case of the FQ508N filter, we retrieved these values from \cite{Calamida2021}, with PHOTBW = 42.37 and PHOTFLAM = 2.9976e-18. Next, we scaled down the F547M flux to match the much narrower bandpass of the [\ion{O}{3}] filters. This is required so we are only subtracting the amount of continuum under [\ion{O}{3}] within the narrowband filter, and was accomplished by multiplying the F547M flux by the ratio of the PHOTBW for the narrow-band divided by PHOTBW for the F547M.

\begin{deluxetable}{ll}[htb!]
\tabletypesize{\normalsize}
\tablecaption{{\normalsize \textsc{AstroDrizzle} Parameters}}
\setlength{\tabcolsep}{0.3in}
\tablehead{
\colhead{Parameter} & \colhead{Value}}
\startdata
\textsc{updatewcs} & False \\
\textsc{stepsize} & 1 \\
\textsc{skysub} & True \\
\textsc{skymethod} & globalmin+match \\
\textsc{combine\_type} & minmed \\
\textsc{final\_wht\_type} & IVM \\
\textsc{final\_pixfrac} & 0.6 \\
\textsc{final\_units} & cps \\
\textsc{final\_scale} & 0.04 \\
\textsc{final\_rot} & 0.0 \\
\enddata
\tablecomments{The \textsc{AstroDrizzle} parameters used to create the continuum-subtracted [\ion{O}{3}] emission line images, which are shown in Figure~\ref{fig:images}. See \S\ref{ssec:driz} for a detailed description.}
\label{tab:driz}
\vspace{-3em}
\end{deluxetable}

In theory, this is the only correction that is required. However, in practice there are two complicating factors. First, the intrinsic SED of the AGN may not be perfectly flat even over a relatively small wavelength range. If the flux density increases at longer wavelengths, then the F547M will sample a higher flux than the narrow-band. Similarly, despite their close proximity in wavelength, the narrow-band filters will suffer minutely more from extinction, such that the continuum can be over-subtracted. Complicating this further is the fact that extinction is generally patchy and varies spatially across the host galaxy disk. We accounted for this by using a simple iterative subtraction scheme. First, we subtracted off the full amount of F547M flux that was scaled down to match the width of the narrow-band filters. We then calculated the mean, median, and standard deviation in several apertures not containing strong [\ion{O}{3}] emission. In general, the values were negative, so we iteratively scaled down the amount of continuum that was subtracted off until the median in these empty regions was effectively zero. In general, these manual scale factors ranged between 0.88 and 0.98, with an average value of 0.96 $\pm$ 0.03, corresponding to a 4\% reduction in the F547M flux to account for extinction in each galaxy.

In addition, the FQ508N images are subject to vignetting, where the edges of the field lose flux in a spatially varying manner. We accounted for this by fitting and subtracting a 2D gradient from the background of the resulting emission-line images. We used the \href{https://photutils.readthedocs.io/en/stable/background.html}{\textsc{photutils background2d}} module for this purpose with a filter size of 11~pixels. This correction reduces the median background level by a factor of $\gtrsim$10 from $\approx$ 4$\times$10$^{-19}$ to $\approx$ 3$\times$10$^{-20}$ erg s$^{-1}$ cm$^{-2}$ arcsec$^{-2}$, which is substantially smaller than the [\ion{O}{3}] fluxes we will measure in our analysis.

We note that these observations are designed to encompass the [\ion{O}{3}] $\lambda$5007~\AA~emission line; however, due to the range of galaxy redshifts, some images also contain contributions from the [\ion{O}{3}] $\lambda$4959~\AA~component. This is a common occurrence in narrowband surveys \citep{Storchi-Bergmann2018}, and the line is always 1/3 the brightness of [\ion{O}{3}] $\lambda$5007~\AA~based on atomic transition probabilities. This provides additional signal-to-noise in the [\ion{O}{3}] images, but can cause minor complications for analyses that compare with models or other observations that are strictly based on the [\ion{O}{3}] $\lambda$5007~\AA~line. In this case, [\ion{O}{3}] $\lambda$4959~\AA~is present in the F502N filter at $z =$~0 and reaches the peak of the filter transmission curve at $z \approx$ 0.004, meaning it is present in our images of NGC~3227 and NGC~6300. In the case of the FQ508N filter, it reaches the peak of the filter transmission curve at $z \approx$ 0.015, meaning it is in our images of NGC~1667 and UGC~1395, and weakly present, if at all, in the remaining images.

The WFC3 data can be accessed using the DOI: \dataset[10.17909/np2s-gd34]{https://doi.org/10.17909/np2s-gd34}. We also provide our calibrated mosaics for use by the community as MAST High Level Science Products (HLSPs) using the DOI: \dataset[10.17909/7ccz-mc93]{https://doi.org/10.17909/7ccz-mc93} and online at \url{https://archive.stsci.edu/hlsp/nlr-agn/}.

\begin{figure*}[ht]
\centering
\includegraphics[width=0.85\textwidth, trim={.2em 0em 0em 0em}, clip]{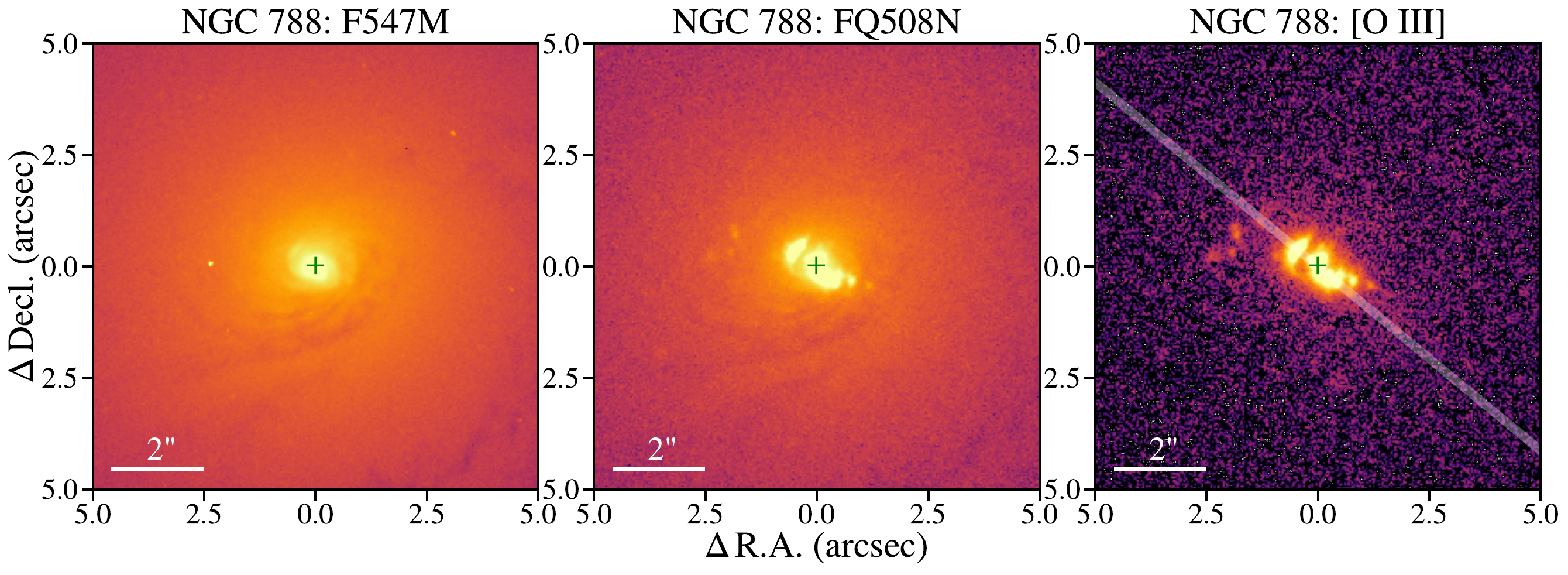}\\
\includegraphics[width=0.85\textwidth, trim={.2em 0em 0em 0em}, clip]{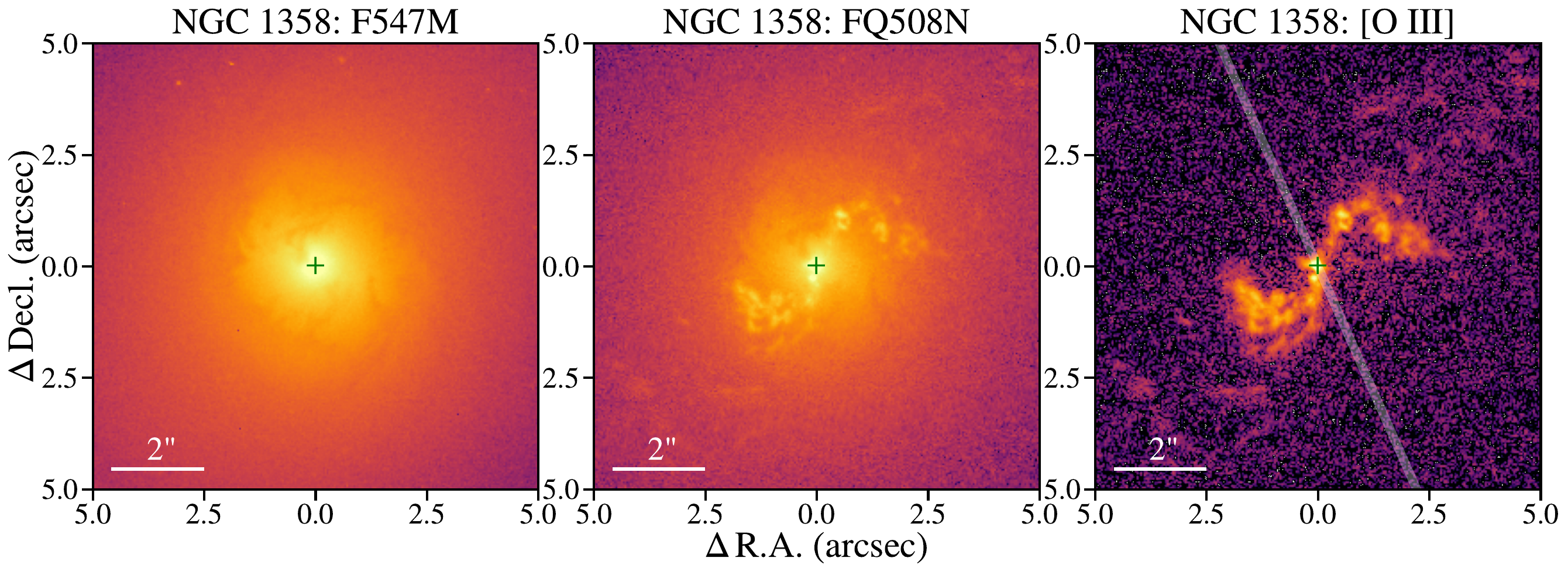}\\
\includegraphics[width=0.85\textwidth, trim={.2em 0em 0em 0em}, clip]{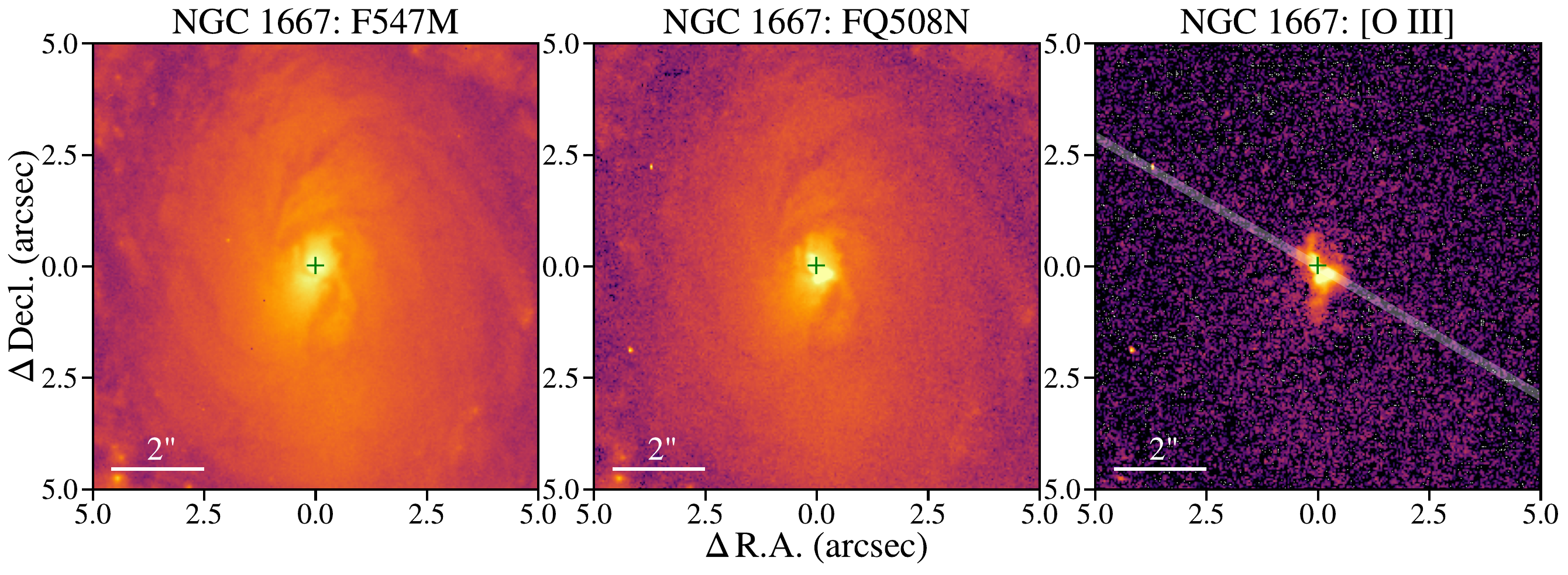}\\
\includegraphics[width=0.85\textwidth, trim={.2em 0em 0em 0em}, clip]{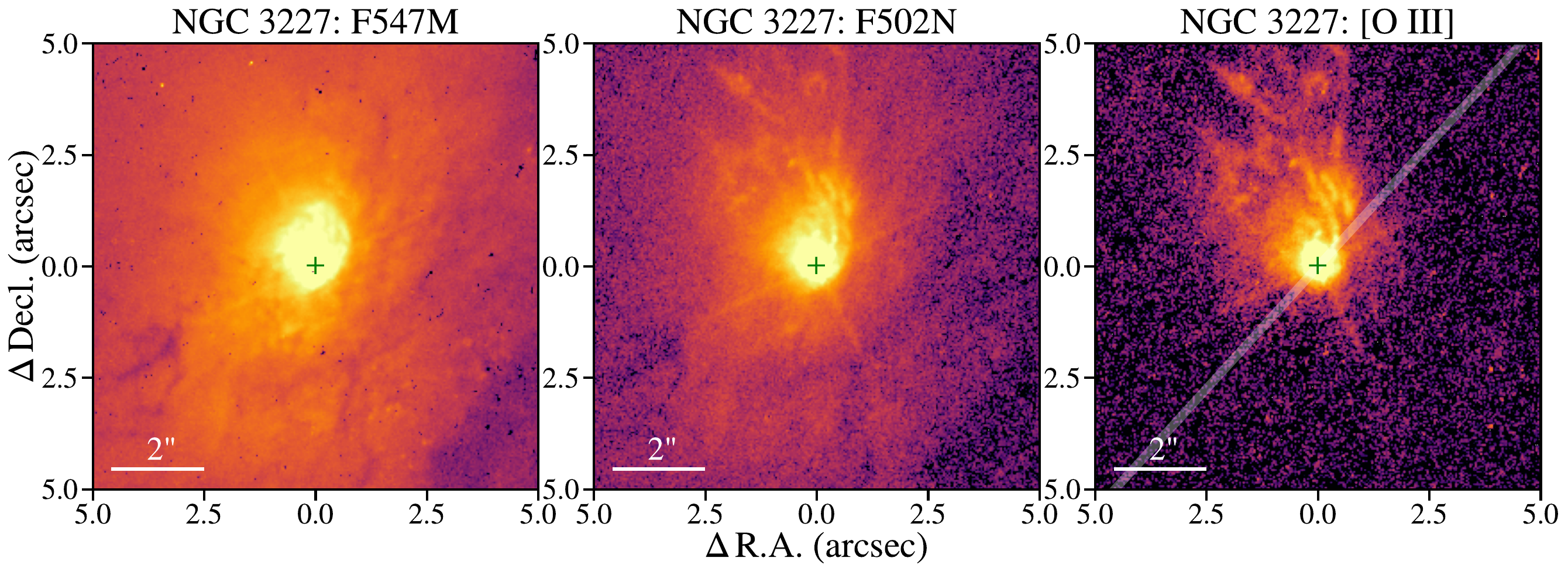}\\
\caption{The HST WFC3 F547M (left), FQ508N or F502N (center) and continuum-subtracted [\ion{O}{3}] images (right) for the 12 AGN with new [\ion{O}{3}] images. The location of the continuum peak from the F547M image is shown by a green cross. The archival HST STIS long-slit position angles are denoted by white shaded lines overlaid on the [\ion{O}{3}] images. All panels show the inner 10$\arcsec$$\times$10$\arcsec$ for each galaxy, with north up and east to the left.}
\label{fig:images}
\end{figure*}
\addtocounter{figure}{-1}
\begin{figure*}[ht!]
\centering
\includegraphics[width=0.85\textwidth, trim={.2em 0em 0em 0em}, clip]{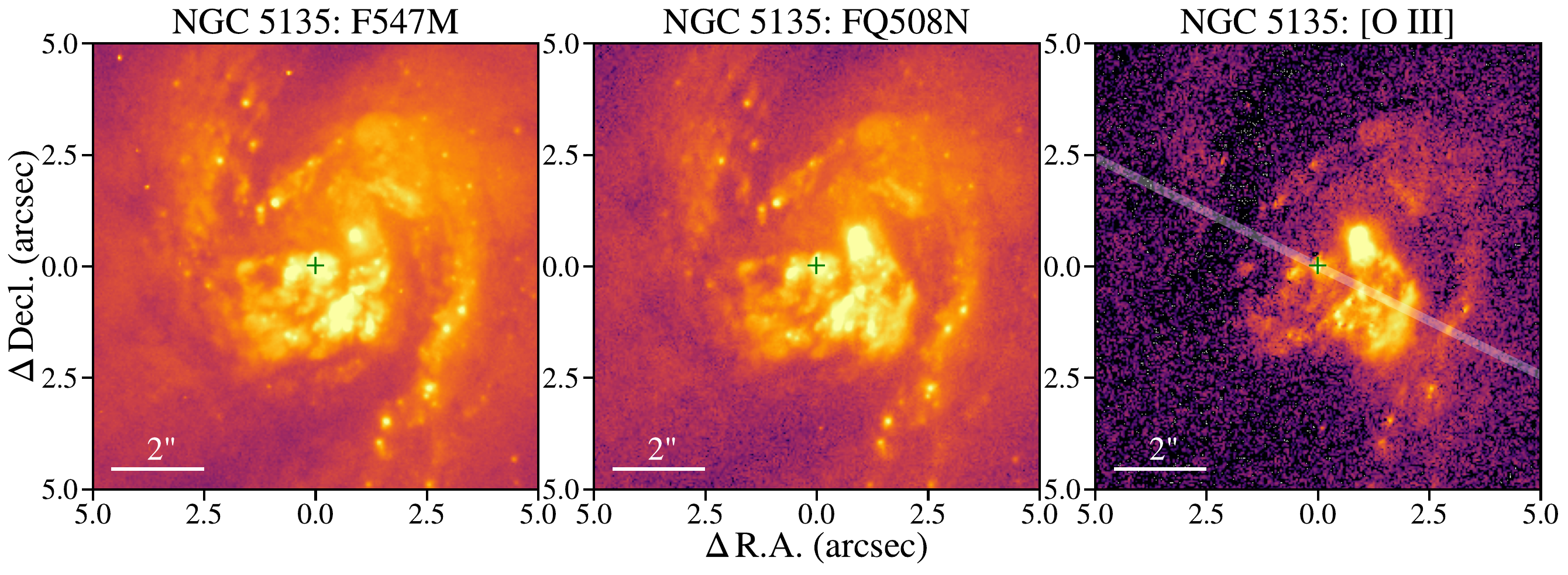}\\
\includegraphics[width=0.85\textwidth, trim={.2em 0em 0em 0em}, clip]{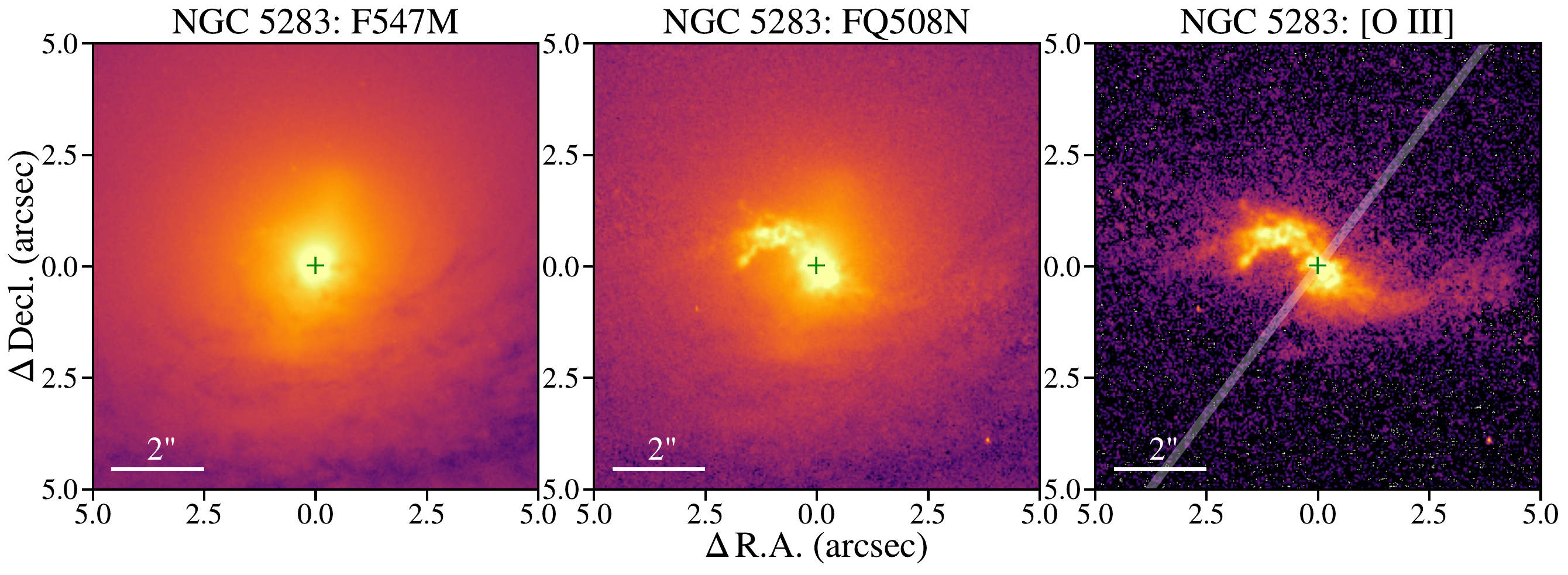}\\
\includegraphics[width=0.85\textwidth, trim={.2em 0em 0em 0em}, clip]{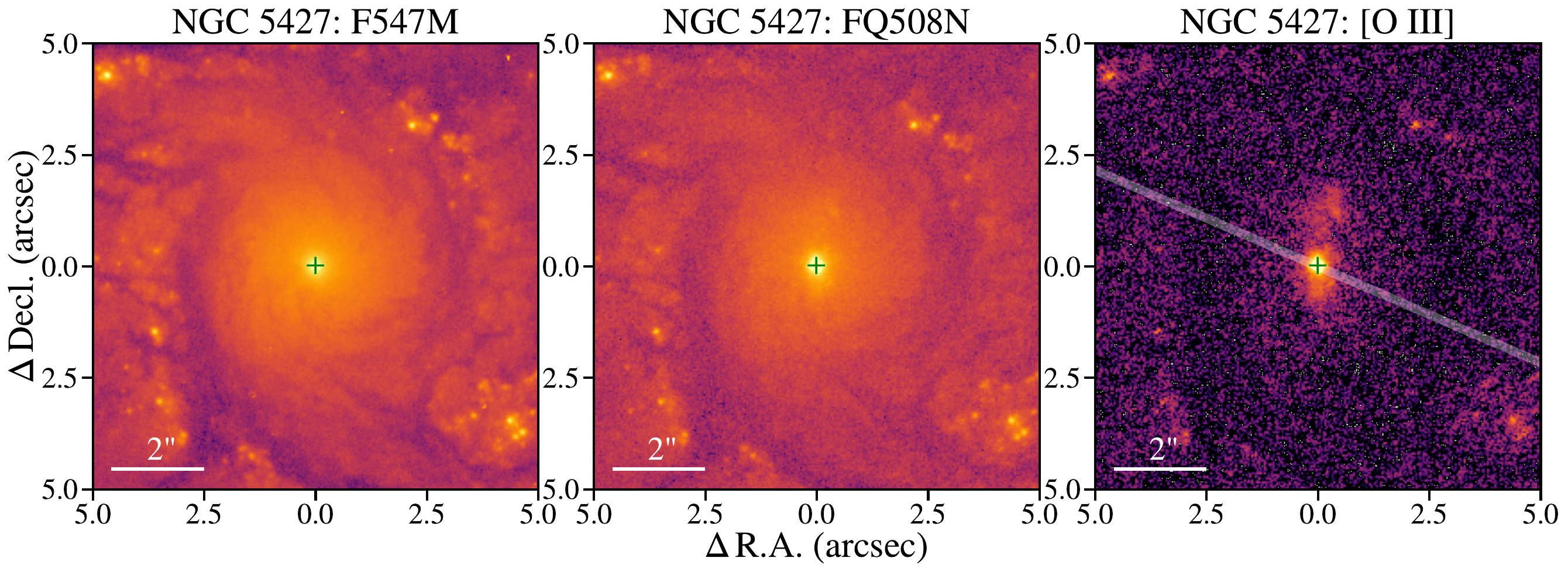}\\
\includegraphics[width=0.85\textwidth, trim={.2em 0em 0em 0em}, clip]{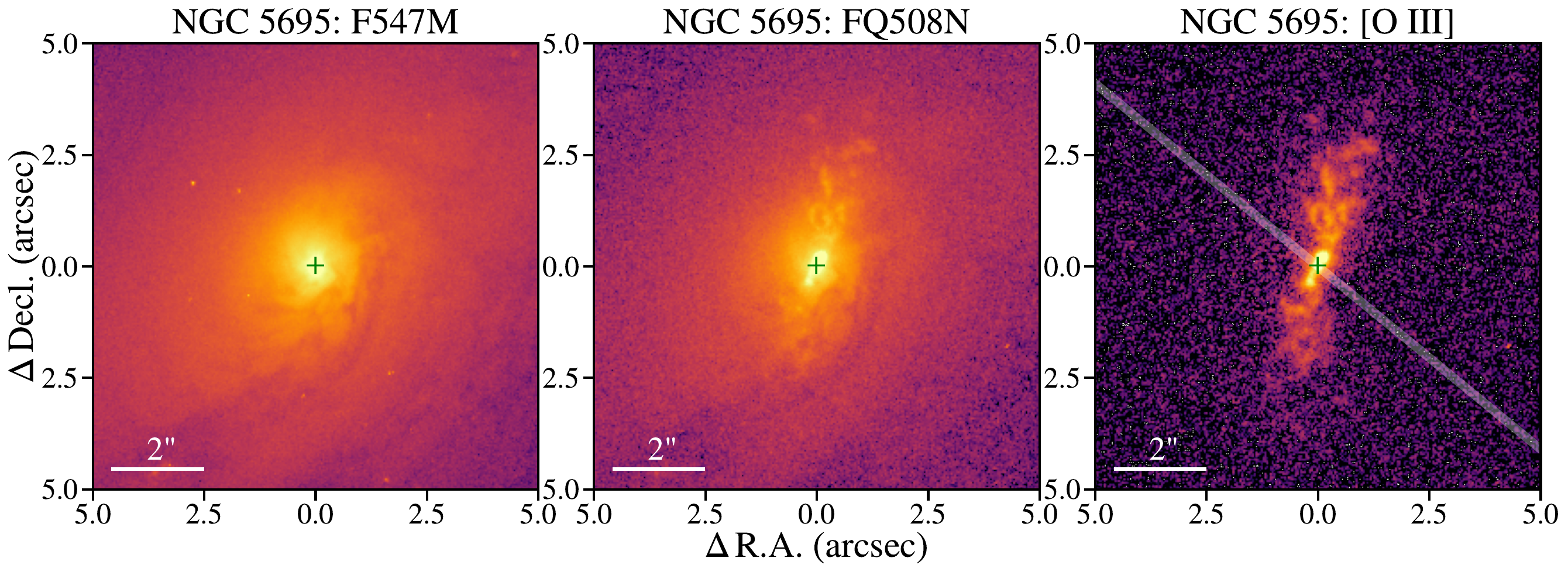}\\
\caption{\textit{continued.}}
\end{figure*}
\addtocounter{figure}{-1}
\begin{figure*}[ht!]
\centering
\includegraphics[width=0.85\textwidth, trim={.2em 0em 0em 0em}, clip]{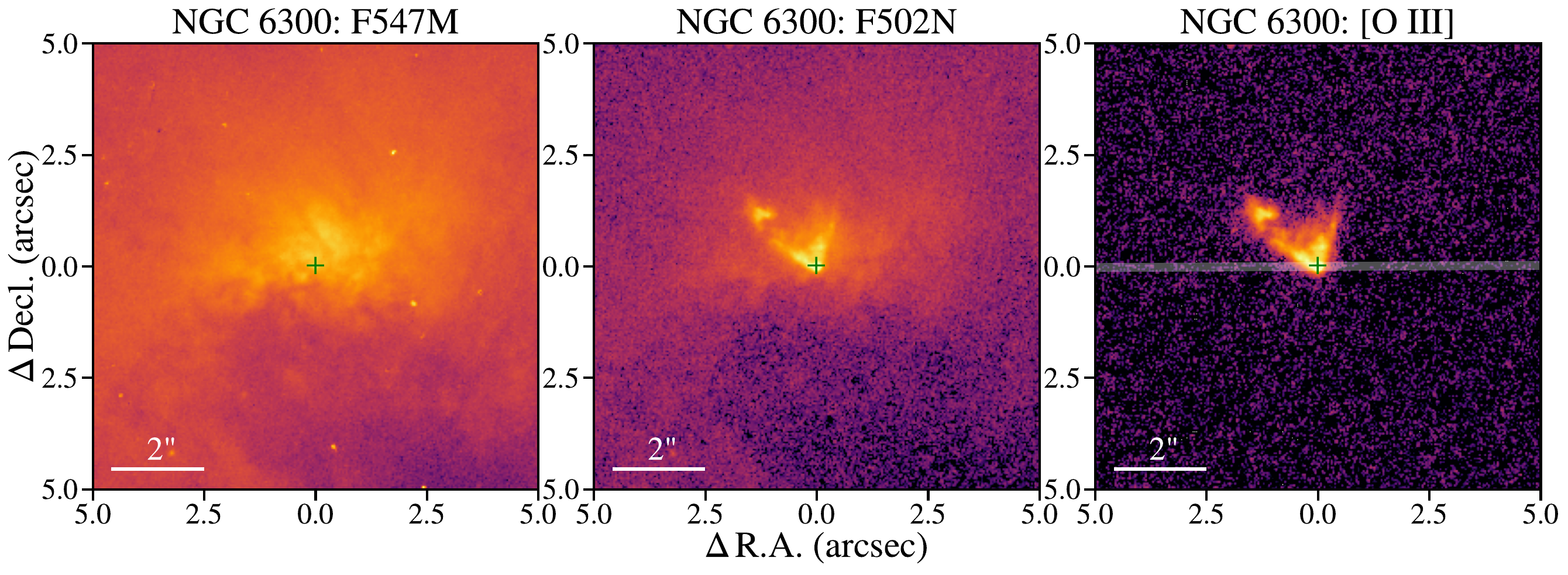}\\
\includegraphics[width=0.85\textwidth, trim={.2em 0em 0em 0em}, clip]{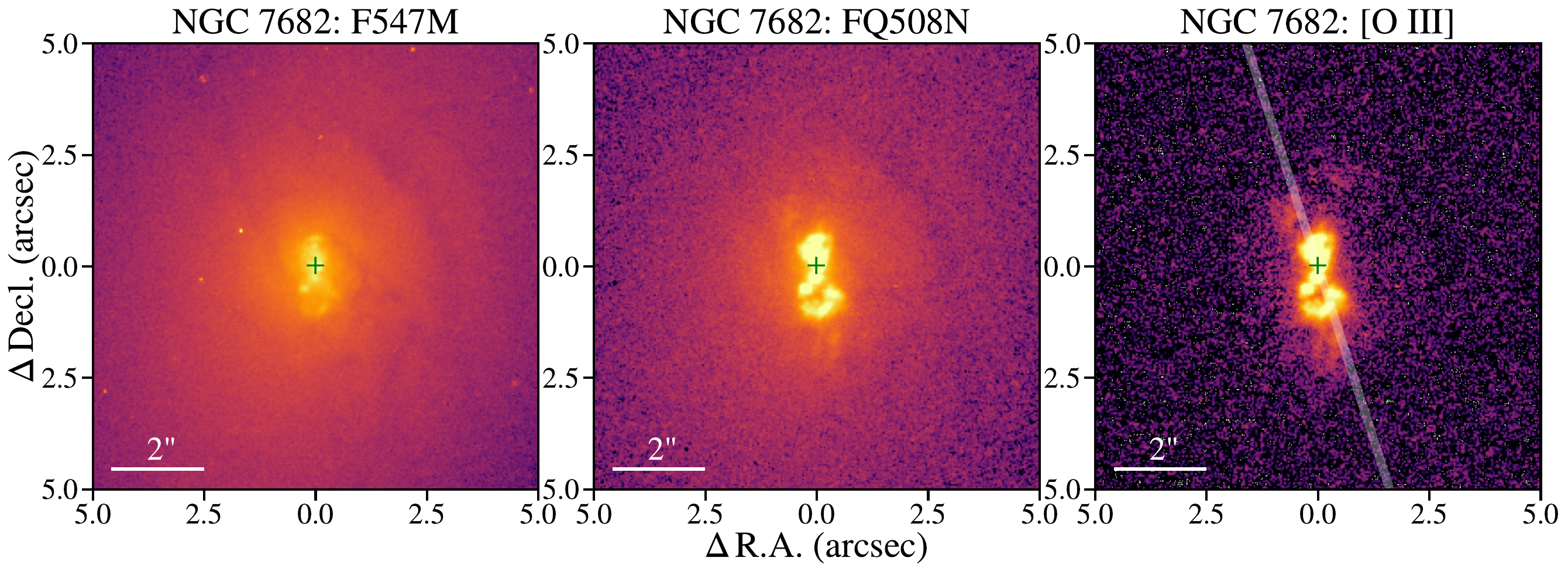}\\
\includegraphics[width=0.85\textwidth, trim={.2em 0em 0em 0em}, clip]{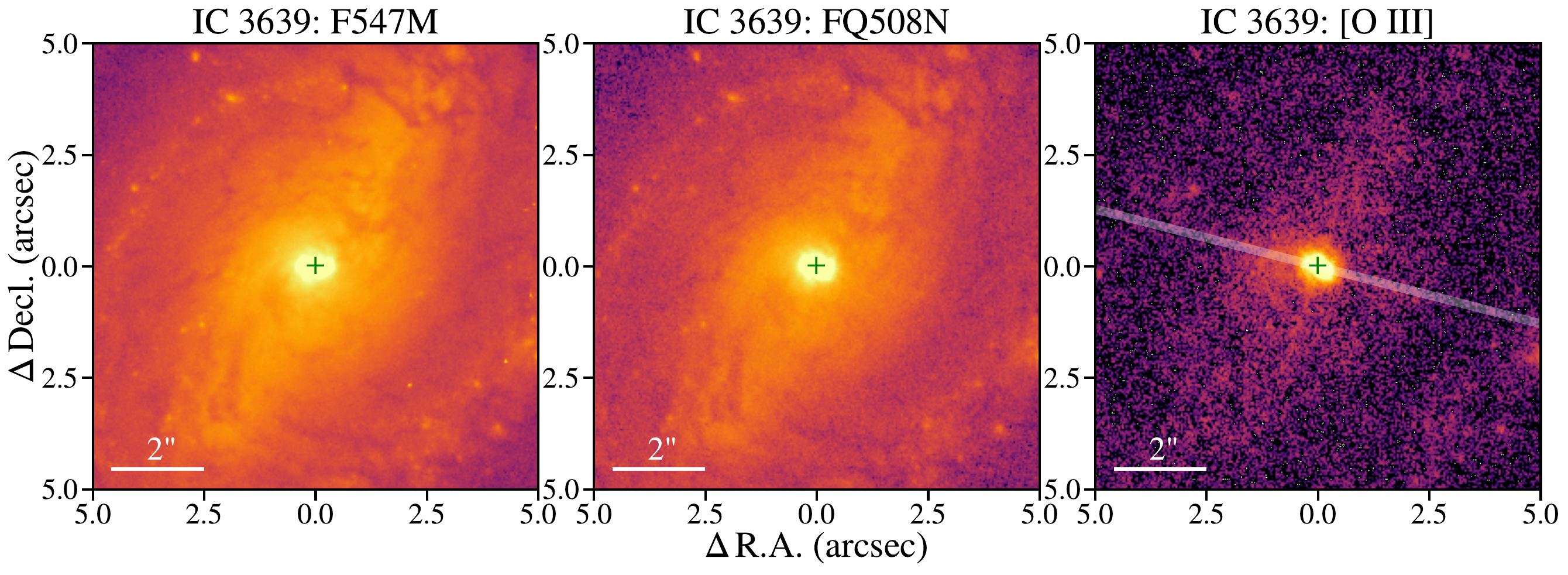}\\
\includegraphics[width=0.85\textwidth, trim={.2em 0em 0em 0em}, clip]{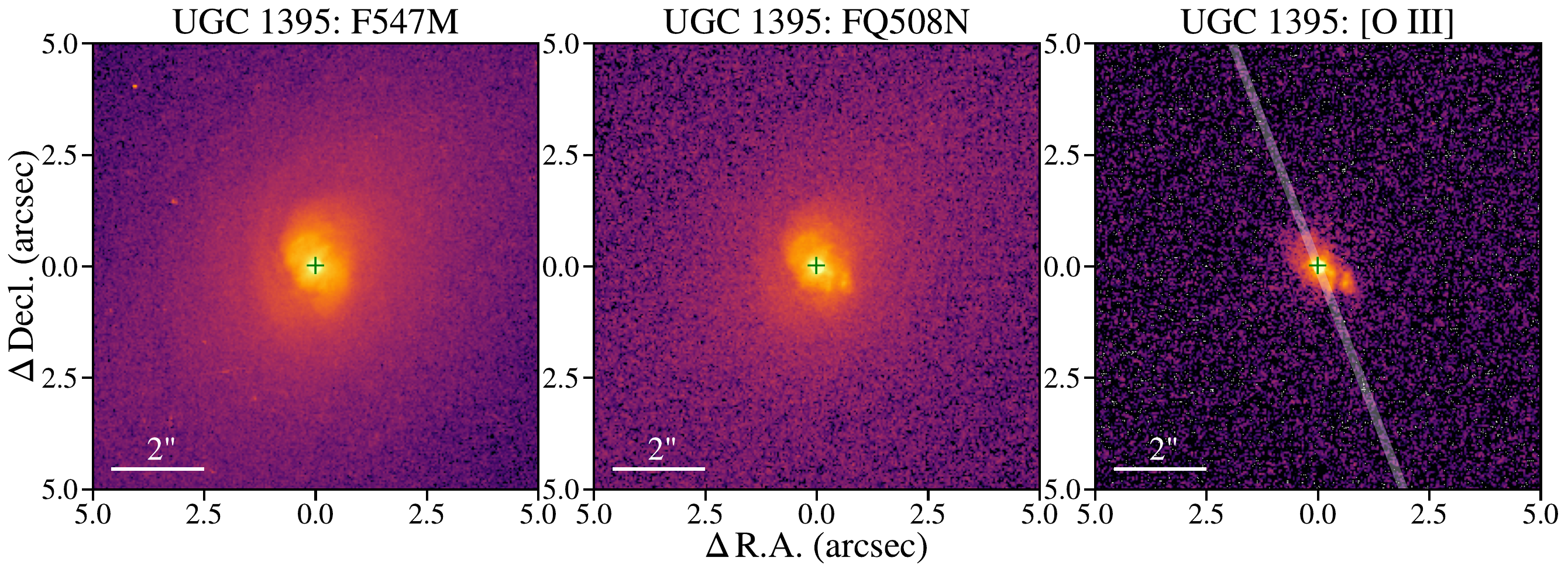}\\
\caption{\textit{continued.}}
\end{figure*}
\clearpage

\subsection{HST Spectroscopy} \label{subsec:Hspec}

The spectral observations for this work use the HST STIS with a 52\arcsec x 0\farcs2 slit. We use the calibrated data originally obtained for \cite{Fischer2017} from the Mikulski Archive at the Space Telescope Science Institute (MAST) portal. Information on the individual exposures can be found in Table \ref{tab:obs}. To investigate the NLR in each galaxy, it is necessary to target prominent AGN emission lines, so we use spectra taken with the G430L ($\lambda / \Delta \lambda$ $\approx$ 500) and G750M ($\lambda / \Delta \lambda$ $\approx$ 5900) gratings to capture the [\ion{O}{3}] $\lambda\lambda$4959,5007~\AA~and \halpha$\lambda$6563~\AA~lines, respectively. The medium dispersion G750M grating is used to determine the ionized gas kinematics, while the lower dispersion G430L grating provides emission line diagnostics for later photoionization models.

\begin{figure*}[htb]
\centering
\includegraphics[width=\textwidth]{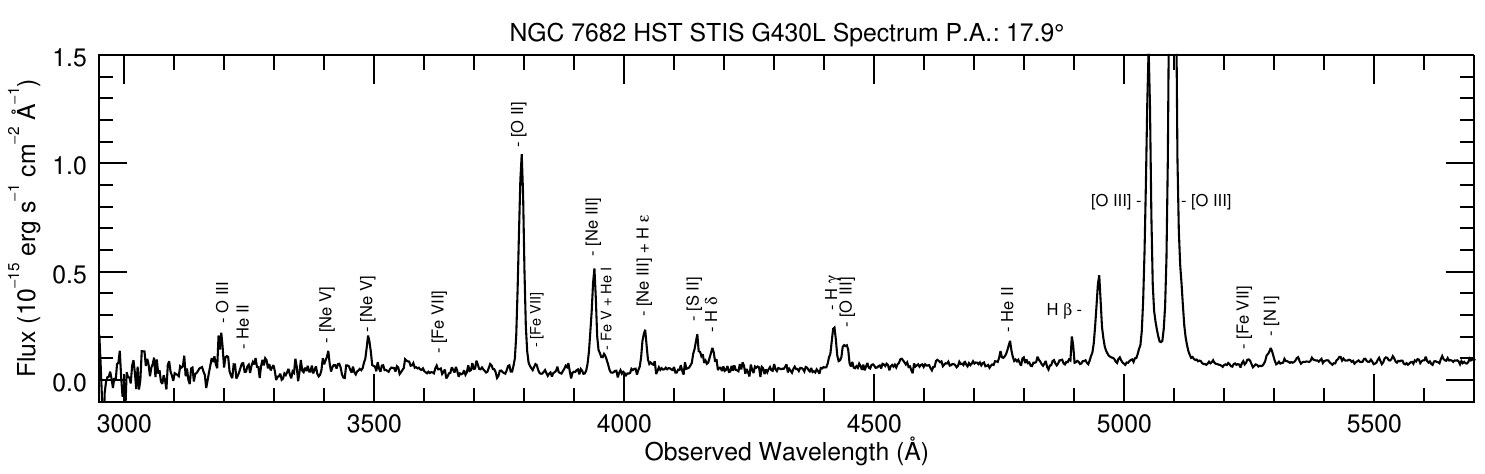}\\
\includegraphics[width=\textwidth]{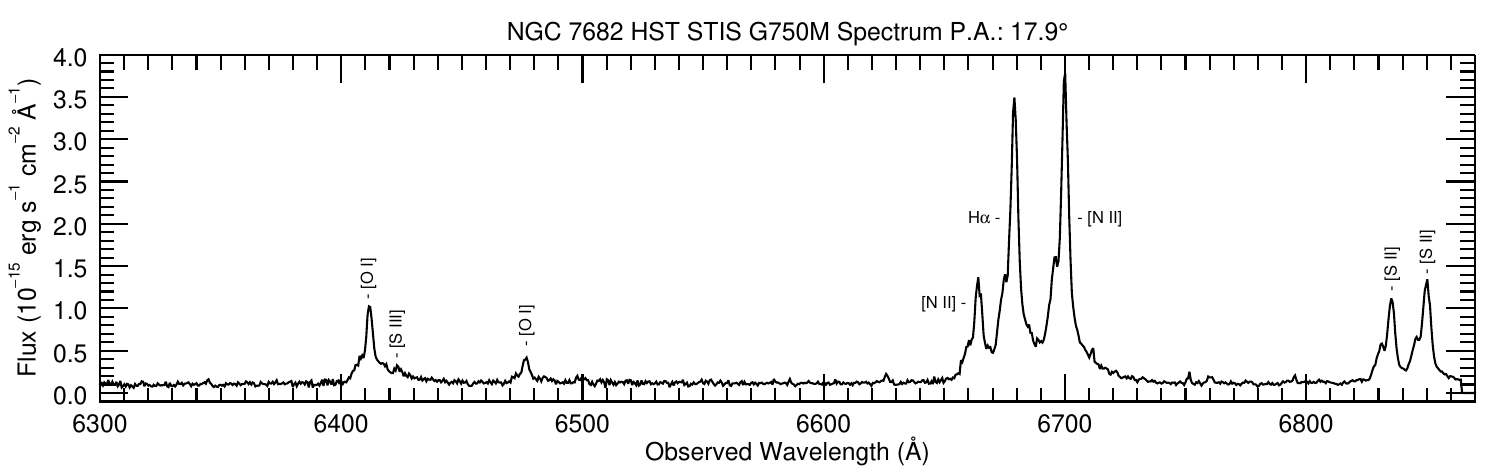}\\
\caption{Example HST STIS archival spectra of NGC~7682, showing the prominent emission lines useful for AGN analysis. The top spectrum shows the G430L filter and the bottom shows the G750M filter. Each plot is a summation of 20 bins along the slit (about 1\arcsec) centered on the AGN nucleus.}
\label{fig:spec}
\end{figure*}

An example spectrum of NGC~7682, using both gratings, is shown in Figure \ref{fig:spec}, with labels affixed to key emission lines. This figure denotes a clear flux difference between the prominent [\ion{O}{2}], [\ion{O}{3}], H$\alpha$, and [\ion{N}{2}] lines, compared to the rest of the distinguishable emission features. The elevated flux values of these lines above the continuum allow for better differentiation of the individual components comprising the ionized gas. We note here that the spectra in Figure \ref{fig:spec} are combinations of a number of rows across the slit, centered over the nucleus of the target, and not of a single row's spectrum.

\section{Analysis} \label{sec:analysis}

\subsection{\texorpdfstring{[\ion{O}{3}]}~~Image Analysis} \label{subsec:ImAn}

Using the [\ion{O}{3}] images shown in the right column of Figure~\ref{fig:images}, we determine how well the archival spectral slits (the overlaid white regions) match up to the position of the visible structure, as well as measure the luminosity ($L_\mathrm{[O~III]}$) and the radial extent ($R_\mathrm{[O~III]}$) of the ionized gas. With these images we are able to estimate the full extent of the [\ion{O}{3}] gas without any dependence on the position angle or resolution of the long slit spectra. It is important to note that using the imaging alone we cannot distinguish between outflow and other [\ion{O}{3}] kinematic variations, such as rotation. We describe the use of spectral observations for outflow identification and disentanglement in Section \ref{subsec:outflow}.

\begin{figure}
\vspace{0.5em}
\centering
\includegraphics[width=\columnwidth]{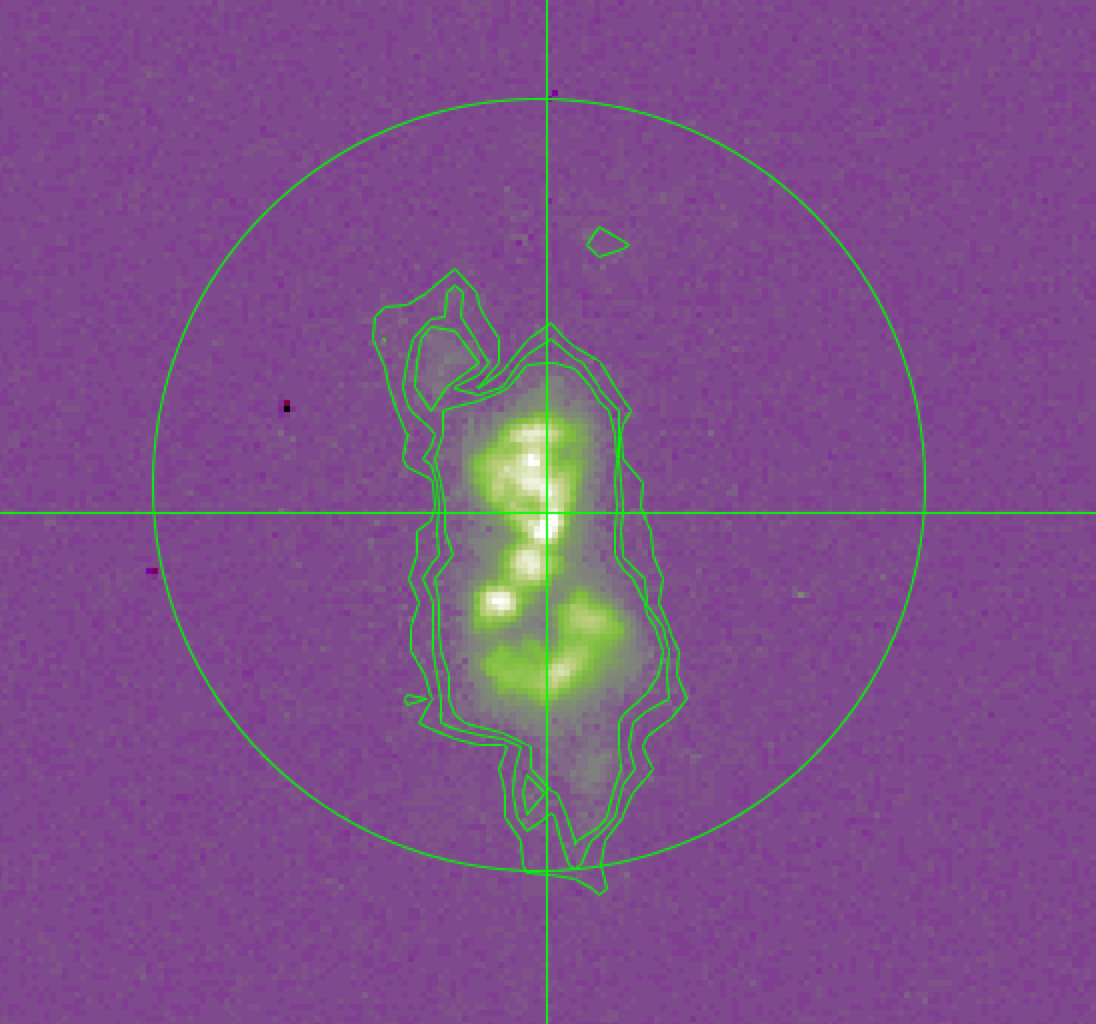}
\caption{A 3\farcs6 $\times$ 3\farcs4 [\ion{O}{3}] image of NGC~7682 with three green contour lines surrounding the ionized gas structure. The middle contour is the measured background 3$\sigma$ value of 2.3 $\times$ 10$^{-18}$ erg s$^{-1}$ cm$^{-2}$ arcsec$^{-2}$ and the two additional contours on either side represent offsets of 1$\sigma$ in continuum brightness. The cross indicates the brightness centroid used as the zero point for radial measurements, and the green circle denotes the radial extent of the middle contour. The recorded extent of the emission is an average of all three contour extents measured this way, with the luminosity being an average of the value contained within each circle.}
\label{fig:imext}
\end{figure}

We measure the background of each exposure to determine a flux level that is 3 standard deviations (3$\sigma$) above the continuum brightness level. This value (given in column 2 of Table~\ref{tab:modelin}) is used to generate a contour, which is overlaid on the image to encapsulate the bounds of the [\ion{O}{3}] emission. A threshold of 3$\sigma$ ensures that the extent of the ionized gas is only measured out to where it is statistically significant. We then measure the farthest point on the contour line radially from the AGN to find the first extent estimate. To minimize errors, we place two additional contours, corresponding to $\pm$1$\sigma$, above and below the original contour level. The mean of the three contour extents in pixels is converted to parsecs using the spatial scale for each object, and results in $R_\mathrm{[O~III]}$. Similarly, the mean of the luminosities within each contour results in the value for $L_\mathrm{[O~III]}$, after being scaled based on each target's distance. An example of this process is shown in Figure \ref{fig:imext}.

We explore the biases introduced by the inhomogeneous depth between our images and other archival data sets by adopting an additional threshold for this analysis. Instead of individual flux thresholds, we apply a contour limit of 2.034$\times$ 10$^{-17}$ erg s$^{-1}$ cm$^{-2}$ arcsec$^{-2}$to every exposure and measure the radial extent ($R_\mathrm{[O~III],fixed}$) and luminosity $L_\mathrm{[O~III],fixed}$ in the same way. This 3$\sigma$ threshold is the average of several values used in the analysis of previous archival data sets \citep{Schmitt2003, Storchi-Bergmann2018, Trindade2021} and is adopted in order to test our method for observational effects.

\subsection{AGN Host Galaxy Parameters} \label{subsec:morph}

Our kinematic outflow models make the assumption that the outflows of ionized gas are purely radial and constrained to the plane of the galactic disk. This assumption is based on the finding that most AGN NLRs show ambiguous or complex kinematics that are not easily fit with a simple biconical outflow model \citep{Fischer2013}, and yet are often consistent with radially outflowing, ionized gas in the intersection between the bicone and the galactic plane \citep{Fischer2017, Fischer2018}. Further evidence for this assumption comes from examination of Figure \ref{fig:images}, which shows a wide variety of loops, arcs, and spirals indicative of the original disk gas that was ionized by the AGN. Adopting this outflow model means that we must account for projection effects in the observations in order to accurately analyze the kinematics of the gas in the disk. To do so, we must first determine each galaxy's morphological parameters, such as the inclination and the position angle of the major axis. 

We begin with the assumption that spiral galaxy disks are intrinsically circular and have an elliptical shape due to their orientation relative to our line of sight. Accurate geometric models require values of position angle and ellipticity as functions of distance from the nucleus. We calculate these disk values by fitting isophotal ellipses to images of the entire galaxy on the sky. For all but three of our targets, we use ground-based images obtained through the Pan-STARSS-1 Image Access Portal, selecting the \textit{i}-band with high signal-to-noise (S/N) that traces the majority of the older stellar populations within the disk and bulge. The field of view for each image is 100\arcsec $\times$ 100\arcsec, except for NGC~1358, which is 125\arcsec $\times$ 125\arcsec~so as to fit the entirety of the suspected disk. For NGC~3393 and NGC~7682 we use the \textit{z} and \textit{g}-bands respectively, as their \textit{i}-band images contain artifacts that hinder the modeling process. NGC~5643, NGC~6300, and IC~3639 are southern sky objects, and so require images from separate sources, listed in Table~\ref{tab:modeling} \citep{DSS,6300image, ICimage}.

We use the \textsc{Ellipse} package from Astropy \citep{AstropyCollaboration2013, AstropyCollaboration2018, AstropyCollaboration2022} to fit concentric ellipses to the shape of each galaxy, beginning with a small, initial ellipse over the brightness centroid and moving outward in bins of equivalent surface brightness. Each iteration places isophotal ellipses of increasing radii out to the maximum extent of the target's morphology, which is determined by the flux differential calculated between each step. The position angle (PA), surface brightness (SB), and ellipticity (f, defined as (1 - $b / a$)) for each ellipse are given as a function of radius. The outputs are taken as an average of the outer 10$\%$ of the ellipses in order to account for uncertainties in flux estimates at the galaxy ``edges". These values are then used to calculate the orientations and the inclinations (i = arccos(1 - f)) of the disks. This method does not take into account any warping or varying inclination of the inner disk containing the NLR.

\begin{figure*}[ht]
\centering
\includegraphics[width=0.95\textwidth,clip]{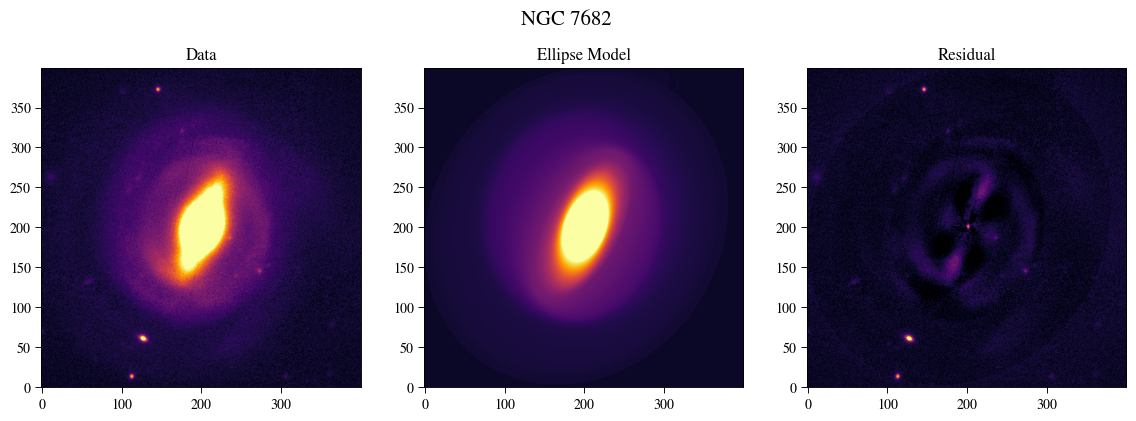}\\
\caption{Elliptical modeling of the ground-based image of NGC~7682 on a false color scale to highlight detailed structure. The left panel shows the ground-based imaging, the center panel shows the reconstructed model of the galaxy using the isophotal ellipses fit to the image, and the third panel shows the residual from the model subtraction. Each image shows a 100\arcsec $\times$ 100\arcsec~field of view with the axes given in pixels}
\label{fig:models}
\vspace{1em}
\end{figure*}

\begin{figure}[h!]
\centering
\includegraphics[width=\columnwidth]{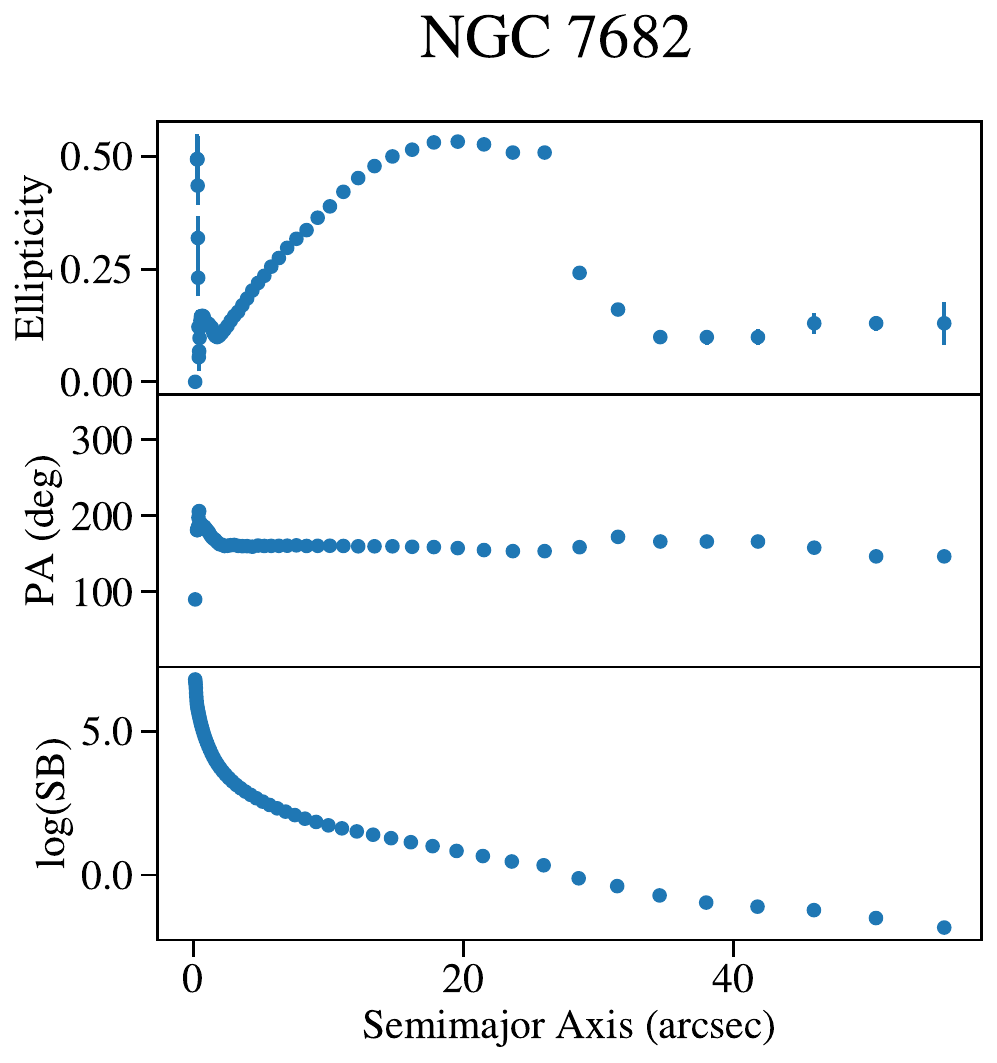}
\caption{Example output of NGC~7682 from the isophotal ellipse modeling. The x-axis gives the semimajor axis of the isophotal ellipse in arcseconds, while the y-axis shows the model ellipticity, major axis position angle in degrees, and surface brightness in log counts.}
\label{fig:modeloutputs}
\end{figure}

An example result of our elliptical fitting for NGC~7682 is shown in Figure \ref{fig:models}. The three panels from left to right are: the initial ground-based image, the model fit using the measured isophotal ellipses, and the residual image after model subtraction. Some artifacts remain in the residuals due to non-elliptical features in the images; however, the general shapes at each radial bin are well represented by the models. The modeling outputs of the position angles and ellipticities are provided in Table~\ref{tab:modeling}, while Figure~\ref{fig:modeloutputs} shows the output values from the example of NGC~7682, presenting the measured properties as a function of semi-major axis size. The remaining model images and model output results are provided in Appendix~\ref{app:ellipse}.

\begin{deluxetable}{lcccc}
\vspace{-0.5em}
\tabletypesize{\normalsize}
\setlength{\tabcolsep}{0.085in} 
\tablecaption{Isophotal Modeling}
\tablehead{
\colhead{Catalog} & \colhead{Image} & \colhead{Image} & \colhead{PA} & \colhead{Ellipticity\vspace{-0.5em}}\\
\colhead{Name} & \colhead{Source} & \colhead{Filter} & \colhead{(deg)} & \colhead{(unitless)\vspace{-0.5em}}\\
\colhead{(1)} & \colhead{(2)} & \colhead{(3)} &\colhead{(4)} & \colhead{(5)}
}
\startdata
NGC 788  & PanSTARRS-1 & i & 116 & 0.15 \\
NGC 1358 & PanSTARRS-1 & i & 191 & 0.31 \\
NGC 1667 & PanSTARRS-1 & i & 163 & 0.30 \\
NGC 2273 & PanSTARRS-1 & i & 245 & 0.40 \\
NGC 3393 & PanSTARRS-1 & z & 230 & 0.17 \\
NGC 5135 & PanSTARRS-1 & i & 124 & 0.46 \\
NGC 5283 & PanSTARRS-1 & i & 118 & 0.14 \\
NGC 5347 & PanSTARRS-1 & i & 105 & 0.42 \\
NGC 5427 & PanSTARRS-1 & i & 191 & 0.18 \\
NGC 5643 & DSS & i & 158 & 0.33 \\
NGC 5695 & PanSTARRS-1 & i & 139 & 0.31 \\
NGC 6300 & ESO & r & 144 & 0.37 \\
NGC 7682 & PanSTARRS-1 & g & 160 & 0.32 \\
IC 3639  & CTIO & i & 52 & 0.16 \\
UGC 1395 & PanSTARRS-1 & i & 148 & 0.35 \\
\enddata
\tablecomments{Columns are (1) target name, (2) database from which the original ground-based image was retrieved, (3) filter of the image (4) major axis position angle, and (5) ellipticity (1 - $b / a$).}
\label{tab:modeling}
\vspace{-2em}
\end{deluxetable}

\subsection{Spectral Fitting} \label{subsec:SpecFit}

We model the kinematics of our targets using archival HST STIS long slit spectra. Every 2 rows of each slit are binned in the cross-dispersion direction to increase the S/N and match HST's spatial resolution. We utilize a fitting routine developed by \cite{Fischer2013}, that fits Gaussians to specified emission lines in each bin. While a Gaussian shape does not perfectly describe a line profile, a superposition of several Gaussians has been shown to accurately model the line profiles of prominent emission features \citep{Revalski2018b}. Using a Bayesian statistical algorithm from MultiNest \citep{multinest}, the Gaussians are fit to each emission line to determine the number of kinematic components comprising the line profile at each bin across the slit. 

The simplest configuration with the least number of Gaussians is prioritized, as long as the algorithm determines additional components are statistically irrelevant (see \citealp{Fischer2017} and \citealp{Falcone2024} for an explanation of the Bayesian likelihood calculations). We limit the number of components to three, because attempting to place four or more components consistently yields results below the likelihood threshold for targets in our redshift range. The Gaussian characteristics (height above the continuum, width, and position) are allowed to vary, while a maximum and minimum wavelength range is fixed for each emission line. The program adds additional components until it determines that additional Gaussians no longer substantially improve the best-fit solution. The resulting output gives the wavelength centroid, width, and peak flux of each Gaussian component, for each emission line.

\begin{figure}
\vspace{0.5em}
\centering
\includegraphics[width=\columnwidth]{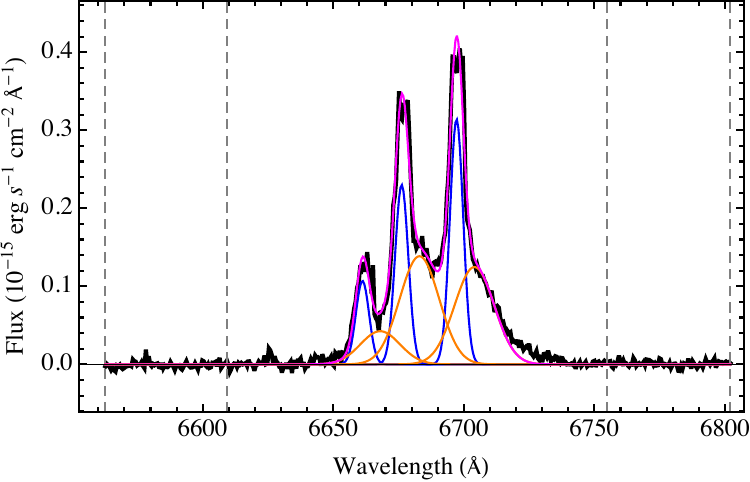}
\vspace{0.5em}
\caption{An example of the Gaussian fitting results for the H$\alpha$ $\lambda$6563~\AA~and [N~II] $\lambda\lambda$6548,6583~\AA~emission lines for one bin along the G750M observation for NGC~7682. Observed wavelengths are on the x-axis and flux is on the y-axis. Two sets of gray vertical dashed lines indicate the regions used to determine the continuum level (horizontal gray line). The thick black line indicates the original spectral data, while the magenta line shows the combined fit for all emission line components. In this case, the fitting routine found that two kinematic components were optimal to fit each line profile: a wider, redshifted component with a lower peak flux (orange) and a narrower, blueshifted component with a higher peak flux (blue).}
\vspace{1em}
\label{fig:gauss}
\end{figure}

Figure \ref{fig:gauss} shows an example fit for the H$\alpha$ $\lambda$6563~\AA~and [N~II] $\lambda\lambda$6548,6583~\AA~lines for one bin of the NGC~7682 G750M observation. In this example, each line profile is comprised of two separate components, with distinct positions, widths, and heights. The horizontal positions of the Gaussians, relative to the systemic velocity of the target, are used to calculate the velocity of the gas, while the widths are used to calculate the velocity dispersion. The continuum flux level is the average flux contained within the vertical dashed line sections to either side of the emission lines. The heights of the Gaussians above the continuum line give the peak fluxes. The Gaussian peaks must reach a minimum S/N $>$ 3 above the continuum for a component to be accepted as physically realistic.

The Gaussians from the \halpha fits are used as templates to fit the rest of the emission lines in both gratings, due to the G750M spectra having higher resolution and smaller residuals compared to the G430L. Using scaled \halpha profiles preserves intrinsic velocity widths and centroids, needing only to account for the line spread function of the instrument gratings to fit the rest of the spectra. Through this process, the less prominent emission lines can be characterized for use in future photoionization modeling.

\subsection{Kinematics}

We create kinematic plots of the gas components using the output parameters from the spectral fitting routine. We fit spectra from both HST gratings, however we only use the \halpha component fits for this analysis because of the increased spatial resolution they provide over the [\ion{O}{3}] results. The \halpha kinematics essentially trace the same patterns and extents as the [\ion{O}{3}], but are able to distinguish more gas components for the emission lines, and therefore more accurate velocities and FWHM values across the slits. 

\begin{figure*}[ht!]
\centering
\subfigure{
\includegraphics[width=0.325\textwidth]{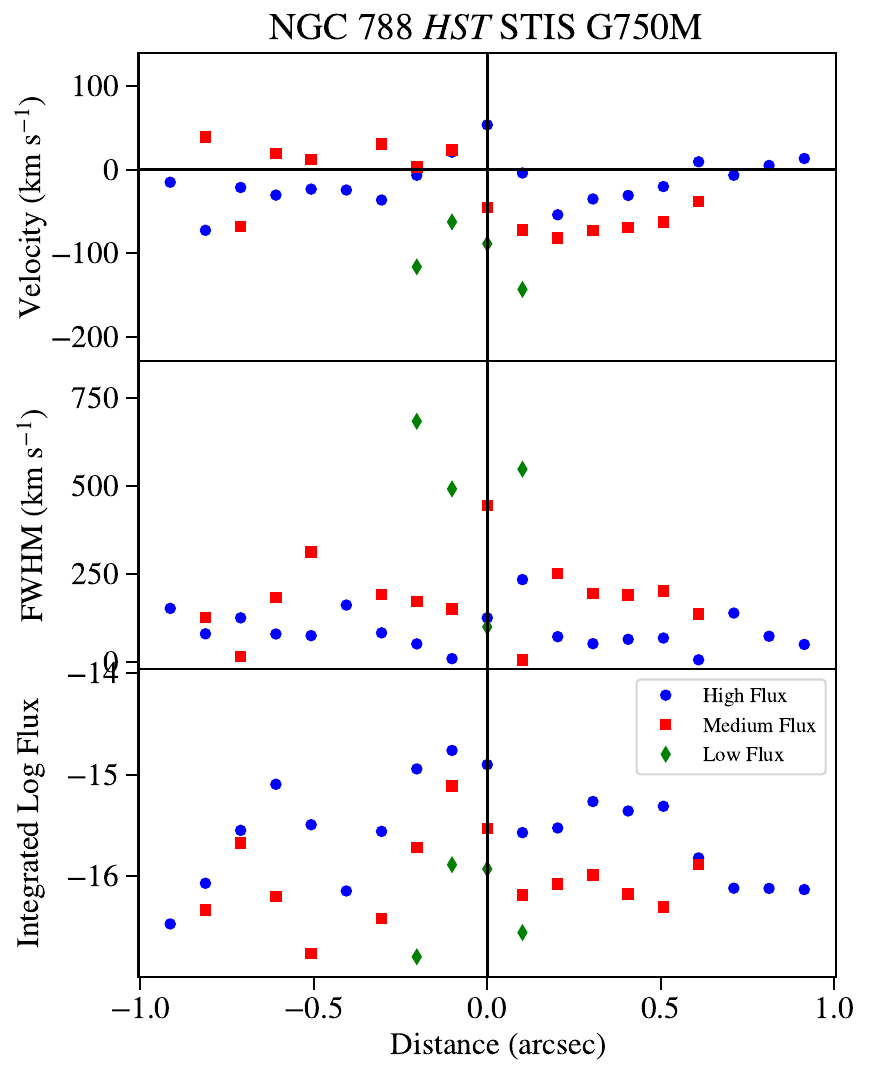}}
\subfigure{
\includegraphics[width=0.325\textwidth]{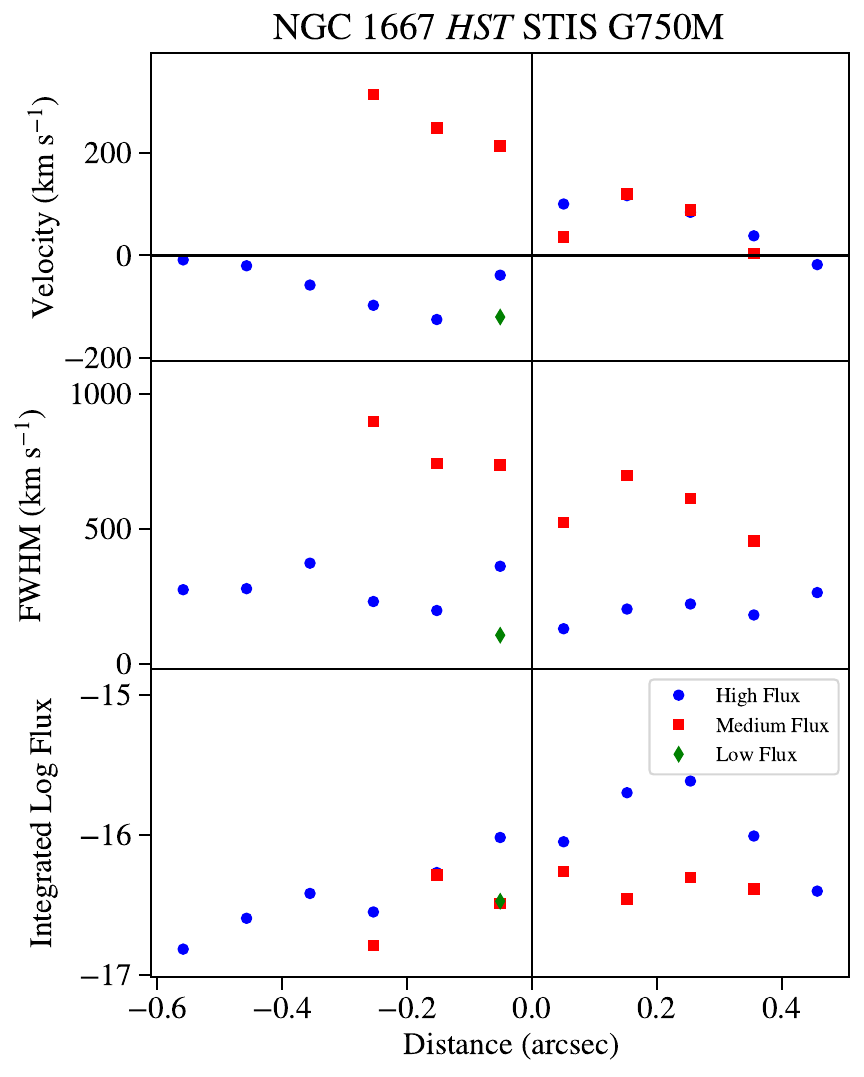}}
\subfigure{
\includegraphics[width=0.325\textwidth]{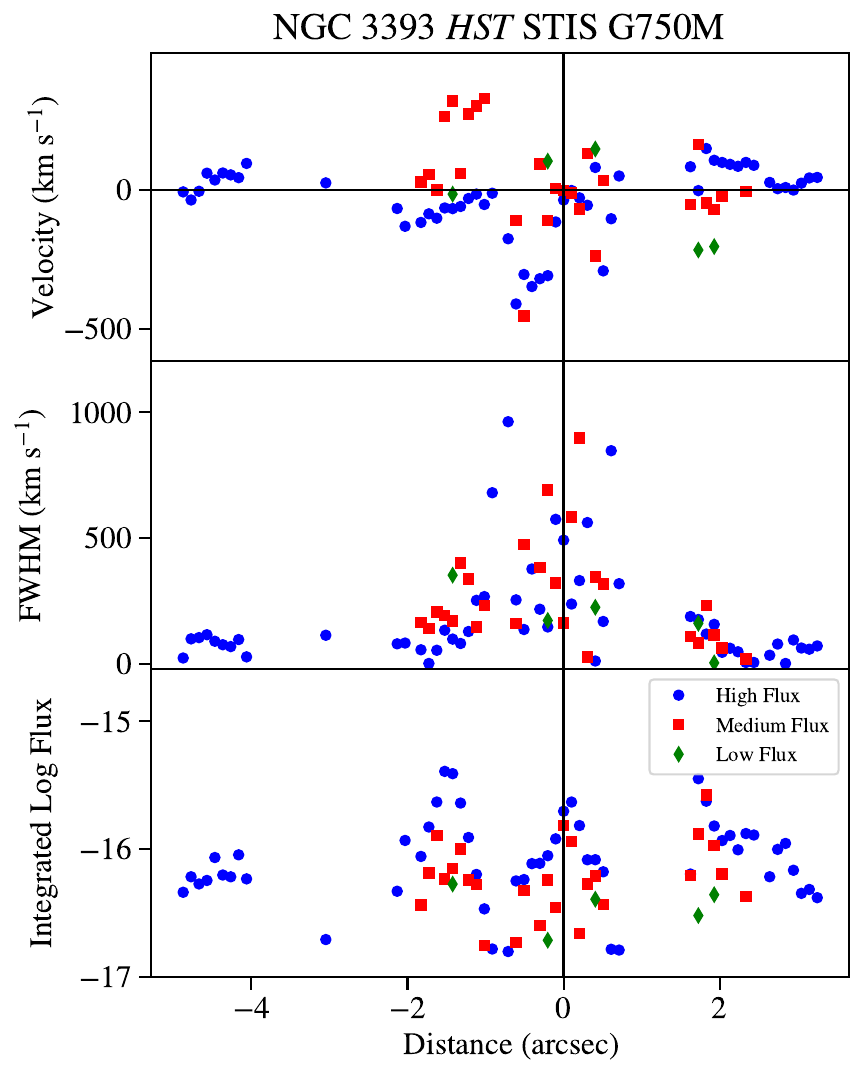}}
\subfigure{
\includegraphics[width=0.325\textwidth]{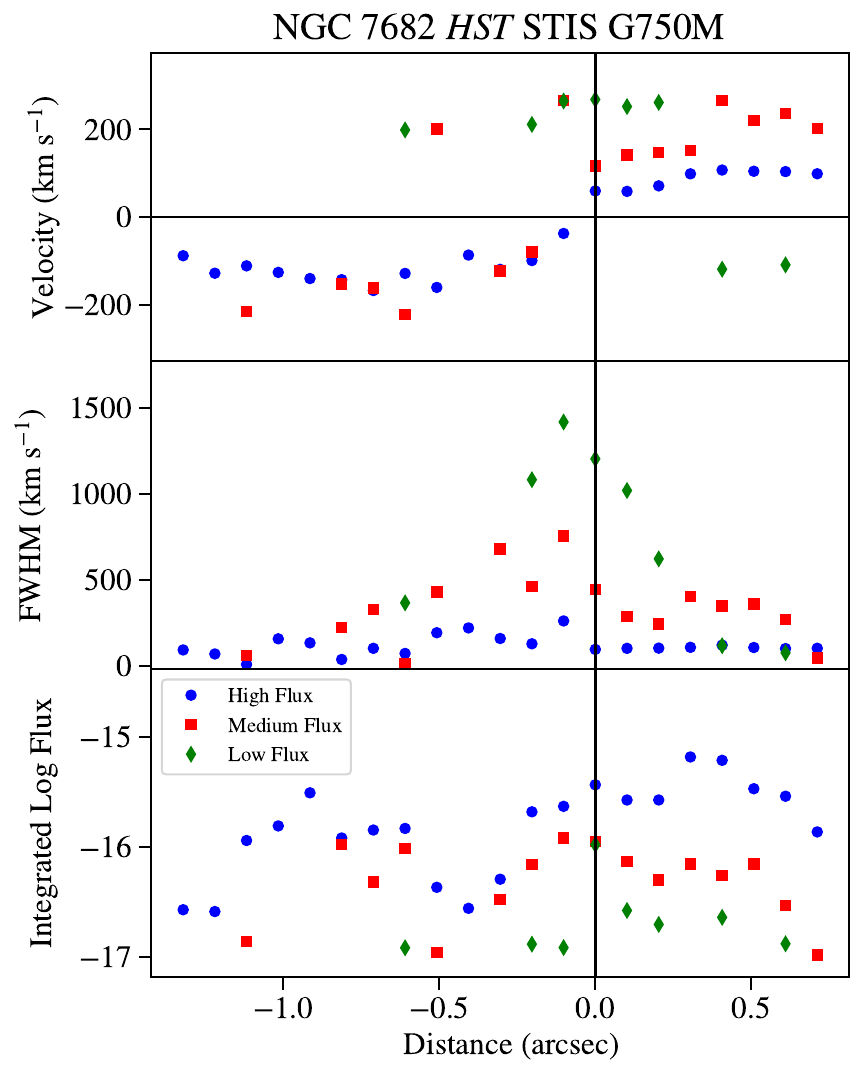}}
\subfigure{
\includegraphics[width=0.325\textwidth]{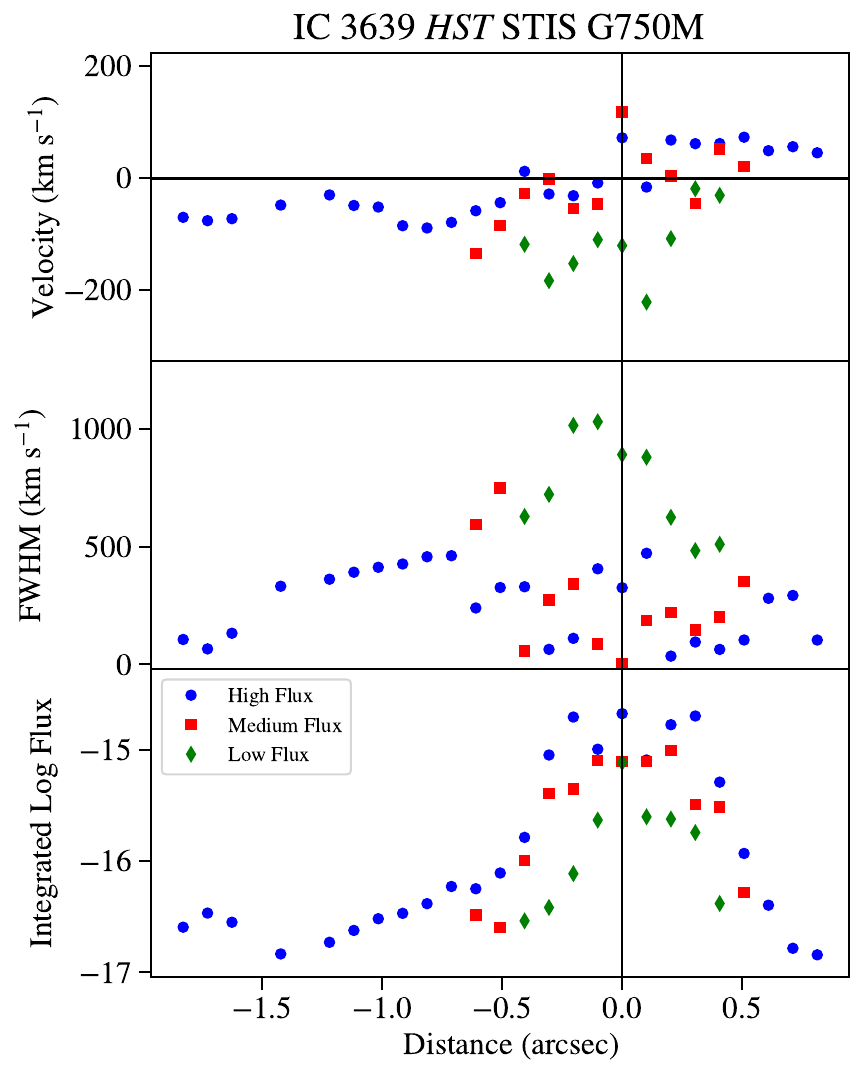}}
\subfigure{
\includegraphics[width=0.325\textwidth]{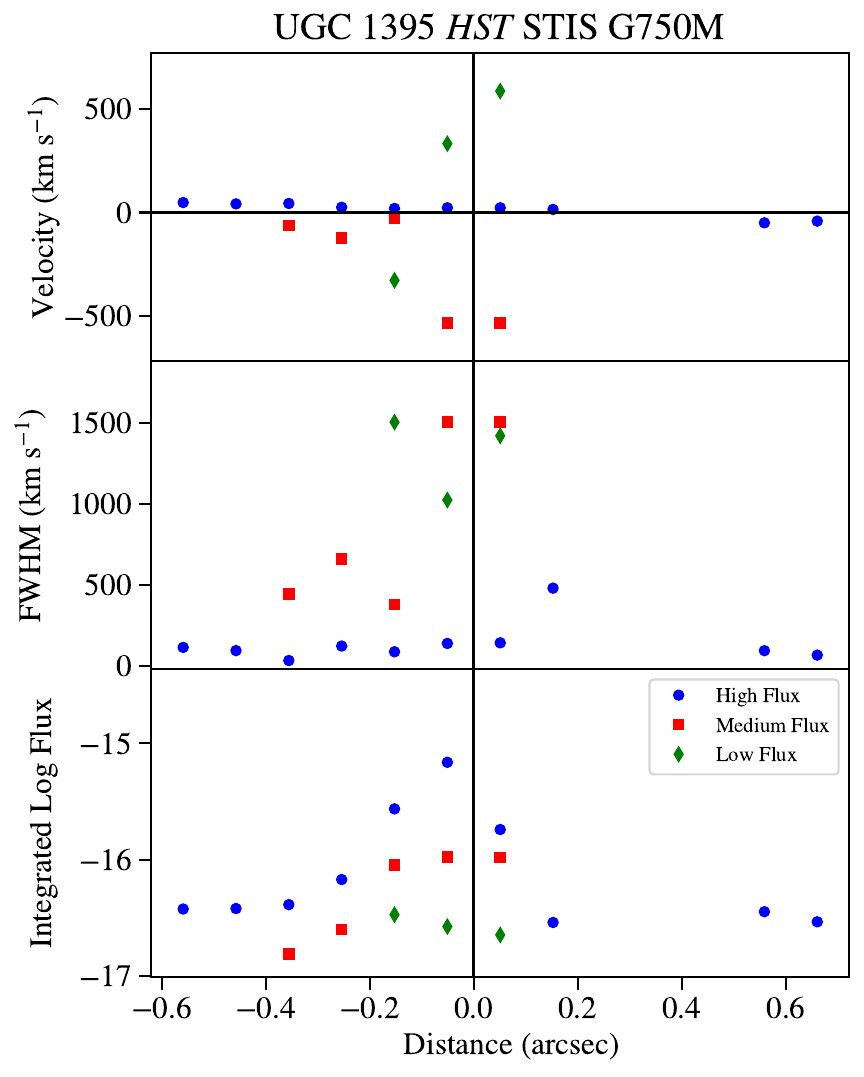}}
\caption{Kinematic plot examples for six targets. Each plot is composed of three vertical sections, velocity (top), FWHM (middle), and flux (Log(erg s$^{-1}$ cm$^{-2}$)) (Bottom). The x-axis shows distance across the slit, with the zero-point established by the brightness centroid of the AGN, and positive x-values indicating direction along of the position angle of the slit. For NGC 788, NGC 1667, and IC 3639 positive x-values indicate a southward direction on the sky, and for NGC 3393, NGC 7682, and UGC 1395 positive x-values indicate a northward direction. Each slit bin has either one, two, or three associated components output from the fitting routine, sorted by flux into high (blue circles), medium (red squares), or low flux (green triangles) components. Positive velocity indicates red-shifted components and negative values indicate blue-shifted components.}
\label{fig:Kinematics_Test}
\end{figure*}

The \halpha kinematic plots are shown in Figure \ref{fig:Kinematics_Test}. They are vertically broken into three sections: radial velocity in the rest frame of the galaxy (top), FWHM (middle), and integrated flux (bottom). Depending on how many kinematic components the program finds to be statistically significant at each point, the plots can include up to three different marker types designating the individual knots of gas. Multiple gas components detected in a single bin indicate multiple clouds or knots traveling at different radial velocities, and are most prevalent in the nuclear regions indicting a combination of different motions including rotation and outflows.
In the examples shown, each target was found to have up to three distinct clouds in a single slit-bin, sorted by integrated flux. Negative velocity values indicate blue-shifted components relative to the galaxy's systemic velocity, while positive values indicate red-shifted components. The x-axis shows distance from the brightness centroid of the galaxy in arcseconds, with positive values moving in the direction of the position angle of the slit. The zero-position for this axis is determined from the continuum emission peak, located in the nucleus of the galaxy.

\subsection{Outflow Prevalence}\label{subsec:outflow}

A key goal of our analysis is the determination of outflow presence, as well as the extents to which outflows reach in their host-galaxy disks. As discussed in section \ref{subsec:ImAn}, imaging alone cannot differentiate between rotating and non-rotating gas components, but can only measure the ionized gas extent as a whole. A quantitative distinction must be made to establish whether the ionized gas at each point is outflowing from the AGN (typically high velocities and high FWHM values) or simply rotating in the disk of the galaxy (low velocities and low FWHM values). There is also a third possibility, that is kinematically disturbed gas (identified by low velocities and high FWHM values), which is distinct from simple rotation though not necessarily outflowing. With this goal in mind, we adopt two classification techniques to categorize the gas components based on their kinematics.

Our first method assumes that the true velocity for each component is either equal to, or greater than, the observed velocity. This means we are either seeing the maximum velocity of the gas cloud or only some fraction of it along the line of sight. In the latter case, the value may be increased after accounting for projection effects. This method gives an absolute lower limit of our outflow detection estimates, because we are using the lowest possible velocity values to categorize the components. We classify components as ``disturbed" if they are measured with FWHM values large enough that they might be outflows depending on what fraction of the true velocity we are observing.

The second method we employ gives an upper limit, by assuming that every component with a velocity above the ``true" rotation value is outflow. This method utilizes the models from Section~\ref{subsec:morph} to deproject velocities to the galaxy disks. Deprojecting the outflows gives the absolute maximum velocities that the ionized gas may be able to achieve. The combination of these two methods, when applied to our kinematic plots, isolate possible outflow components, ranging from the most conservative estimates to the most optimistic. 

The extents of the boundaries are largely based on the number of bins in which our routine is able to determine Gaussian components. An important detail to note when dealing with spectrally-derived kinematics, is that the chosen position angle of the slit may not encompass the full span of the ionized gas, and so only the central bins may have enough flux to measure meaningful parameters. By comparing the offsets of the slit alignments with the angles of the structures in our [\ion{O}{3}] images, we find only six of our targets have their slits oriented such that the majority ($>$50\%) of the [\ion{O}{3}] structure was traced (containing the majority of the emission can also be quantified based on extent comparisons in section \ref{sec:res}). Only these targets (NGC~788, NGC~1667, NGC~3393, NGC~7682, IC~3639, and UGC~1395) are chosen for kinematic deprojection, to refrain from defining outflow boundaries using incomplete data. The two methods used for kinematic classifications on these six targets are as follows:

\subsubsection{Observational Boundaries}\label{subsubsec:obs}

The first method uses a series of boundary definitions established by \cite{Fischer2018} for the identification of AGN outflows based on observed kinematic measurements. These boundaries are defined for spatially resolved ionized gas, and thus apply to our targets. The classifications, with the velocities defined relative to the target's systemic, are:

\begin{itemize}
    \centering
    \item Outflow: Observed Velocity $>$ 250 km~s$^{-1}$ and FWHM $>$ 250 km~s$^{-1}$
    \item Disturbed: Observed Velocity $<$ 250 km~s$^{-1}$ and FWHM $>$ 250 km~s$^{-1}$
    \item Rotation: Observed Velocity $<$ 250 km~s$^{-1}$ and FWHM $<$ 250 km~s$^{-1}$
\end{itemize}

These conditions are based on empirical measurements of host galaxy kinematics.\footnote{Some studies argue that bins along the slit displaying more than one component are inherently either disturbed or outflowing gas, based on the assumption that galaxy rotation cannot exhibit more than single-component clouds. This additional constraint is not taken into account during our analysis.} For instance, 250 km s$^{-1}$ is typically the maximum velocity of rotation in galactic disks \citep{Robinson2021}, and it is used as the upper boundary for rotating gas components. However, observed values are highly subjective, depending not only on the inclinations of the AGN, but also on the orientation of the slit and its placement over the disk. Therefore, we incorporate a separate system of boundary conditions utilizing ``true" space velocities for each galaxy to deal with the projection effects.

\subsubsection{Deprojected Boundaries}\label{subsubsec:deproj}

An assumption of our biconical outflow model is that the ionized gas we observe resides primarily in the disk of the host galaxy. For this reason, our second outflow identification method deprojects the measurements to an assumed disk, using the inclinations and ellipticities from our isophotal modeling. This converts the observed velocities into spatial velocities, and the observed arcsecond extents into actual extents, converted into parsecs. We first take the established maximum velocity for rotation and deproject it using:

\begin{equation}
\mathrm V_{proj}(\textit{i},\varphi) = \frac{(250~km~s^{-1})}{sin(\textit{i})cos(\varphi)}
\end{equation}

Where \textit{i} is the inclination of the galaxy, and $\varphi$ is the angular offset of the slit from the galaxy major axis. This changes the observed maximum rotation velocity, $V_{proj}$, into a target-specific maximum for rotation. Components with velocities exceeding $V_{proj}$ are assumed to be outflowing gas, while components with velocities below this limit may be either rotation or disturbed gas. 

The components that fit the criteria for outflow are then deprojected to the disk as well, in order to find their maximum outflowing velocities. These intrinsic radial velocities, $V_{int}$, and the deprojected spatial scales, $R_{int}$, are estimated using these two equations:

\begin{equation}
\mathrm V_{int}(\textit{i},\varphi) = \frac{V}{sin(\textit{i})sin(\varphi)}
\end{equation}

\begin{equation}
\mathrm R_{int}(\textit{i},\varphi) = \sqrt{ R^{2} \left(cos^{2}(\varphi) + \frac{sin^{2}(\varphi)}{cos^{2}(\textit{i})}\right)}
\end{equation}

Where V is the observed velocity and R is the observed distance from the center of the slit in parsecs \citep{Revalski2018a, Revalski2018b}. While $V_{int}$ may be an overestimation for the velocity of some of the components, it is useful for detecting potential regions of outflow that observed values may miss.

\section{Results} \label{sec:res}

\subsection{Imaging Extents}

Table~\ref{tab:modelin} lists the extents and luminosity of the [\ion{O}{3}] emission using both of our image analysis methods. Results from the first method, using each targets' background 3$\sigma$ value as the contour limit, is given in column 3 (hereby $R_\mathrm{[O~III]}$), with each 3$\sigma$ value listed in column 2. The second method uses a fixed 3$\sigma$ value averaged from the literature, of 2.034$\times$$10^{-17}$ erg s$^{-1}$ cm$^{-2}$, in order to examine how differences in image parameters affect the outcome of our extents and luminosity relationships. This method's resulting extents, $R_\mathrm{[O~III],fixed}$, are listed in column 5. The [\ion{O}{3}] luminosities provided in column 6 ($L_\mathrm{[O~III]}$) are measured using the individual 3$\sigma$ flux limits. Additionally included in this table are five published targets from \cite{Revalski2018b} with values updated by this work using the first imaging method.

The fixed contour flux limit is around an order of magnitude greater than any of the individual continuum flux measurements of our images. Increasing the contour threshold decreases the range of the contour across the image and effectively decreases the measured radial extents of the targets. While the idea that using a different flux threshold to measure images changes the results is not a novel concept, it is an extremely important takeaway from this process and is further discussed in the next section.

Using imaging for extent measurements comes with the limitation of being unable to distinguish between our three categories of rotation, outflow, and disturbed gas. The imaging methods utilize luminosity to isolate the ionized gas from the image background, but cannot further segregate the ionized components and therefore no longer specifically describe ``outflow". The extents and luminosities depicted in Table \ref{tab:modelin} are measurements of the [\ion{O}{3}] gas as a whole and may include components that are simply rotating in the disk.

\begin{deluxetable}{lccccc}
\vspace{-0.5em}
\setlength{\tabcolsep}{0.0in} 
\tabletypesize{\small}
\tablecaption{[\ion{O}{3}] Extents}
\tablehead{
\colhead{Catalog} & \colhead{3$\sigma$ Flux} & \colhead{$R_\mathrm{[O~III]}$} & \colhead{$L_\mathrm{[O~III]}$} & \colhead{$R_\mathrm{[O~III],fixed}$} & \colhead{$L_\mathrm{[O~III],fixed}$} \vspace{-0.5em}\\
\colhead{Name} & \colhead{($10^{-18}$ erg s$^{-1}$} & \colhead{(pc)} & \colhead{(log erg s$^{-1}$)} & \colhead{(pc)}  & \colhead{(log erg s$^{-1}$)\vspace{-0.5em}}\\
\colhead{} & \colhead{cm$^{-2}$  arcsec$^{-2}$)} & \colhead{} & \colhead{} & \colhead{}  & \colhead{\vspace{-0.5em}}\\
\colhead{(1)} & \colhead{(2)} & \colhead{(3)} &\colhead{(4)} & \colhead{(5)} & \colhead{(6)}
}
\startdata
\hline
This Study \\
\hline
NGC 788  & 3.30 & 568 & 40.65 & 268 & 40.62 \\
NGC 1358 & 2.70 & 711 & 40.20 & 487 & 40.07 \\
NGC 1667 & 2.76 & 351 & 40.05 & 179 & 40.01 \\
NGC 3393 & 2.20 & 1197 & 41.33 & 1074 & 41.31 \\
NGC 5135 & 2.41 & 1343 & 40.89 & 520  & 40.81 \\
NGC 5283 & 1.32 & 882 & 40.37 & 399 & 40.32 \\
NGC 5427 & 2.15 & 236 & 39.33 & 40  & 39.21 \\
NGC 5695 & 2.10 & 828 & 40.17 & 190 & 39.93 \\
NGC 6300 & 2.59 & 132 & 38.74 & 23  & 38.73 \\
NGC 7682 & 2.30 & 875 & 40.76 & 519 & 40.73 \\
IC 3639  & 2.83 & 654 & 40.54 & 115 & 40.51 \\
UGC 1395 & 2.35 & 346 & 40.16 & 121 & 40.09 \\
\hline
Published \\
\hline
Mrk 34 & 10.00 & 2478  & 42.6 & ... & ...\\
Mrk 78 & 34.00 & 2612 & 42.4 & ... & ...\\
Mrk 573  & 62.00 & 1887 & 41.99 & ... & ...\\
NGC 1068 & 25.00 & 826 & 40.16 & ... & ...\\
NGC 4151 & 39.00 & 501 & 40.4 & ... & ...\\
\enddata
\tablecomments{Columns are (1) target name, (2) 3$\sigma$ level of the target's continuum used for plotting the [\ion{O}{3}] contours (3) ionized gas extent using the individual 3$\sigma$ level (4) ionized gas extent using the fixed contour value (5) [\ion{O}{3}] Luminosity contained within the fixed extent radius after subtracting off the background continuum. We also include six archival targets from \cite{Revalski2021}, below the horizontal line. There are no fixed extent measurements for the published targets as those images are shallower than those analyzed in this study.}
\label{tab:modelin}
\vspace{-1em}
\end{deluxetable}

\subsection{Correlations with Luminosity}

\begin{figure*}[htb!]
\vspace{0.5em}
\centering
\includegraphics[width=0.99\textwidth]{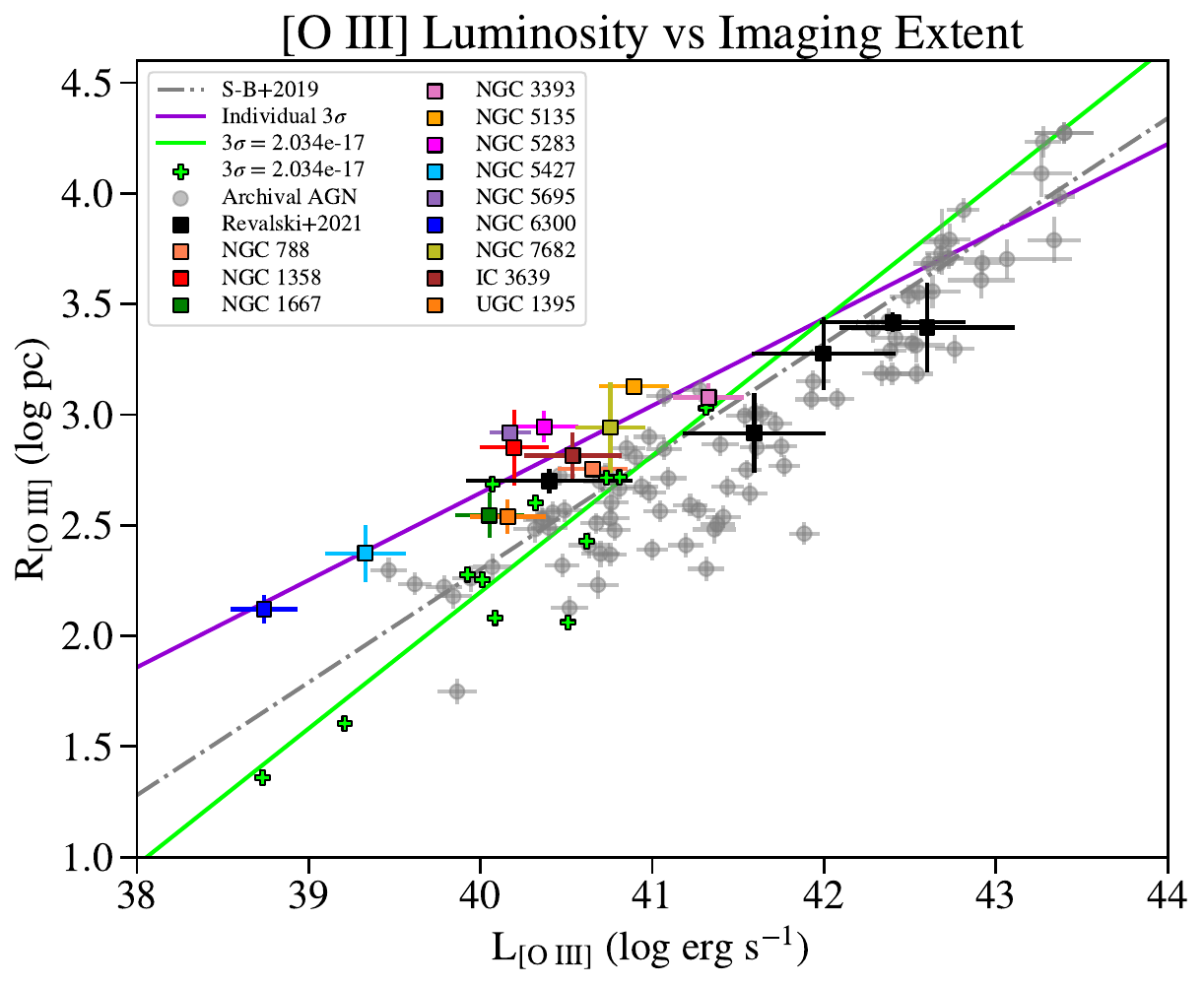}
\caption{The radial extent of the [\ion{O}{3}] gas ($R_\mathrm{[O~III]}$) vs [\ion{O}{3}] luminosity ($L_\mathrm{[O~III]}$) for the 12 targets in this study (colored squares), the five targets from \cite{Revalski2021} (black squares), and the archival targets used to determine a fixed threshold for image analysis (gray circles). The green plusses indicate the values of our 12 targets that were measured using the standardized 3$\sigma$ flux contour of 2.034$\times$$10^{-17}$, demonstrating where these AGN would lie in this parameter space with a shallower threshold closer to those used by the literature studies. The three lines show the best fits to the targets in our study using their individual 3$\sigma$ thresholds (solid purple, 0.39), the same targets using the fixed contour limit (solid green, 0.62), and the archival AGN using an adopted line equation from \citep{Storchi-Bergmann2018} (gray dashed, 0.51). As discussed in the text, the radial extents for our targets are larger because the deep imaging reaches fainter emission than some earlier studies.}
\vspace{1em}
\label{fig:result}
\end{figure*}

Figure \ref{fig:result} shows the relationship between the [\ion{O}{3}] imaging extent and [\ion{O}{3}] luminosity for AGN across multiple samples. The purple trend line represents only our 12 targets, which are not available from the literature due to the previous lack of suitable images, and depicts a relationship between the two variables defined by Log($R_\mathrm{[O~III]}$) = (0.39$\pm$0.05)$L_\mathrm{[O~III]}$ - (13.12$\pm$1.11). The green plusses and green trend line represent the target values measured using the fixed contour limit, with the trend line defined as Log($R_\mathrm{[O~III],fixed}$) = (0.62$\pm$0.09)$L_\mathrm{[O~III],fixed}$ - (22.46$\pm$1.25). The gray, dashed trend line represents the relationship for the literature AGN, and is adopted from \cite{Storchi-Bergmann2018}, defined as Log($R_\mathrm{[O~III]}$) = (0.51$\pm$0.03)$L_\mathrm{[O~III]}$ - (18.12$\pm$0.98). We include these plot elements to show the luminosity range for targets that have been analyzed for this relationship, as well as to provide visualization of the dramatic difference that can be caused by a change in the depth of the analyzed images.

\subsection{Kinematic Outflows} \label{subsec:deproject}

\begin{figure*}[htb!]
\centering
\includegraphics[width=0.49\textwidth]{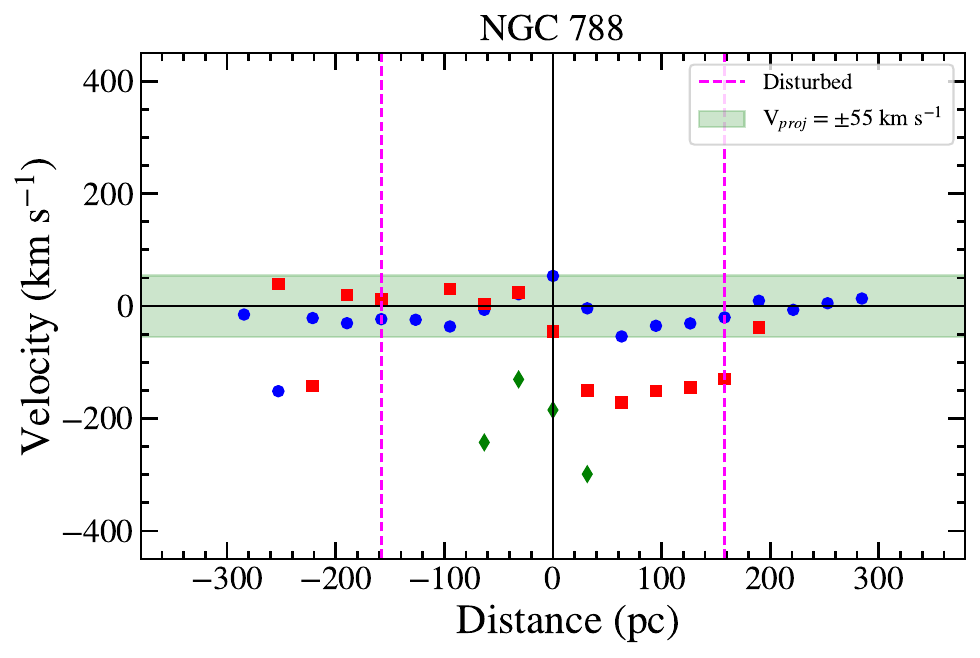}
\includegraphics[width=0.49\textwidth]{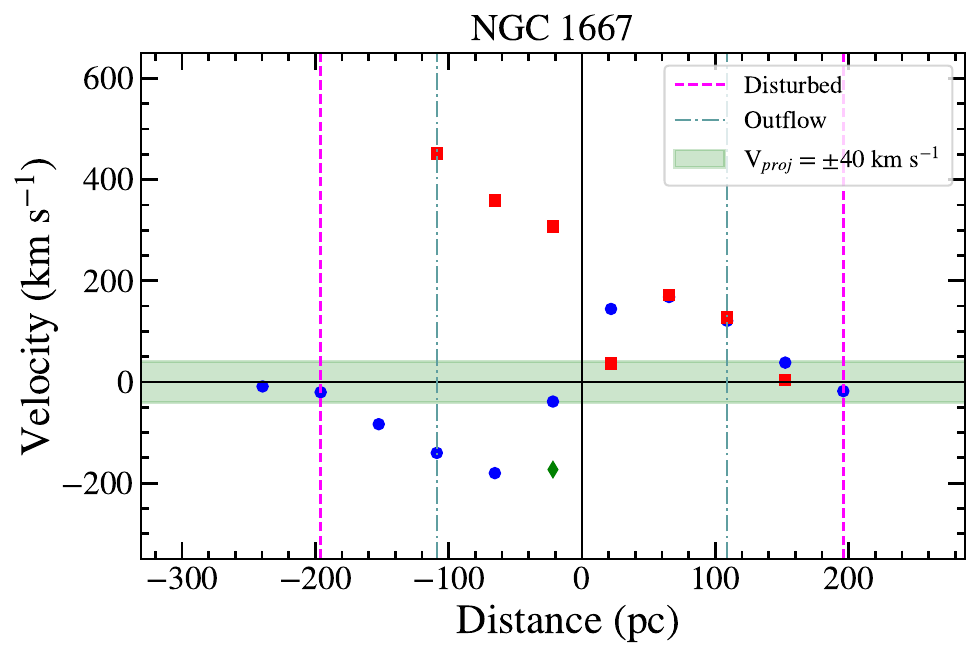}
\includegraphics[width=0.49\textwidth]{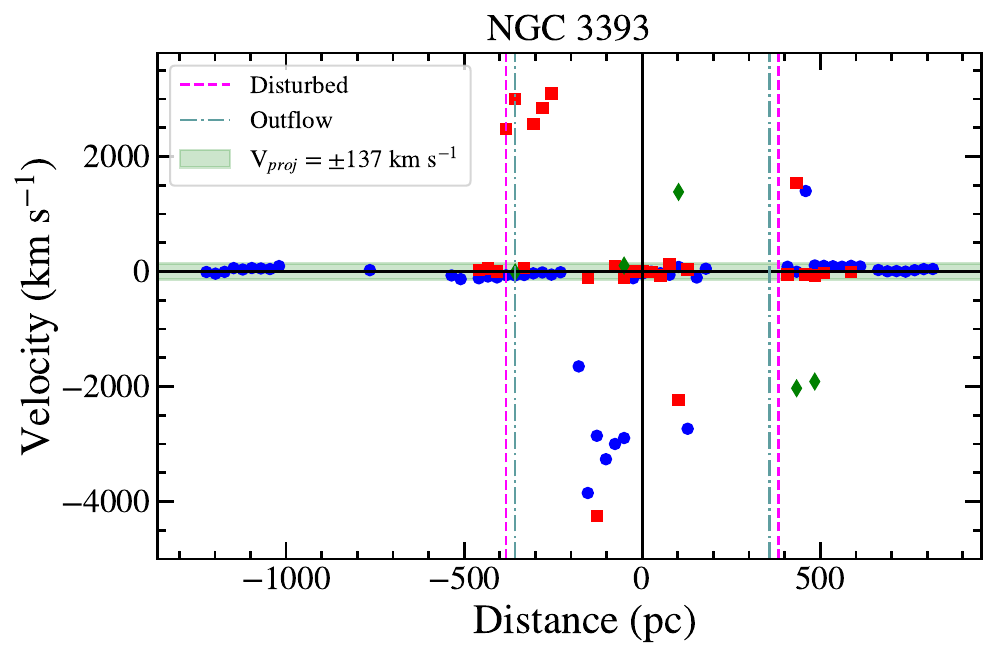}
\includegraphics[width=0.49\textwidth]{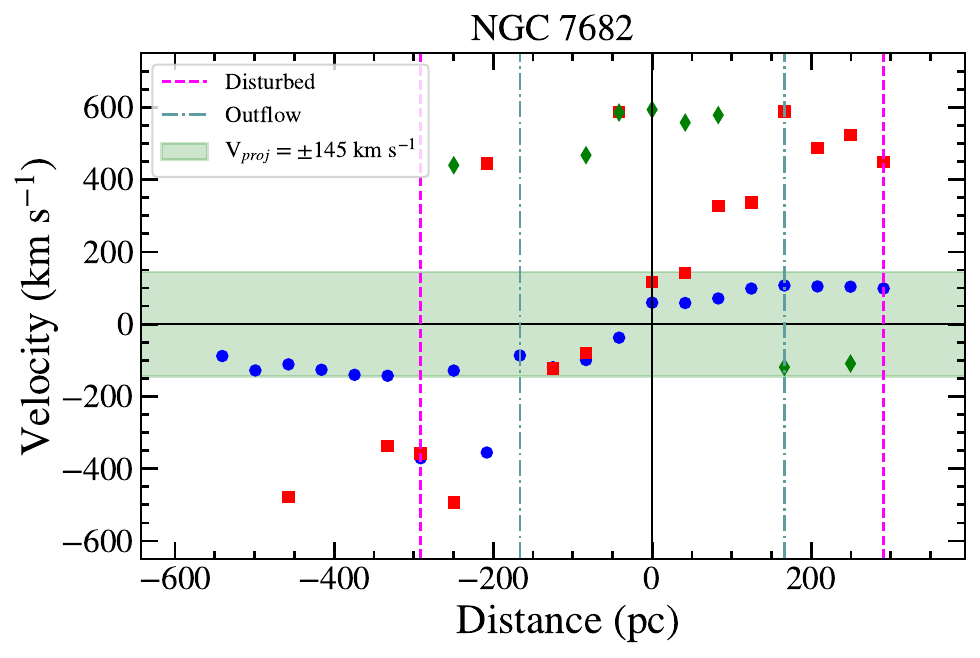}
\includegraphics[width=0.49\textwidth]{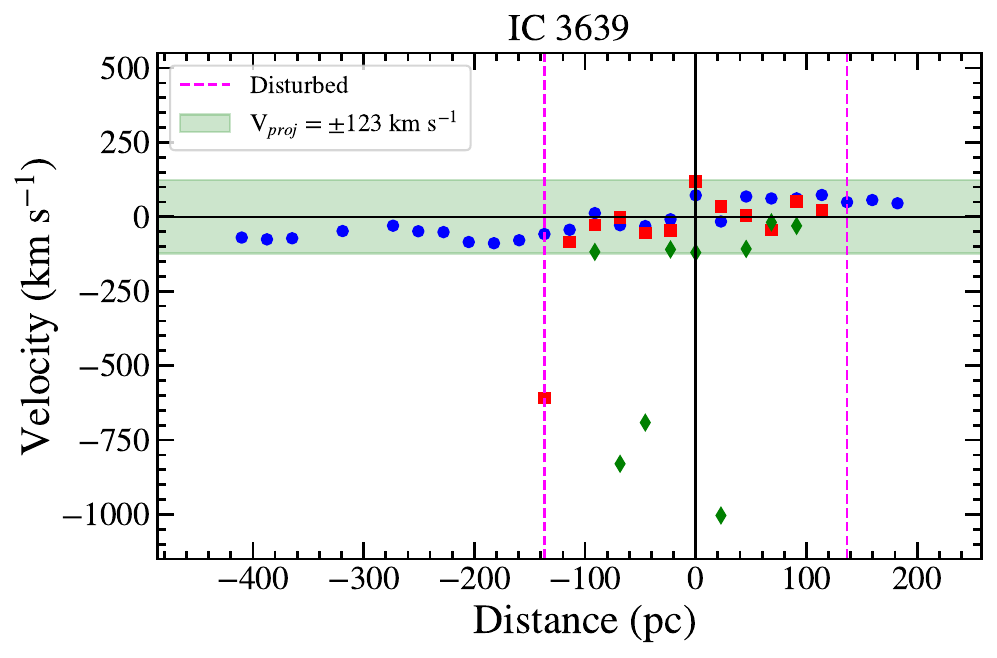}
\includegraphics[width=0.49\textwidth]{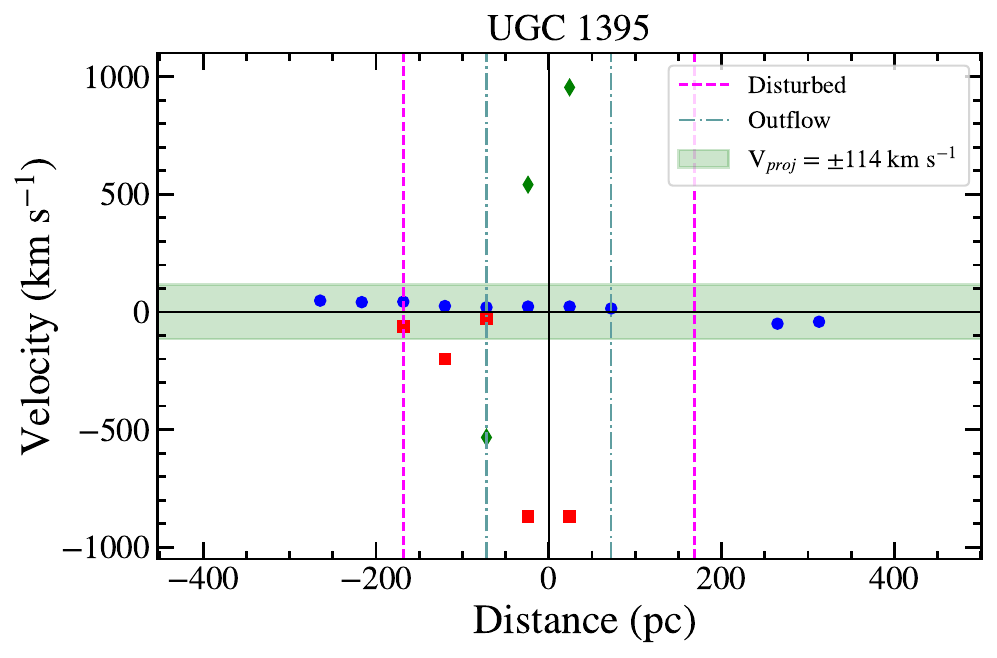}
\vspace{1em}
\caption{\halpha velocity plots for the six targets whose slits encompass the majority of their ionized gas structures. The x-axis for each target is converted to deprojected distances in parsecs within the disk, using the spatial scales from Table~\ref{tab:sample}. Both boundary methods have been applied. The boundaries based on observed velocities and FWHM (\ref{subsubsec:obs}) are symmetric vertical lines, with magenta indicating the extent of the disturbed gas and cyan indicating the extent of the outflow components, if present. The horizontal green band represents the bounds of $V_{proj}$. Components in this region are observed values as they are considered rotational, while components outside are deprojected to their ``true" spatial velocities assuming they are outflowing within the disk. See Equations 2 and 3 for the deprojection methodology.}
\vspace{1em}
\label{fig:bands}
\end{figure*}

Deprojected velocity plots of the six, well-aligned targets are shown in Figure \ref{fig:bands}. These are essentially identical to the top section of the kinematic plots from Figure \ref{fig:Kinematics_Test}, with the boundary extents and deprojection scalings applied. The x-axis gives the scaled distances in parsecs. The vertical magenta and cyan lines indicate the results of the observation-based method (\ref{subsubsec:obs}), marking the farthest extents of the disturbed and the outflowing gas, respectively. We find that all six targets have components with FWHM values indicating disturbed gas, with the average disturbed extent around 222 pc from the AGN. Out of the six targets, four were found containing high enough observed velocities and FWHM to indicate outflow, while the other two, NGC~788 and IC~3639, have only the magenta line separating disturbed components from rotation. In all cases, the ionized-gas extents increase from outflow (if present) to disturbed to pure rotation.

The green, horizontal shaded areas represent the second method's classification boundary, $V_{proj}$ (\ref{subsec:deproject}). Each target has its own $V_{proj}$, which sets the upper limit on rotation in the disk after the scale factor is taken into account. Velocities inside this limit may potentially be rotation, so points in this region remain unchanged from the observed values. Velocities outside of this region are considered too large to be rotation and are deprojected, giving the maximum true outflow velocity for these components based on our kinematic models and the slit position angles.
 
The combination of the two boundary methods gives both a conservative case for outflow estimation (observed values) as well as an optimistic case (deprojected points). The observational method results in outflow detection in four of the six targets, with an average outflow extent of 195 pc. The deprojection method finds outflow in all six targets, where the observational velocities were below the observed rotation limit, but greater than $V_{proj}$. In all six cases, deprojecting the velocities resulted in outflows detected farther out than the outflow extent estimates from the observed method, now with an average outflow extent of 275 pc. Three of the targets show deprojected outflow components extending farther out than the disturbed boundary extents, but only by one or two points. These patterns indicate the general consistency of the disturbed gas boundaries tracing the components with velocities greater than a given target's rotation.

\begin{deluxetable*}{lccccccc}
\vspace{-0.5em}
\setlength{\tabcolsep}{0.15in} 
\tabletypesize{\small}
\tablecaption{[\ion{O}{3}] Kinematic Extents}
\tablehead{
\colhead{Catalog} & \colhead{$R_{obs}$} & \colhead{$R_{out}$} & \colhead{$R_{dist}$} & \colhead{$R_{tot}$} & \colhead{$R_{out} / R_{tot}$} & \colhead{$R_{out} / R_\mathrm{[O~III]}$} & \colhead{$R_{out} / R_\mathrm{[O~III],fixed}$}\\
\colhead{Name} & \colhead{(pc)} & \colhead{(pc)} & \colhead{(pc)} & \colhead{(pc)} & \colhead{} & \colhead{} & \colhead{}\\
\colhead{(1)} & \colhead{(2)} & \colhead{(3)} &\colhead{(4)} & \colhead{(5)} & \colhead{(6)} & \colhead{(7)} & \colhead{(8)}
}
\startdata
NGC 788  & 0  & 253 & 158 & 284 & 0.89 & 0.45 & 0.94 \\
NGC 1667  & 109 & 152 & 196 & 240 & 0.63 & 0.43 & 0.85 \\
NGC 3393  & 357 & 484 & 382 & 1224 & 0.39 & 0.40 & 0.45 \\
NGC 7682  & 166 & 458 & 291 & 541 & 0.85 & 0.52 & 0.88 \\
IC 3639  & 0  & 137 & 137 & 410 & 0.33 & 0.20 & 1.19 \\
UGC 1395  & 72 & 120 & 168 & 165 & 0.45 & 0.32 & 0.99 \\ \hline
Mean & 117 & 275 & 222 & 477 & 0.59 & 0.39 & 0.88 \\
\enddata
\tablecomments{Columns are (1) target name, (2) outflow extent from observed velocity, if found (3) outflow extent from deprojected velocity, if found (4) disturbed region extent based on FWHM, (5) total extent of kinematic components across the slit (6) The ratio of outflow extent to total gas extent measured by kinematics. (7) The ratio of outflow extent to total gas extent measured by the targets 3$\sigma$ imaging threshold. (8) The ratio of outflow extent to total gas extent measured by the fixed imaging threshold. The average for each column is given below the target list.}
\label{tab:kinext}
\vspace{-1em}
\end{deluxetable*}

\section{Discussion}\label{sec:disc}

We show in Figure \ref{fig:result} that using a fixed 3$\sigma$ threshold that is similar to prior literature analyses produces results that coincide with the previous relationships, even at the lowest luminosity regime. The same AGN are shown to produce results differing by almost a full dex, depending on the chosen 3$\sigma$ flux threshold. The targets in this work that were imaged in the same observing campaign have similar exposure times, somewhat standardizing the image analysis. This is not true for all studies, and our result highlights the necessity of using such uniform criteria within a sample, as well as keeping standardization when comparing to targets of other samples. The reduction of our targets' imaging extents when comparing $R_\mathrm{[O~III]}$ with $R_\mathrm{[O~III],fixed}$ indicates that AGN may be ionizing at much larger radii than those from previous, shallower observations. In order to understand the full extent of ionized gas and AGN influence in host galaxies, we emphasize that deeper imaging is needed for these types of analyses.

Results from our deprojected kinematics point toward an additional observational bias. Categorizing gas kinematics purely on observed values may be limiting, or completely missing, the disturbed or outflowing gas regions. The observation-based outflow boundaries placed on the six deprojected targets fail to identify potential outflow components in every case, while the kinematically disturbed boundaries are consistent with the deprojected component extents in only half of the targets. Although the sample we deprojected was small, our results underline the idea that target geometry and/or observation alignment may result in undetected outflow where a deprojection method might have yielded a positive detection. It is possible that such deprojections may give overestimates of the real velocities (such as the extreme values for NGC 3393), components outside of the band can at least be verified if brought to attention. For this reason, it is best to take into account the projection effects of the system when attempting to identify potential outflow locations.

From overlaying the slit position angles on the images, we find that only 6 of the 15 targets have the majority of their [\ion{O}{3}] gas structure falling in line with the spectral observations. We compare the extents estimated by both of our methods in Table \ref{tab:kinext}, combining the kinematic results of these targets with our imaging results. $R_{tot}$ is the farthest extent that our spectral fitting routine was able to detect a gas component from the center, with an average extent of $\sim$477~pc. Comparing this to the farthest range that imaging is able to trace the gas structure, around $\sim$665~pc, shows that only 65\% of the detectable gas extent is able to be analyzed using our kinematic plots.

We can use these results for a crude comparison of outflow size to photoionization size, using the deprojected kinematic components along with the imaging extents. This was performed on separate AGN in an outflow study by \cite{Kim2023}, where they found the average ratio of outflow extent ($R_{out}$) to total ionized gas extent found via imaging ($R_{[O~III]}$) to be 0.72. From the average of $R_{out} / R_{[O~III]}$ for our deprojected targets (column 7 in Table \ref{tab:kinext}), we see that this ratio from the kinematic analysis is 0.39. This marks an interesting comparison to the previous study, but will need a larger sample size in order to determine any reasonable conclusions. \cite{Fischer2018} conducted a similar analysis on nearby type 2 quasars, finding a ratio of $\sim$0.22 on average, suggesting a possible luminosity dependence that needs further study with a large sample.

\subsection{Outflow Classifications}

Our previous analyses on these targets, as well as the rest of our extended sample, were detailed in \cite{Fischer2013}. With our new images and deprojected models we now re-evaluate our prior kinematic categorizations for this works' targets. The AGN kinematics were previously classified based on their observed velocity values, velocity dispersions, and detectable extents as either ``Outflow" (high velocities, multiple components, and/or visible bi-conical structure), ``Ambiguous" (indications of outflow patterns but no strong quantitative evidence to confirm), ``Complex" (large velocity dispersion with no other signatures of outflows or biconical structure), or ``Compact" (asymmetric or unresolvable components, potentially indicating a misaligned slit or minimal emission extent).

Out of the six targets detailed in Section \ref{subsec:outflow}, the \cite{Fischer2013} analysis resulted in strong outflow evidence for only NGC 1667, as the kinematics showed multiple components that agreed with the bi-conical models. NGC 788, NGC 7682, IC 3639, and UGC 1395 were placed in the ambiguous category, with the potential for outflows or disturbed gas but no clear distinction from rotation. The remaining target, NGC 3393, was labeled complex, as its kinematics did not neatly fit our bi-conical models and was recorded as having no evidence for outflows. Our updated fitting routine is better able to untangle the individual components of the line profiles across the slits, increasing the detected data points and radial extents of the kinematics compared to the previous study. Up to three distinct components were found in every target (compared to the previous determinations of only one or two), mainly in the bins surrounding the nuclear regions, bringing attention to high velocity knots, generally attributed to the lowest flux components.

Using our updated outflow criteria, we have found evidence of outflows in four of the targets with possible outflow evidence in the remaining two (NGC 788 and IC 3639), shifting the categorization of NGC 7682, UGC 1395, and NGC 3393 into the outflow group. Despite the increases in maximum emission velocities, NGC 788 and IC 3639 remain in the ambiguous category, based on a lack of sufficient velocity dispersion values. The implementation of these new observations and methodologies are promising for the identification and classification of outflows, however the lack of fidelity of archival slit positions to the [\ion{O}{3}] structures remains a major limitation. A larger sample of well-aligned long-slit observations, including new STIS observations based on our recent images, will be crucial for the verification of these results.

While a qualitative analysis of each AGN is beyond the scope of this paper, we can use the new images (referring mostly to the right-most column of Figure \ref{fig:images}) to generally categorize the targets based on the morphology of their emission. We see several compact sources: NGC 788, NGC 1667, NGC 5427, IC 3639, and UGC 1395. These AGN have the majority of their visible [\ion{O}{3}] structure residing within an arcsecond of their brightness centroids and their measured extents comprising some of the lowest values in the sample. The remaining targets all show characteristics in line with our biconical model, displaying identifiable outflow patterns of at least one side of a biconical shape. NGC 3227, NGC 5135, and NGC 6300 are presumably showing only one side of their bicone, while the rest show at least partially symmetric outflow on both sides of the nucleus. NGC 1358, NGC 3393, and NGC 5283 show signs of spiral structure in their ionized gas, eventually falling in line with the spiral structure and the dust lanes in the host disk. The kinematics of these targets (Fig.\ref{app:Kinematics_App}) are shown to be dominated by low velocity, low FWHM components indicative of rotation in the disk (although this may be biased, given the offset slit placements over the disks).

\section{Conclusion}\label{sec:conc}

We obtained deep [\ion{O}{3}] and continuum images of 11 nearby Seyfert galaxies, adding four additional targets from archival imaging to create our sample. We modeled the geometries of the AGN and fit the emission lines of their HST long-slit spectra to obtain the number of gas components at each point as well as their velocities, FWHM, and fluxes. We chose six AGN with spectra that sample the majority of their NLRs, to evaluate the presence of outflows and, if found, estimate the radial extents of the ionized gas using the observed kinematics and observation-based classification criteria. We then deproject those velocities to the host galaxy disks, according to our models, estimating the outflow prevalence using true spatial velocities and comparing to the observational methodology. Finally, we measured the background continua of our [\ion{O}{3}] images and used them to evaluate the radial extents and luminosities of the ionized gas structures, repeating this analysis with a shallower value from the literature to test the effects of image quality and measurement threshold differences on our results. Our conclusions are as follows:

\begin{enumerate}
\setlength{\itemsep}{0pt}

\item Our results for the correlation between extent and luminosity of ionized gas in the NLRs of AGN (R$_\mathrm{[O~III]}$ vs L$_\mathrm{[O~III]}$) are consistent with previous studies, in that the estimated slope in log space is between 0.4 -- 0.5, within the uncertainties \citep{Woo2016, Luo_2021, Kim2023, Storchi-Bergmann2018, Fischer2018}. We show that this correlation extends to lower luminosities, although more observations of low-luminosity AGN are needed for confirmation.

\item By adopting multiple image thresholds we show that these results are strongly dependent on the depths of the images being analyzed. We find that measurements for NLR radial extents, and subsequent relations to AGN properties, depend on the camera, filter, and exposure time of the observations. A uniform criterion should be adopted among all targets in a sample in order to accurately assess the scaling relationships between AGN and the characteristics of the photoionized regions. Our results indicate a possible shallowing of the extent-to-luminosity relationship slope at lower luminosities, based on the deep images from our most recent HST campaign. However, when we adopt a threshold consistent with previous studies, we see that a single power-law fit accurately represents the relationship across $\sim$5 orders of luminosity.

\item High spatial resolution imaging is a critical resource to have before obtaining spectra, so that observers can accurately align the observations to the gas structures. On the other hand, spectra are needed alongside the imaging in order to distinguish between the ionized gas components and separate the outflows from possible rotation or disturbed gas in the disk. We find that the majority of targets in the literature have spectral observations that are aligned to the major axis of the galaxy. We note that the orientation of NLR gas relative to the host galaxy major axis can be random and so major axis alignment is not always sufficient for analyses of this type. Radial extents measured using HST [\ion{O}{3}] images provide results consistent with those from HST spectra-based kinematics, but we will need further observations that better trace outflows in the targets with poorly aligned slits to verify these results. This will also allow for the separation of the ionized gases into outflowing, disturbed, or rotating components in order to further constrain the emission-line kinematics.

\item Using observed kinematics gives the most conservative estimate of outflow prevalence, avoiding false positive detections of outflows, but only establishes a lower limit for the velocities of the gas. This underestimates the true velocities across the slit and may miss outflows given the current criteria for observational analyses adopted by recent literature studies.

\item Scaling AGN kinematics to account for projection effects improves the identification of outflows in previously ambiguous targets or targets that do not exhibit conventional biconical kinematic structures. We found evidence of outflows in all six of the targets on which this analysis was performed, five of which were not identified as having outflow components. Deprojecting the velocities to the disk gives a more realistic measure of outflow extent, but is more model dependent and may overestimate the velocities in a small number of cases. Given the consistent results from the deprojection method, it would seem that either this method alone, or in combination with the observed values, yields the most reliable results for determining outflow prevalence and extents.

\end{enumerate}

\acknowledgments

The authors would like to thank the anonymous reviewer for helpful comments that improved the clarity of this paper. M. Revalski thanks Varun Bajaj for HST WFC3/UVIS quadrant filter flux calibration support.

Based on observations with the NASA/ESA Hubble Space Telescope obtained from the MAST Data Archive at the Space Telescope Science Institute, which is operated by the Association of Universities for Research in Astronomy, Incorporated, under NASA contract NAS5-26555. Support for program numbers 16246 was provided through a grant from the STScI under NASA contract NAS5-26555. These observations are associated with program numbers \href{https://archive.stsci.edu/proposal_search.php?mission=hst&id=16246}{16246}.

\bibliography{refs}{}
\bibliographystyle{aasjournal}

\appendix

\section{NGC 3393}
\begin{figure*}[ht]
\centering
\includegraphics[width=0.5\textwidth,clip]{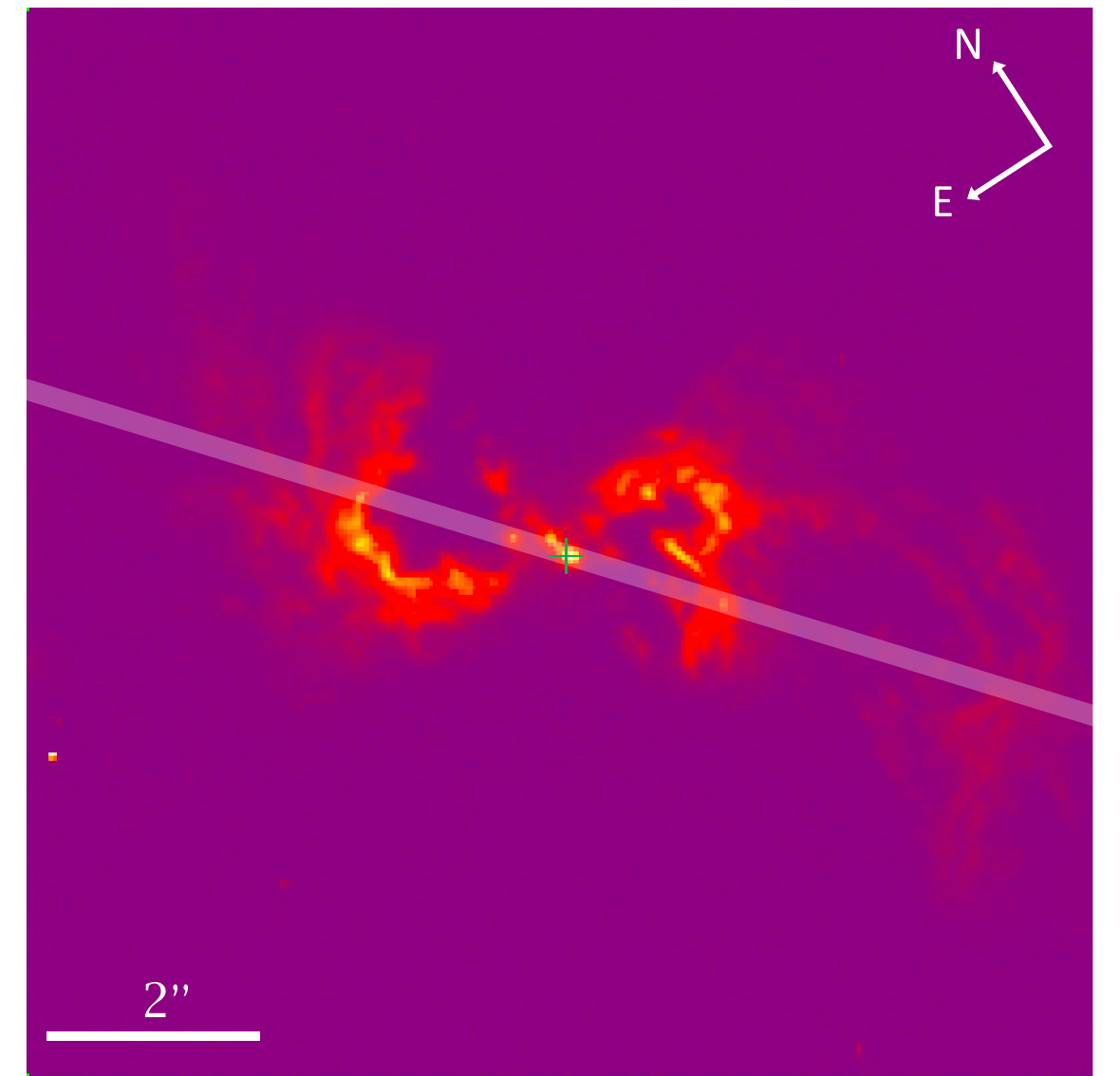}\\
\caption{The NGC 3393 WFC3 continuum-subtracted [\ion{O}{3}] image using data from HST Program ID: 12365 (PI: J. Wang, see also \citealp{Maksym2016, Maksym2017, Maksym2019}). Data reduction was performed using the same methods outlined in Section \ref{ssec:imaging}, and keeps the same 10$\arcsec$ x 10$\arcsec$ aspect ratio as the images from Figure \ref{fig:images}. The HST STIS slit (PA: 39.98$^{\circ}$) is overlaid on the image as the white shaded region running SW to NE.}
\label{fig:3393}
\vspace{1em}
\end{figure*}
\clearpage
\section{Elliptical Fitting}
\label{app:ellipse}

\begin{figure*}[ht]
\centering
\includegraphics[width=0.85\textwidth,clip]{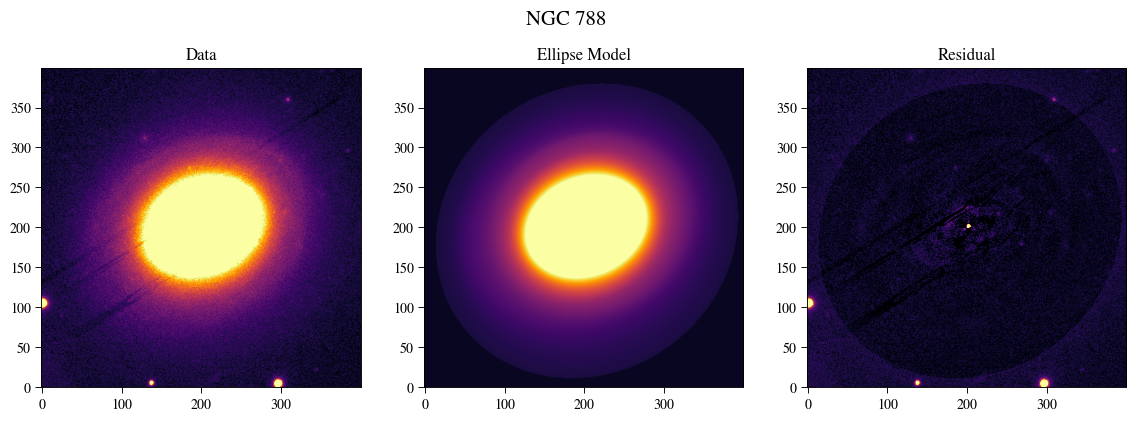}\\
\includegraphics[width=0.85\textwidth,clip]{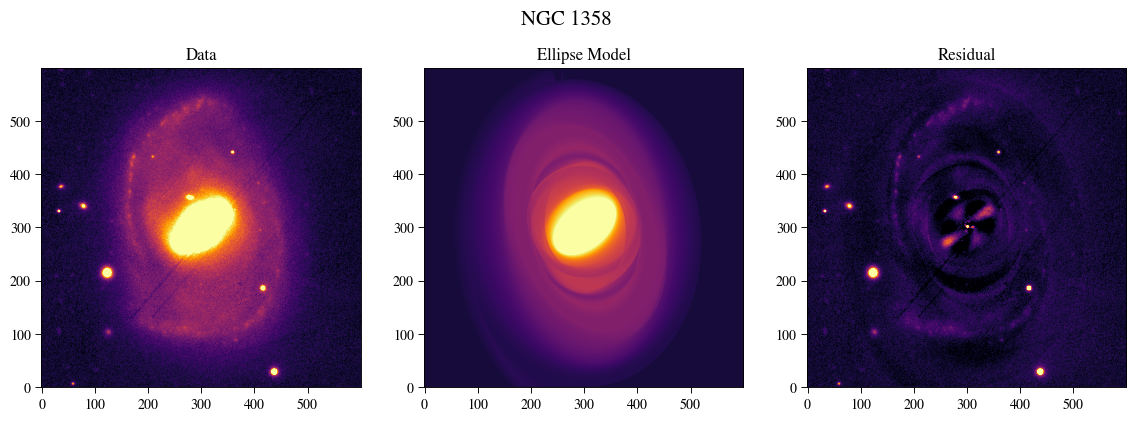}\\
\includegraphics[width=0.85\textwidth,clip]{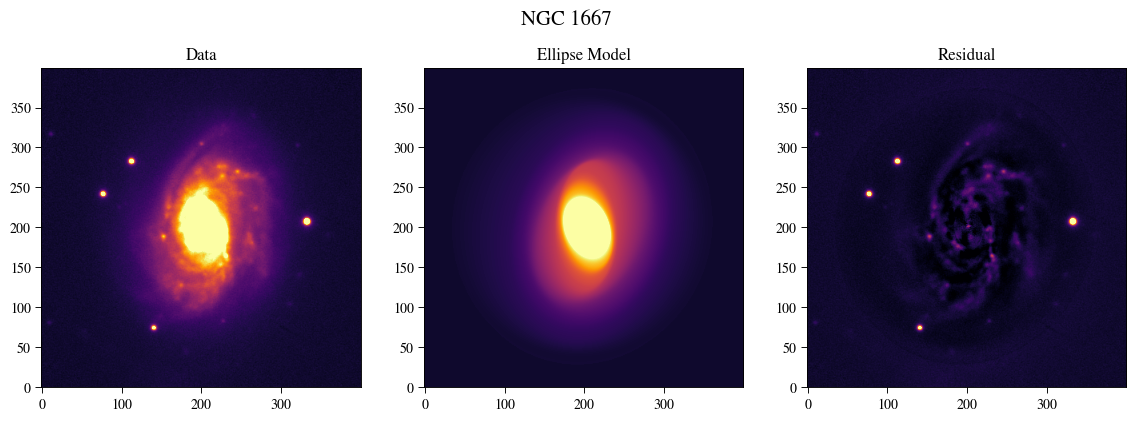}\\

\caption{Results from the elliptical modeling of the ground-based images detailed in Section \ref{subsec:morph}. The left panel shows the ground-based images, the center panel shows the reconstructed model of the galaxy using the isophotal ellipses fit to the image, and the right panel shows the residual from the model subtraction. The axes give the pixel count, with each image showing a 100\arcsec x 100\arcsec field of view, except for NGC 1358, NGC 2273, NGC 3393, and NGC 5427, which have a 150\arcsec x 150\arcsec FOV in order to fit their extended structures.}
\vspace{1em}
\end{figure*}

\addtocounter{figure}{-1}
\begin{figure*}[ht]
\centering
\includegraphics[width=0.85\textwidth,clip]{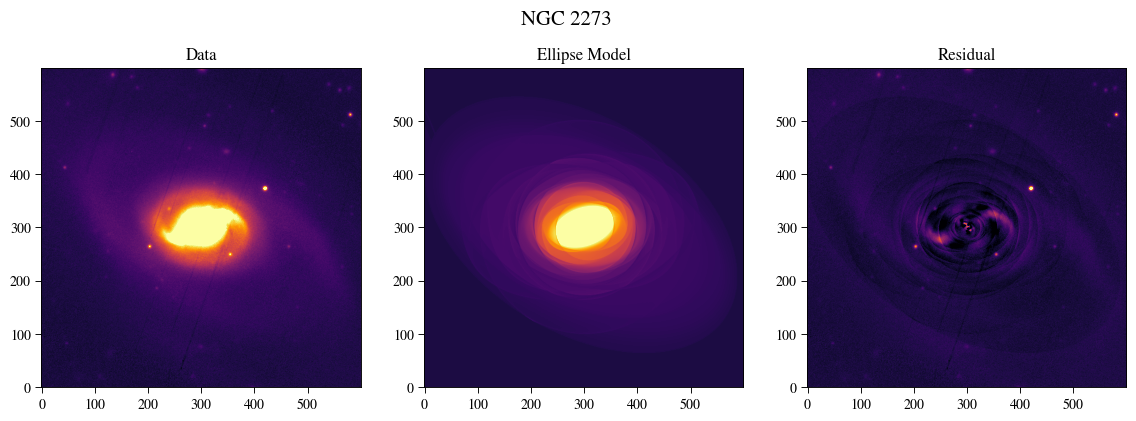}\\
\includegraphics[width=0.85\textwidth,clip]{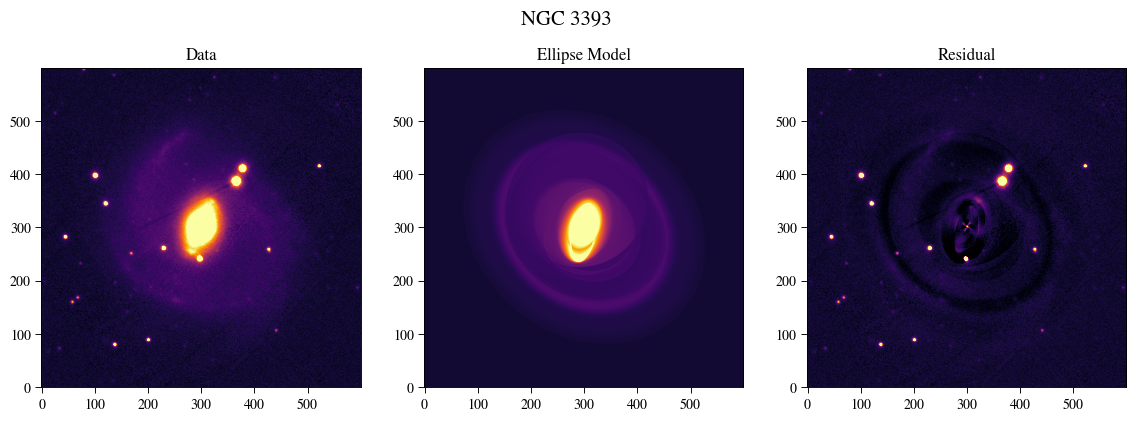}\\
\includegraphics[width=0.85\textwidth,clip]{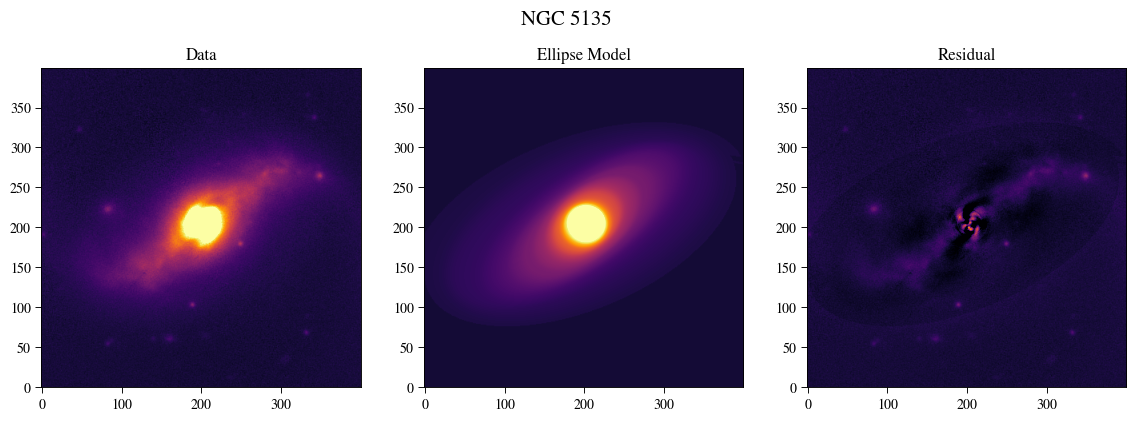}\\
\includegraphics[width=0.85\textwidth,clip]{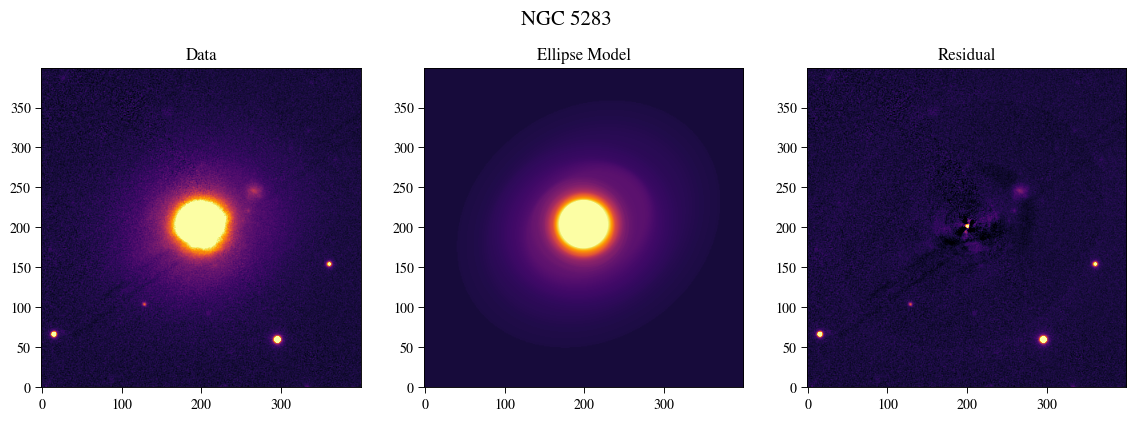}\\
\caption{\textit{continued.}}
\end{figure*}
\addtocounter{figure}{-1}
\begin{figure*}[ht]
\centering
\includegraphics[width=0.85\textwidth,clip]{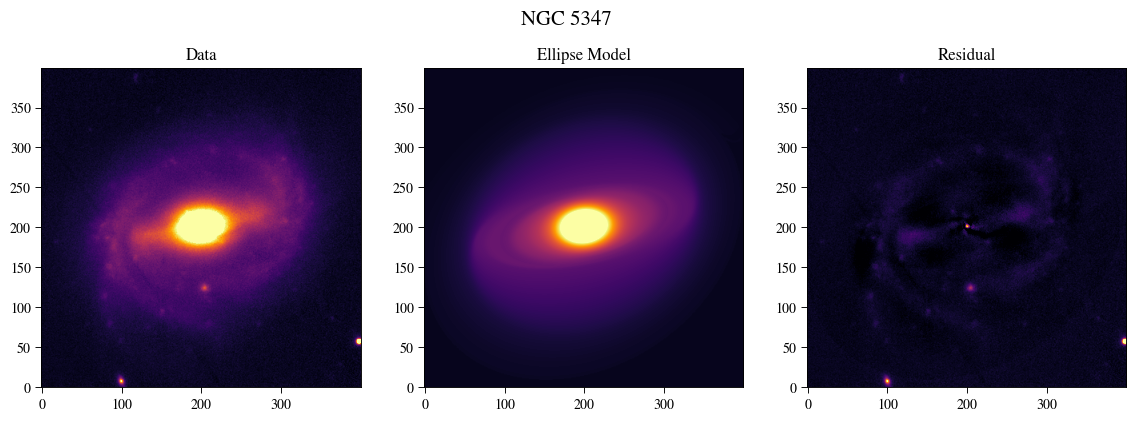}\\
\includegraphics[width=0.85\textwidth,clip]{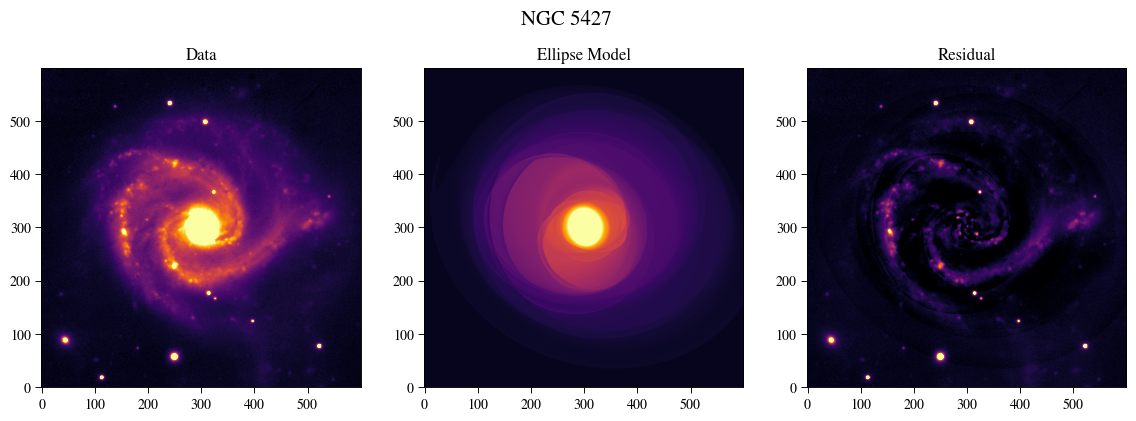}\\
\includegraphics[width=0.85\textwidth,clip]{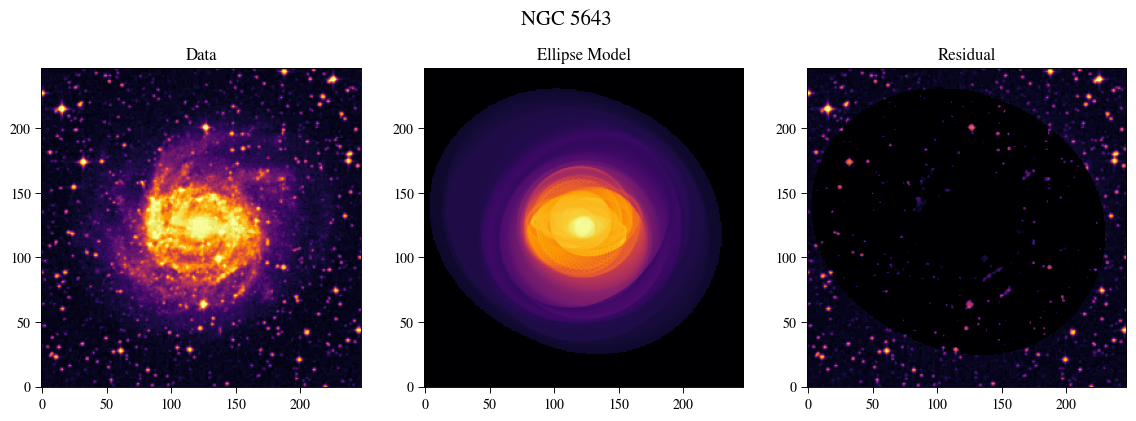}\\
\includegraphics[width=0.85\textwidth,clip]{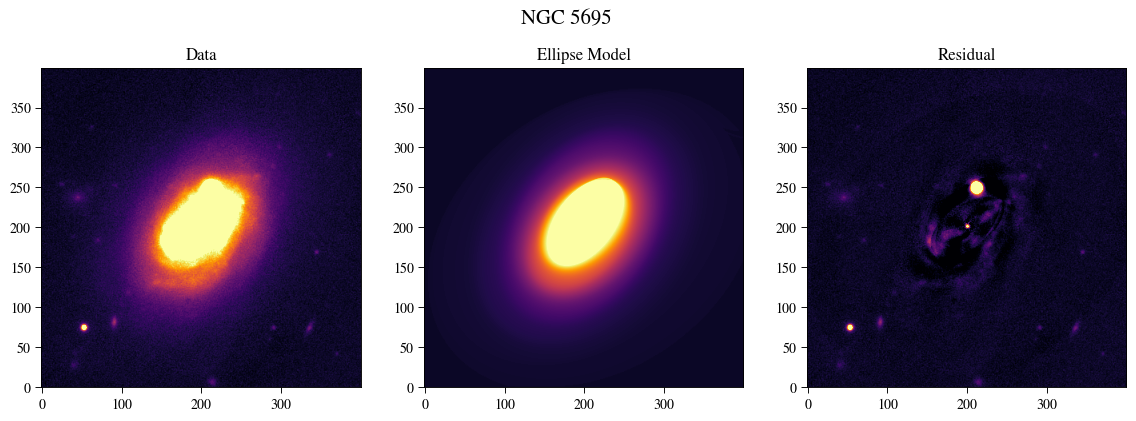}\\

\caption{\textit{continued.}}
\end{figure*}
\addtocounter{figure}{-1}
\begin{figure*}[ht]
\centering
\includegraphics[width=0.85\textwidth,clip]{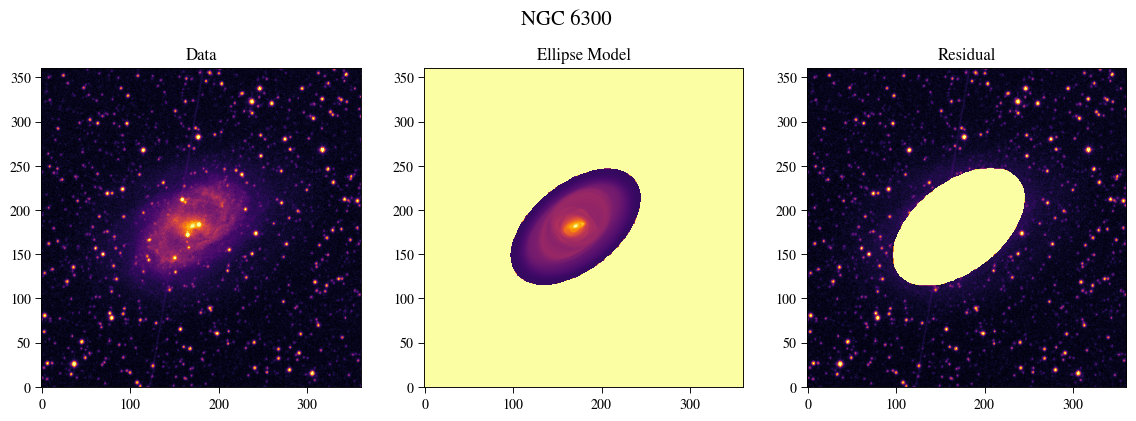}\\
\includegraphics[width=0.85\textwidth,clip]{Plots/ngc7682_efinal3n.png}\\
\includegraphics[width=0.85\textwidth,clip]{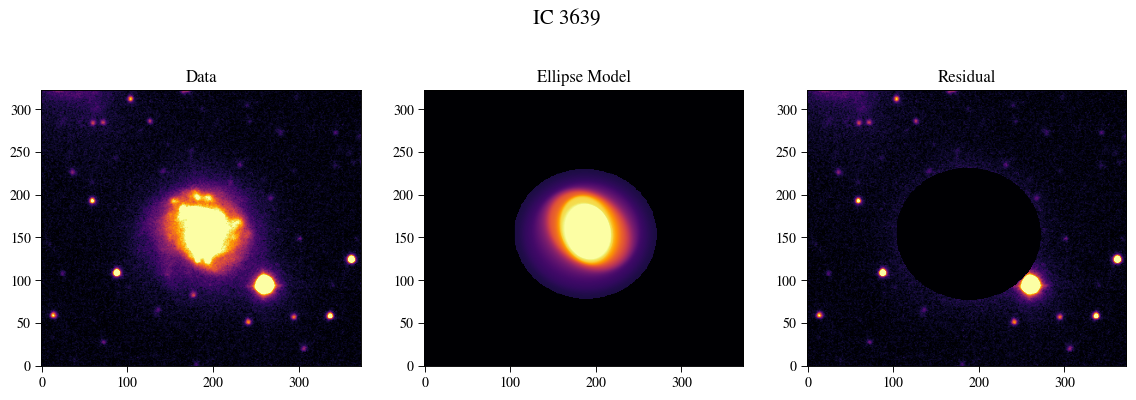}\\
\includegraphics[width=0.85\textwidth,clip]{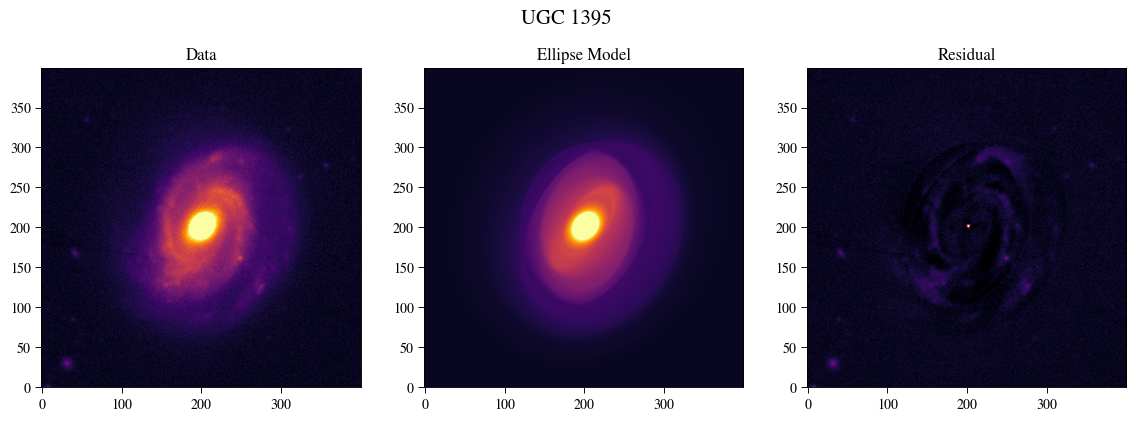}\\
\caption{\textit{continued.}}
\end{figure*}

\begin{figure}[ht]
\centering
\subfigure{
\includegraphics[width=0.305\textwidth]{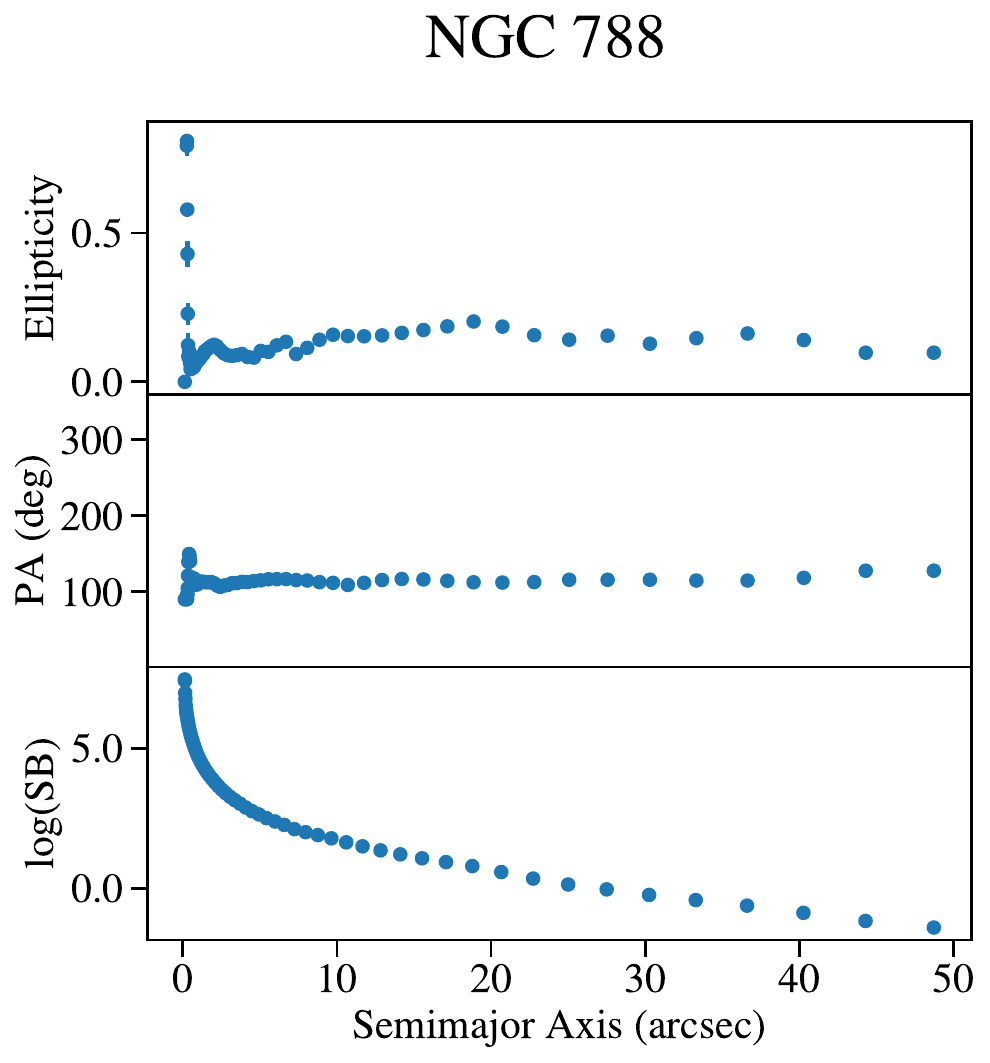}}
\subfigure{
\includegraphics[width=0.305\textwidth]{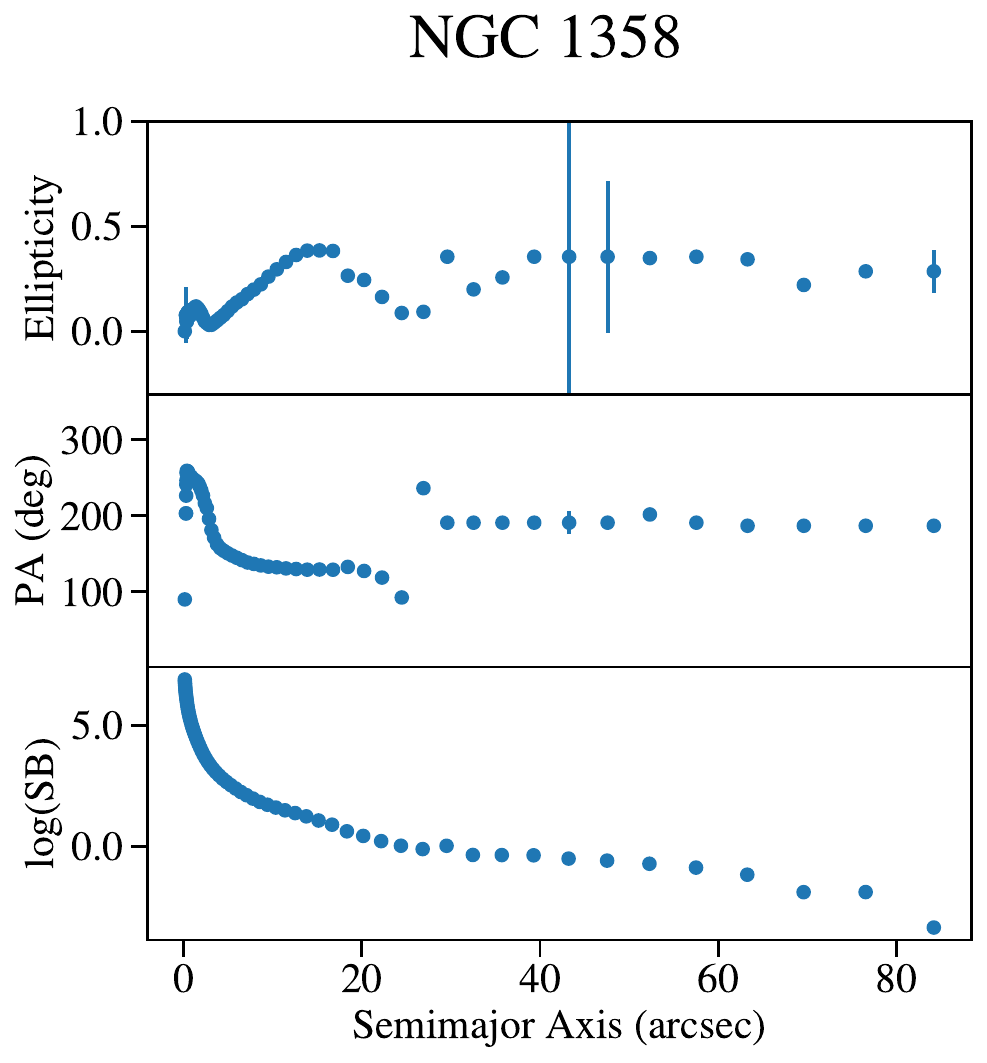}}
\subfigure{
\includegraphics[width=0.305\textwidth]{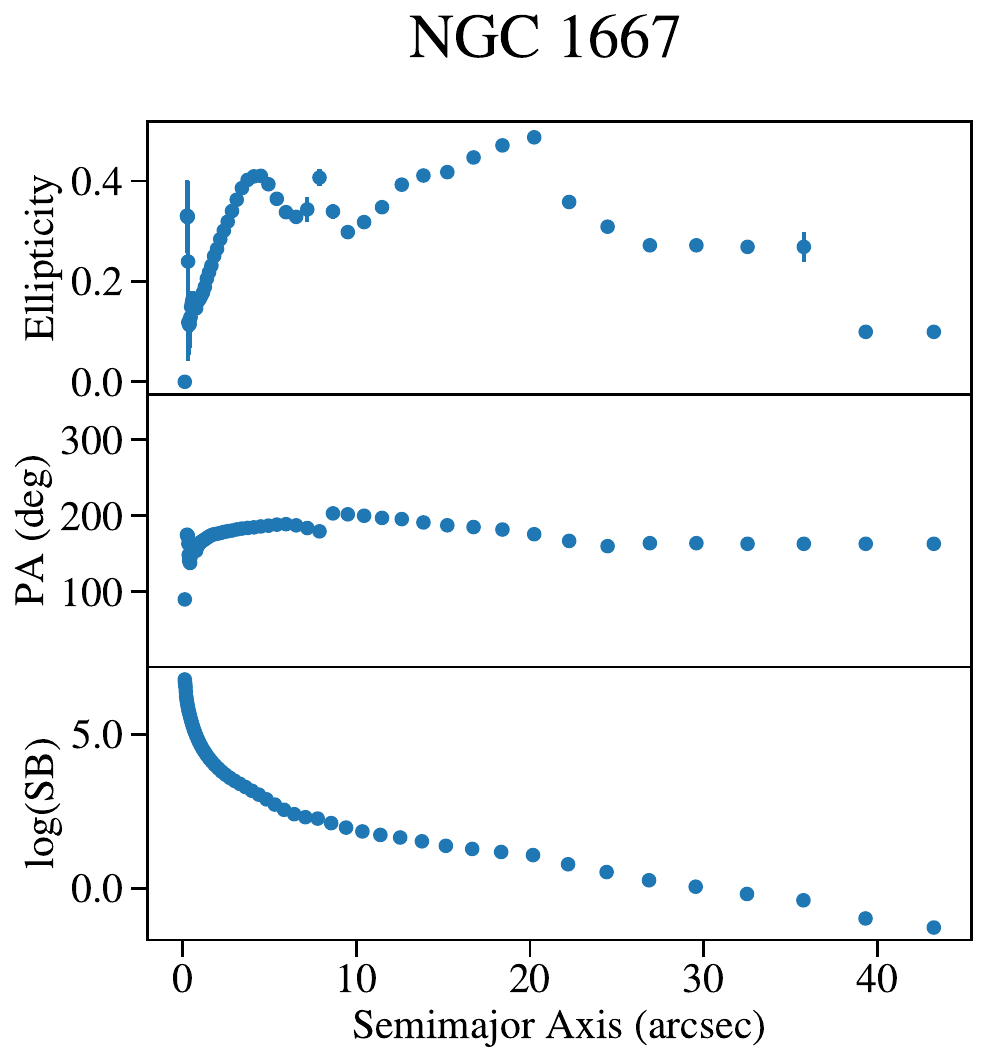}}
\subfigure{
\includegraphics[width=0.305\textwidth]{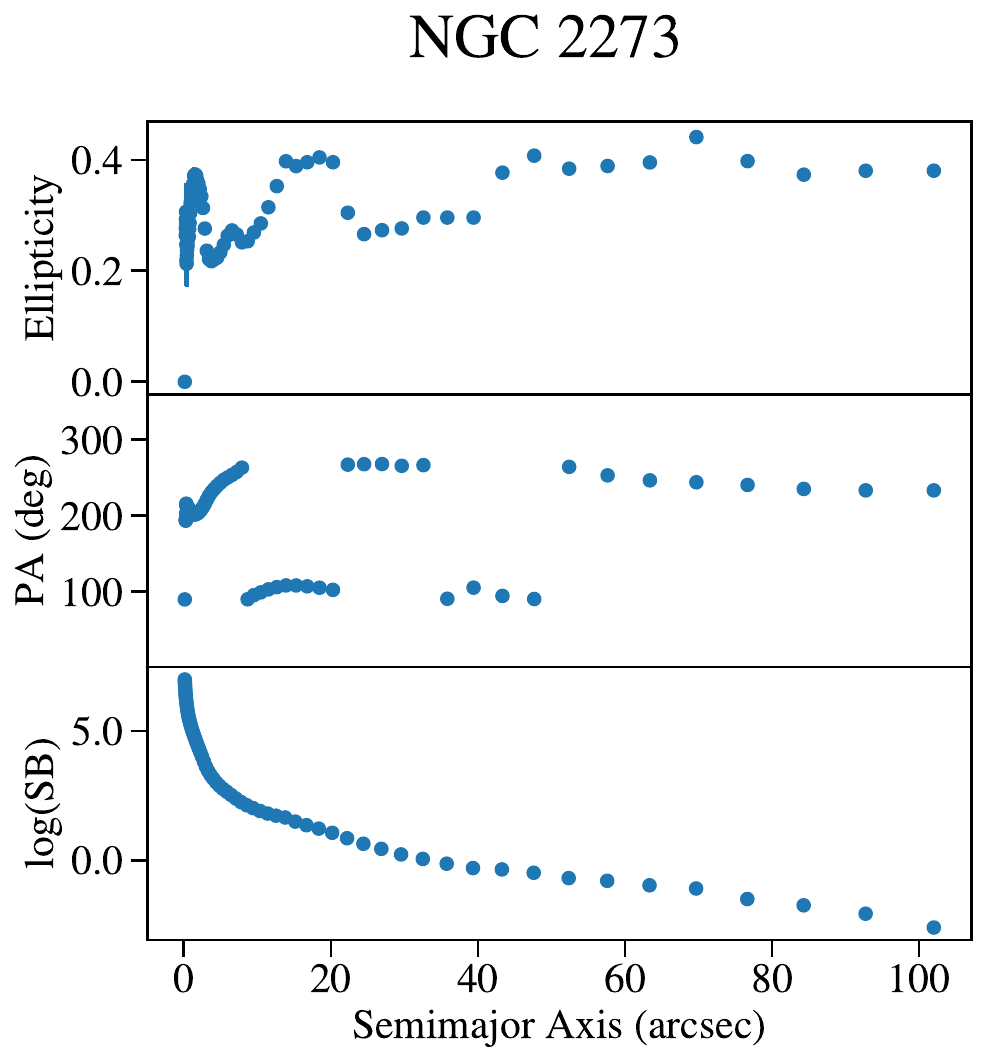}}
\subfigure{
\includegraphics[width=0.305\textwidth]{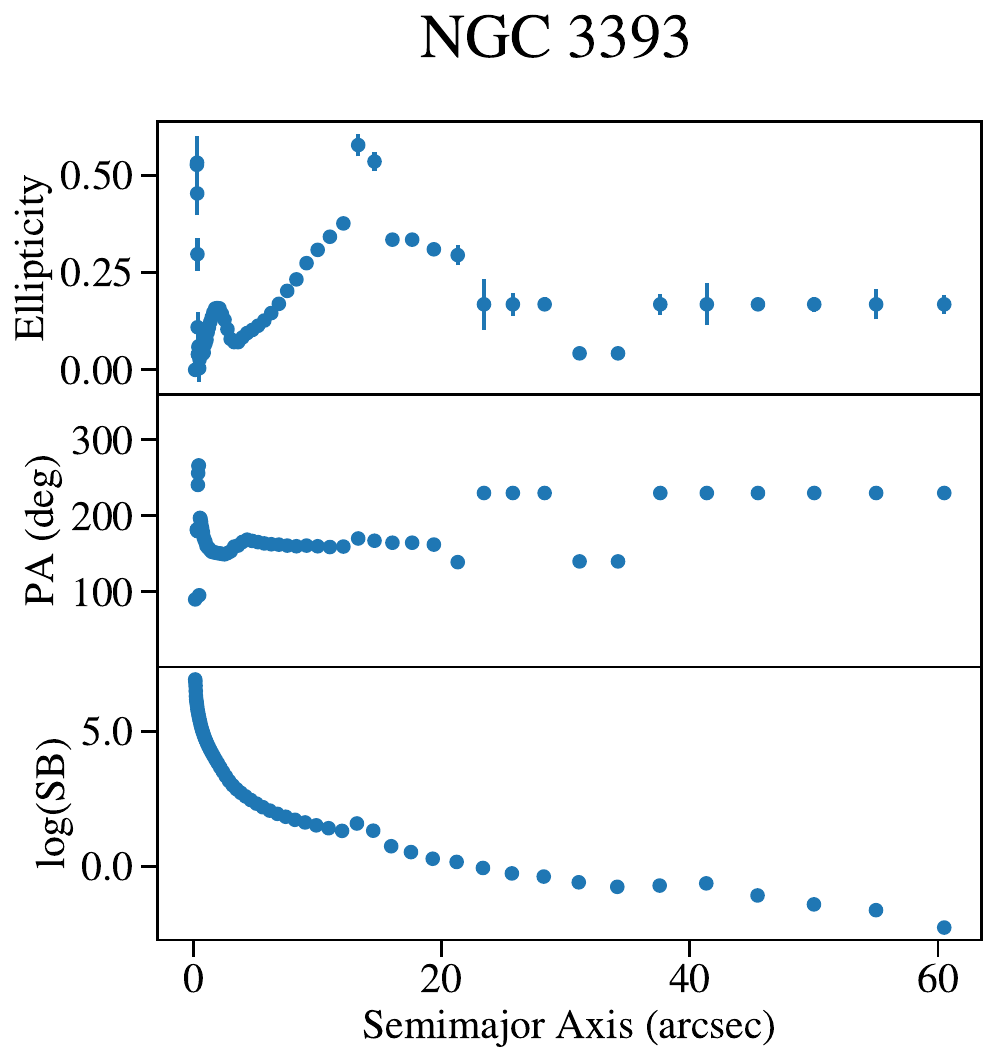}}
\subfigure{
\includegraphics[width=0.305\textwidth]{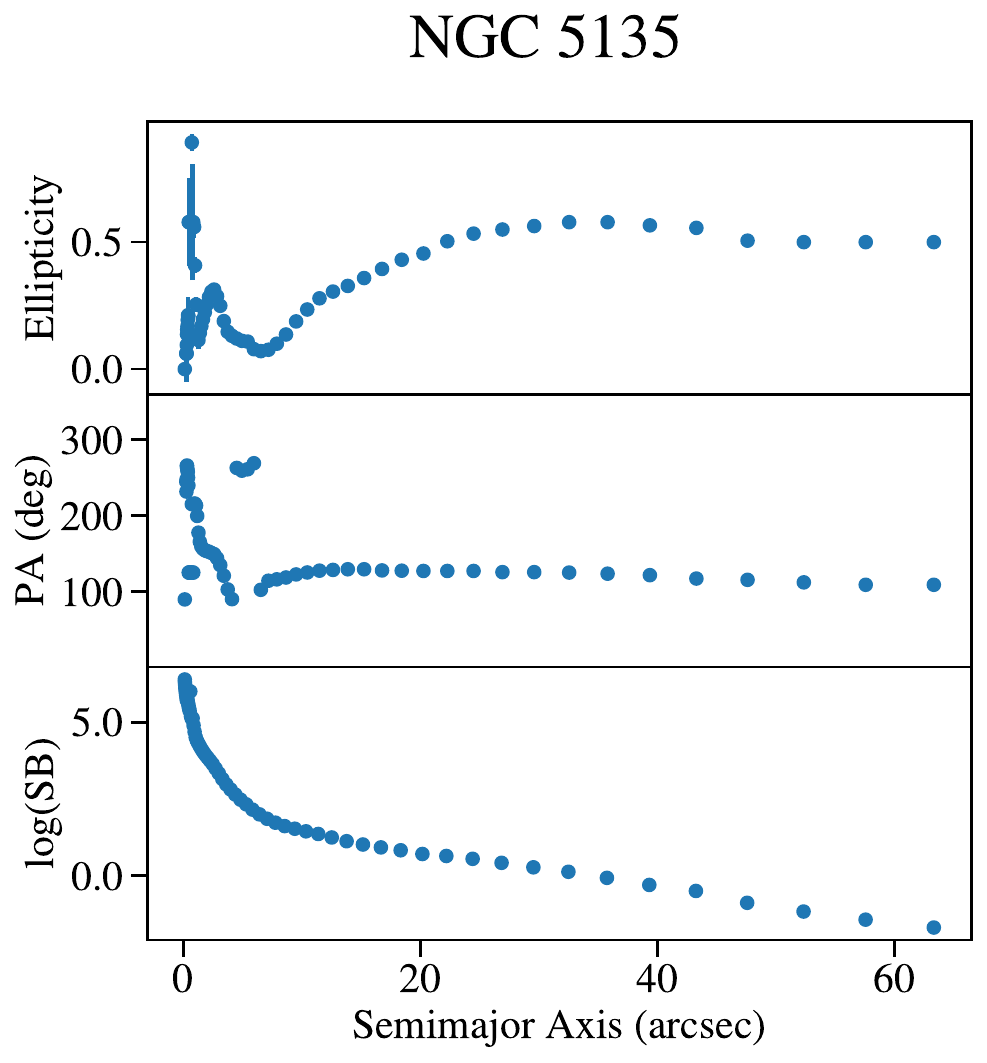}}
\subfigure{
\includegraphics[width=0.305\textwidth]{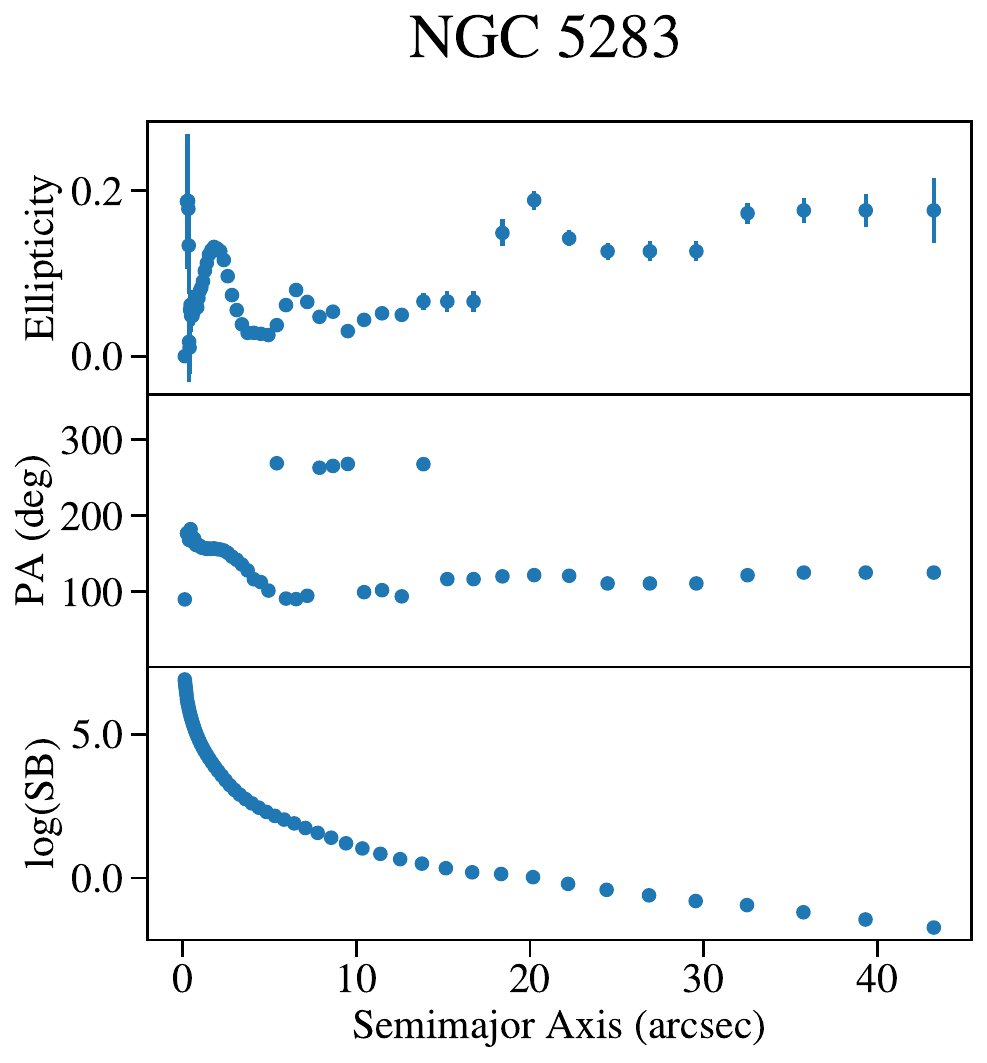}}
\subfigure{
\includegraphics[width=0.305\textwidth]{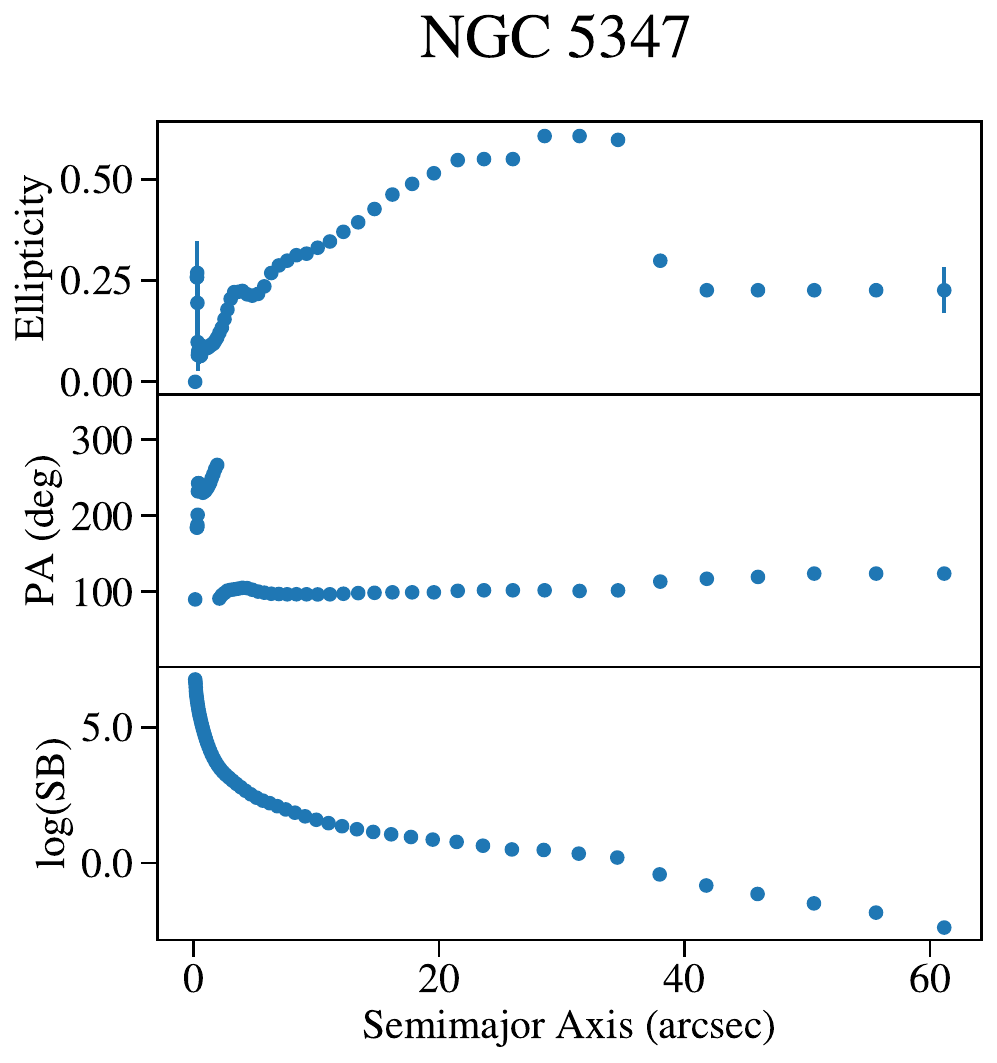}}
\subfigure{
\includegraphics[width=0.305\textwidth]{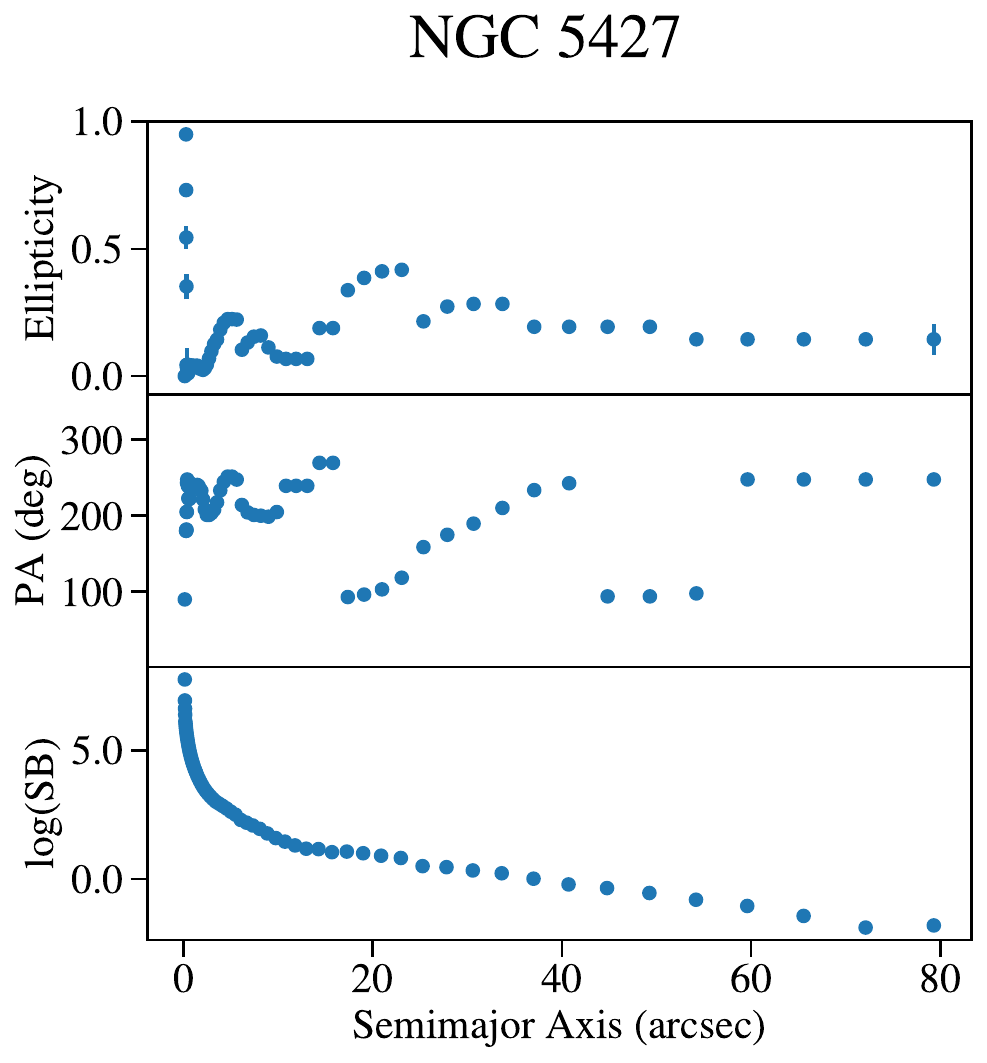}}
\caption{Output plots from the isophotal ellipse models shown in Figure \ref{app:ellipse}. The x-axis gives distance from the nucleus in arcseconds, while the y-axis shows the model ellipticity, major axis position angle in degrees, and surface brightness in log counts for each elliptical isophote.}
\end{figure}
\addtocounter{figure}{-1}
\begin{figure}[ht]
\centering
\subfigure{
\includegraphics[width=0.305\textwidth]{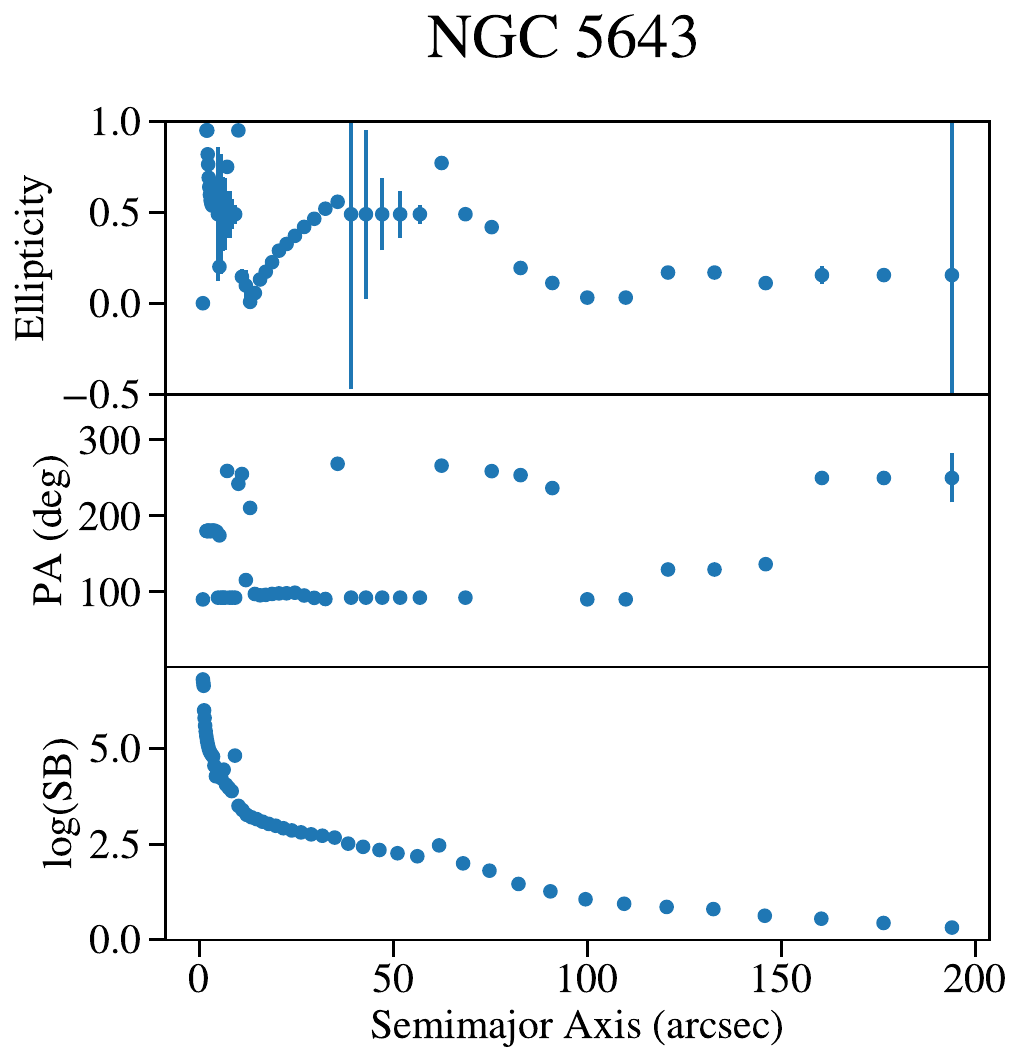}}
\subfigure{
\includegraphics[width=0.305\textwidth]{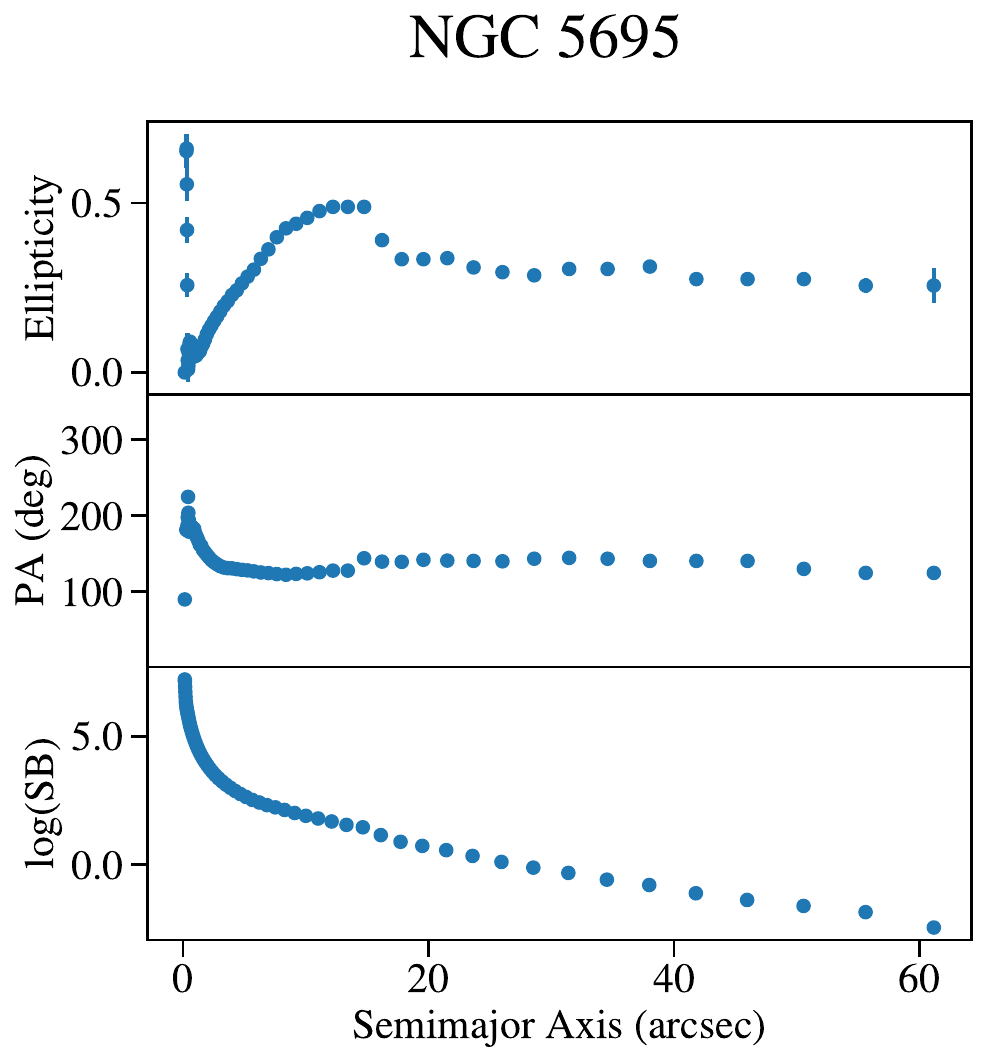}}
\subfigure{
\includegraphics[width=0.305\textwidth]{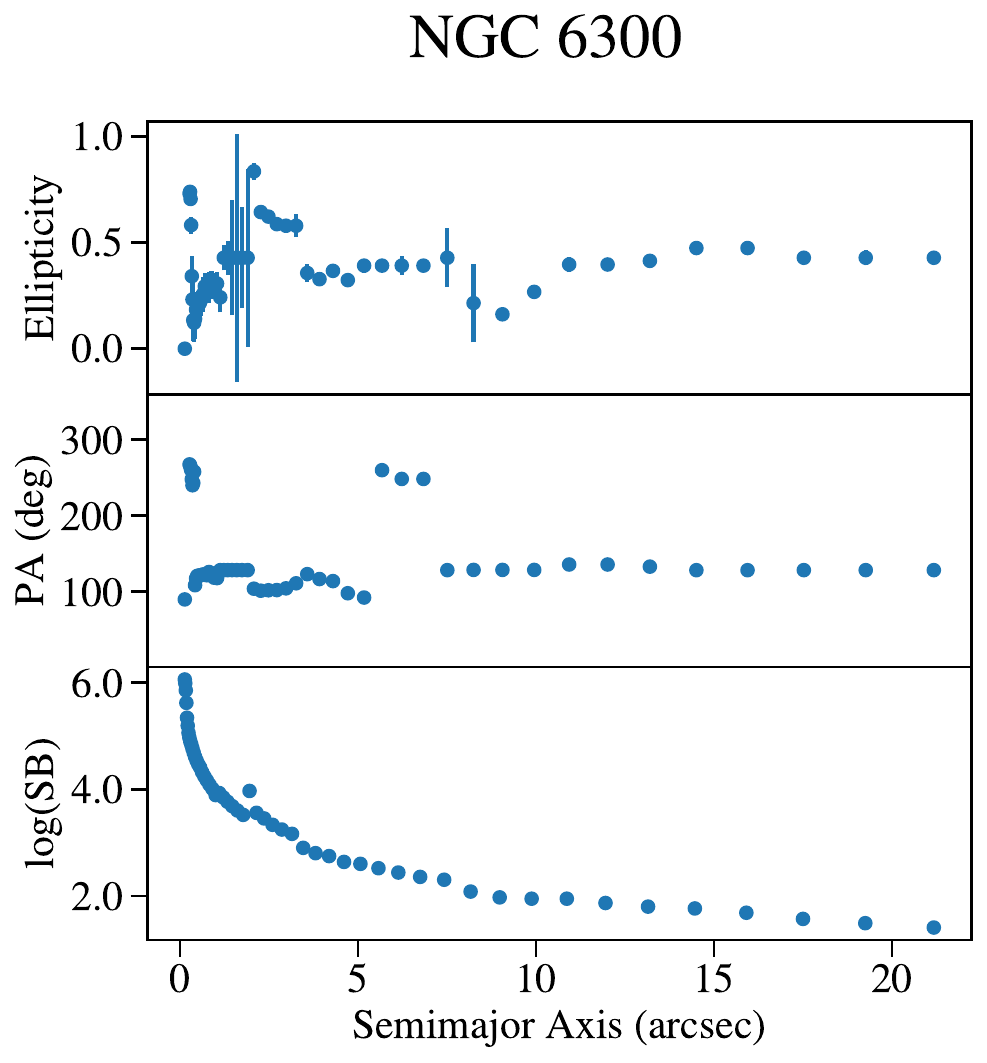}}
\subfigure{
\includegraphics[width=0.305\textwidth]{Plots/ngc7682_modeling.pdf}}
\subfigure{
\includegraphics[width=0.305\textwidth]{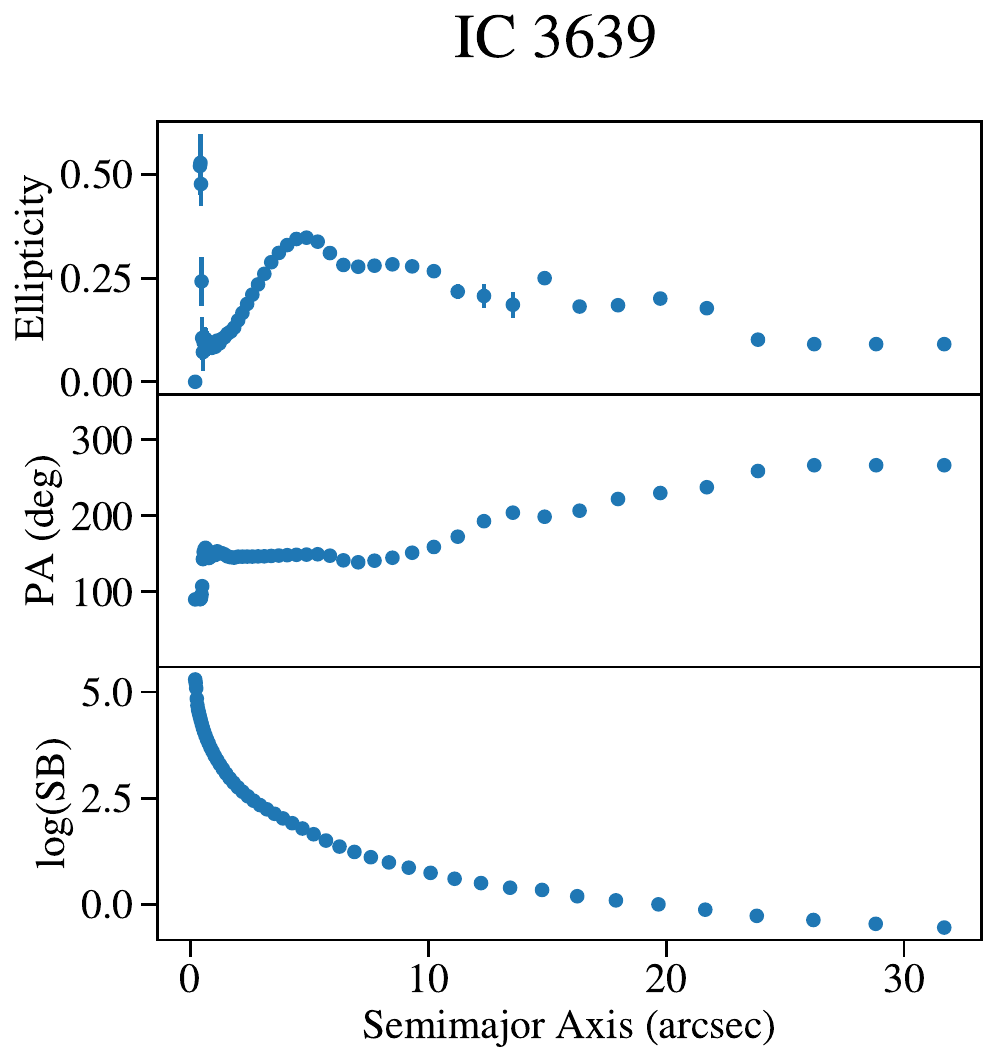}}
\subfigure{
\includegraphics[width=0.305\textwidth]{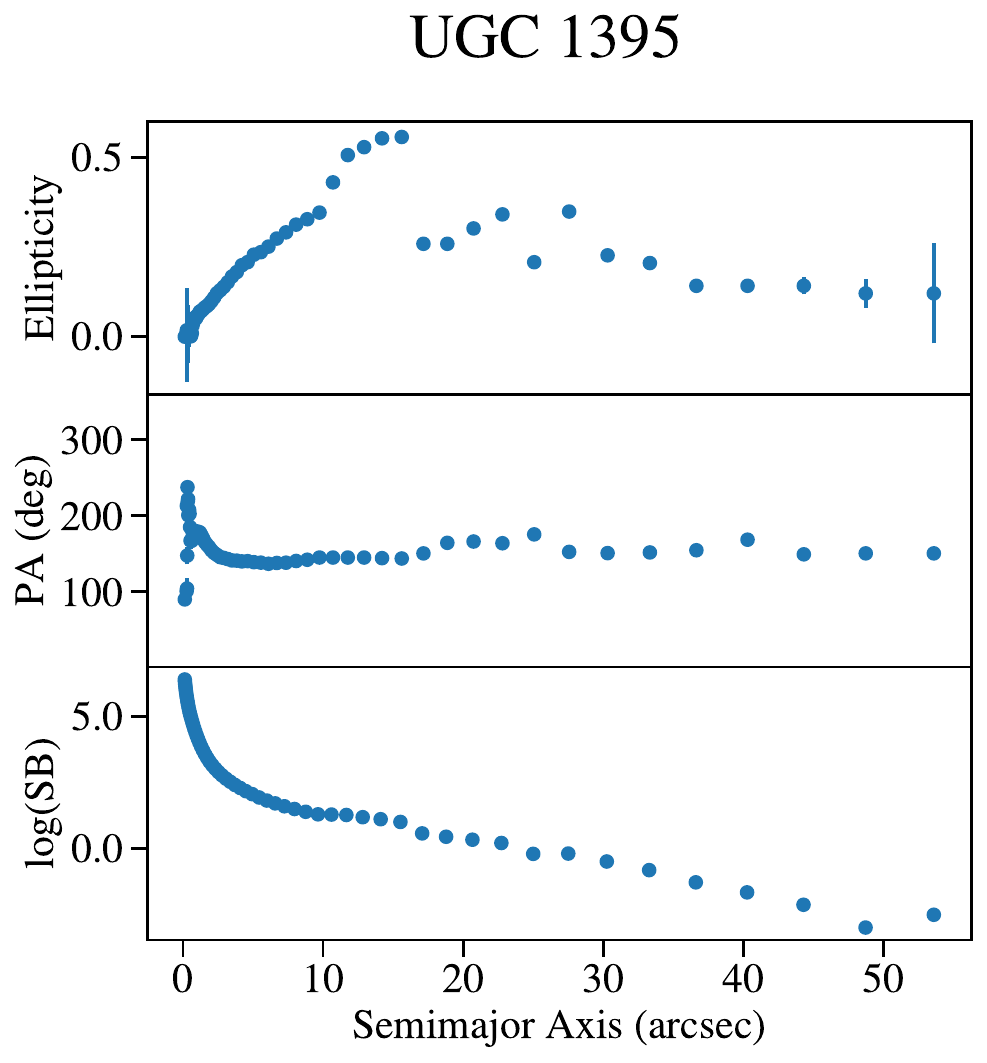}}
\caption{\textit{continued.}}
\end{figure}

\begin{figure*}[ht]
\centering
\subfigure{
\includegraphics[width=0.305\textwidth]{Plots/ngc788_kinematics.pdf}}
\subfigure{
\includegraphics[width=0.305\textwidth]{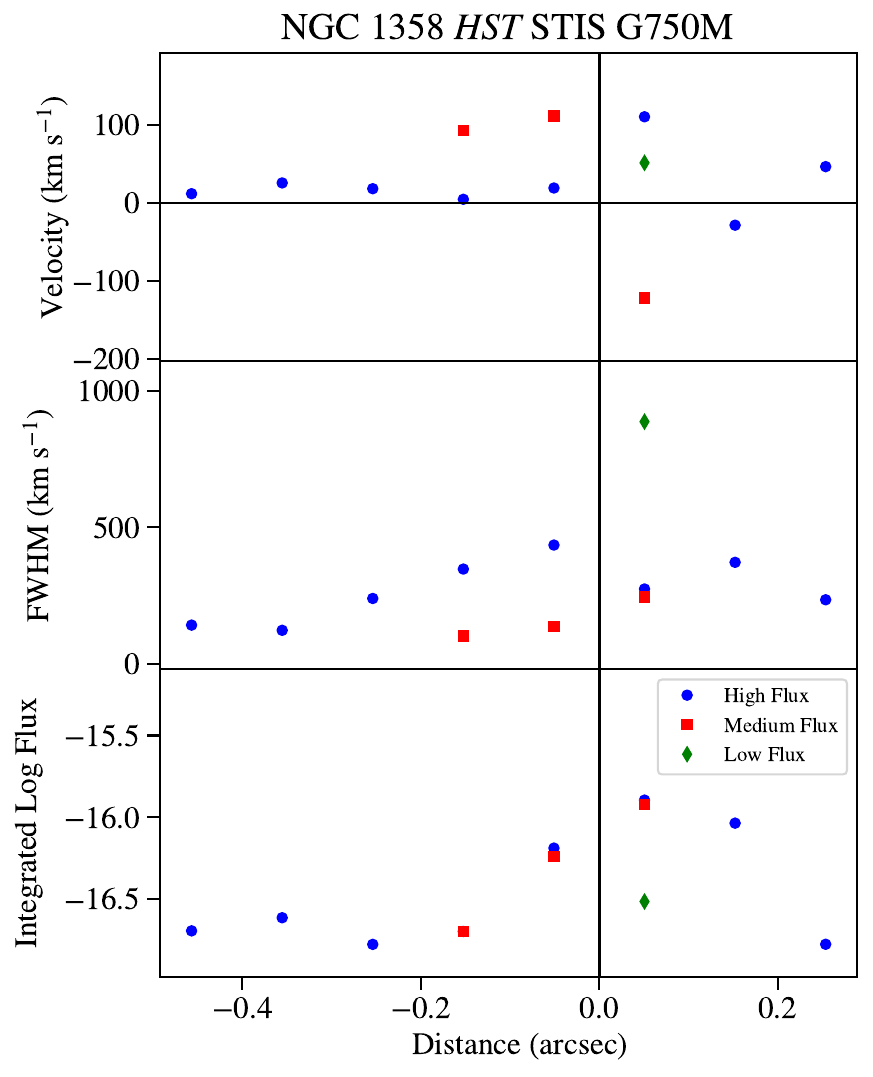}}
\subfigure{
\includegraphics[width=0.305\textwidth]{Plots/ngc1667_kinematics.pdf}}
\subfigure{
\includegraphics[width=0.305\textwidth]{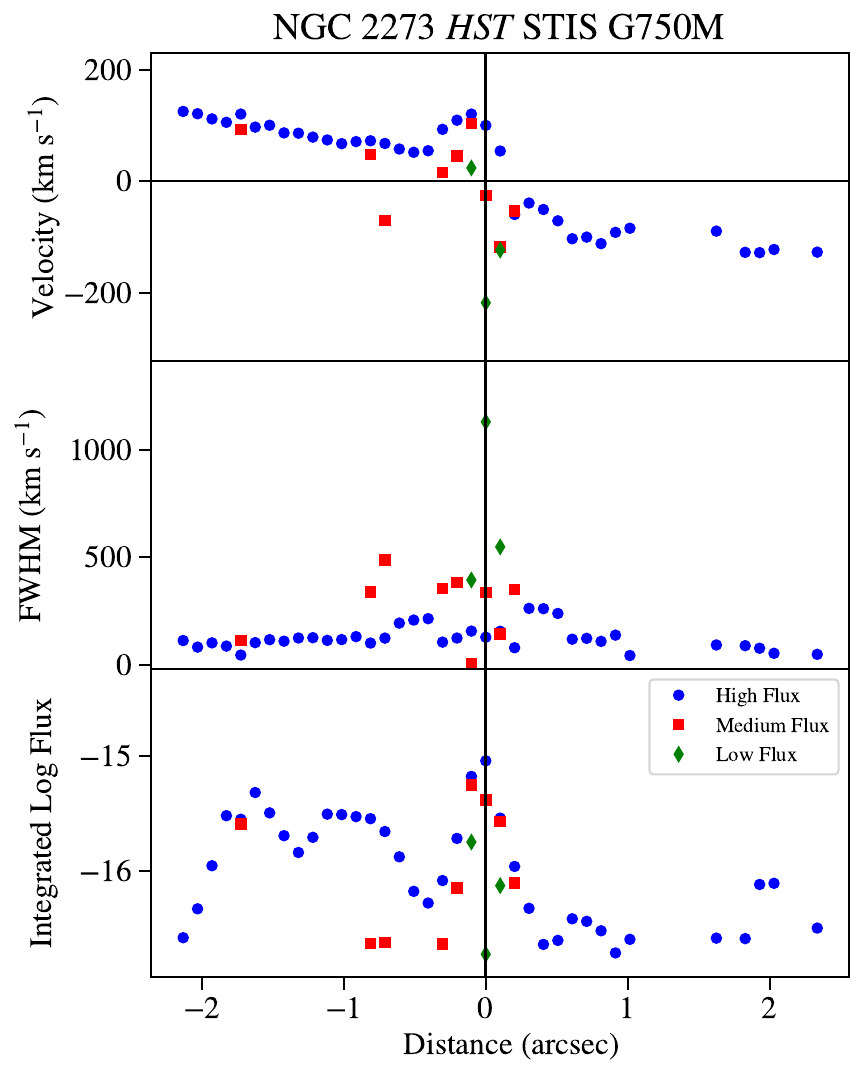}}
\subfigure{
\includegraphics[width=0.305\textwidth]{Plots/ngc3393_kinematics.pdf}}
\subfigure{
\includegraphics[width=0.305\textwidth]{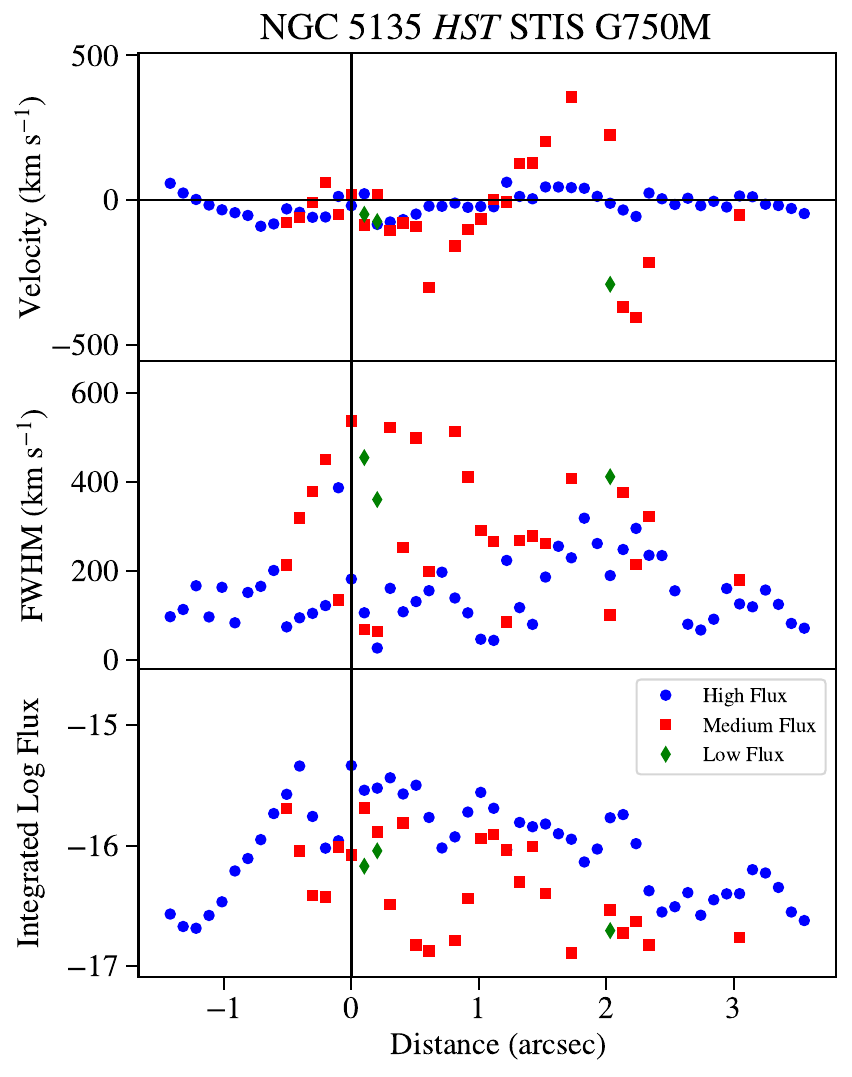}}
\subfigure{
\includegraphics[width=0.305\textwidth]{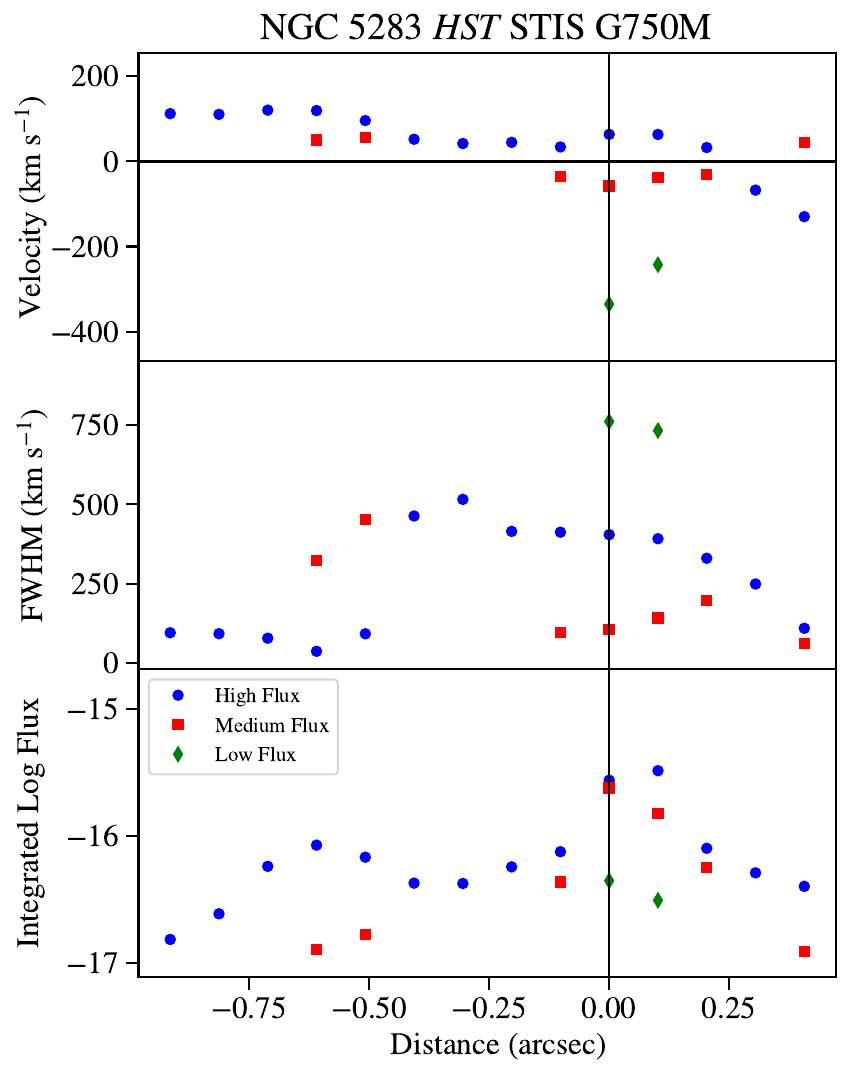}}
\subfigure{
\includegraphics[width=0.305\textwidth]{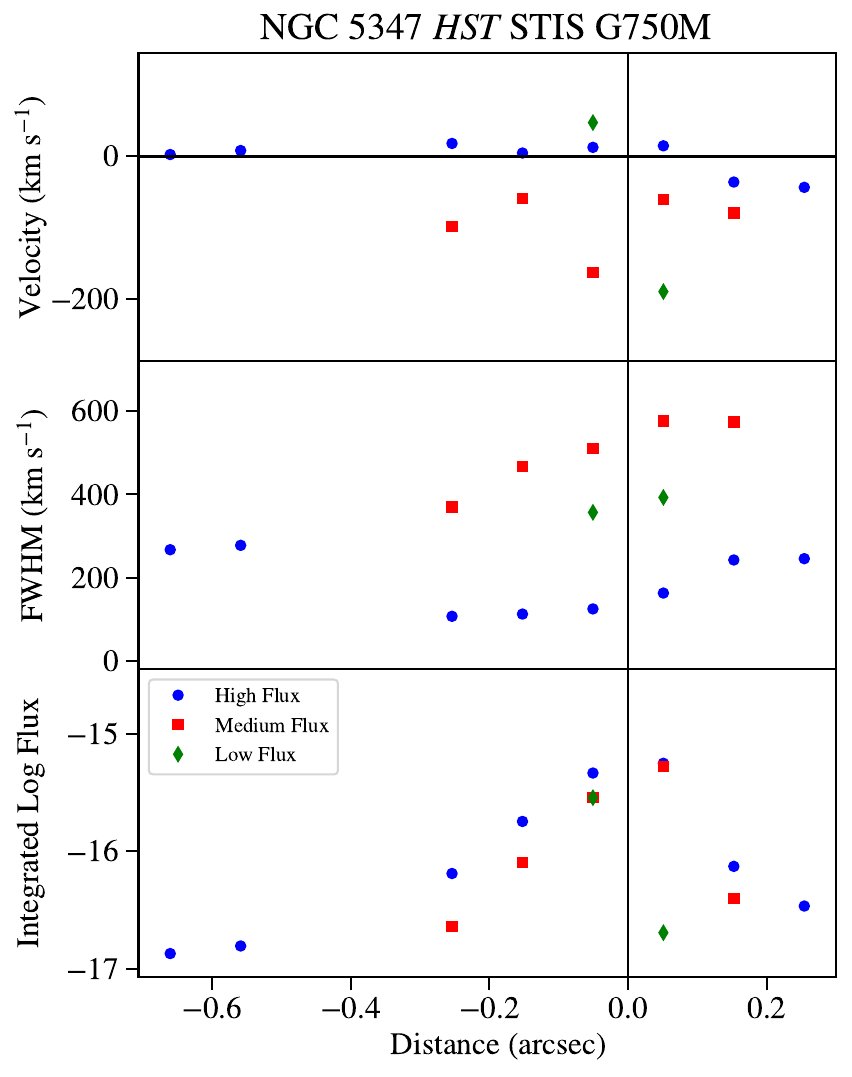}}
\subfigure{
\includegraphics[width=0.305\textwidth]{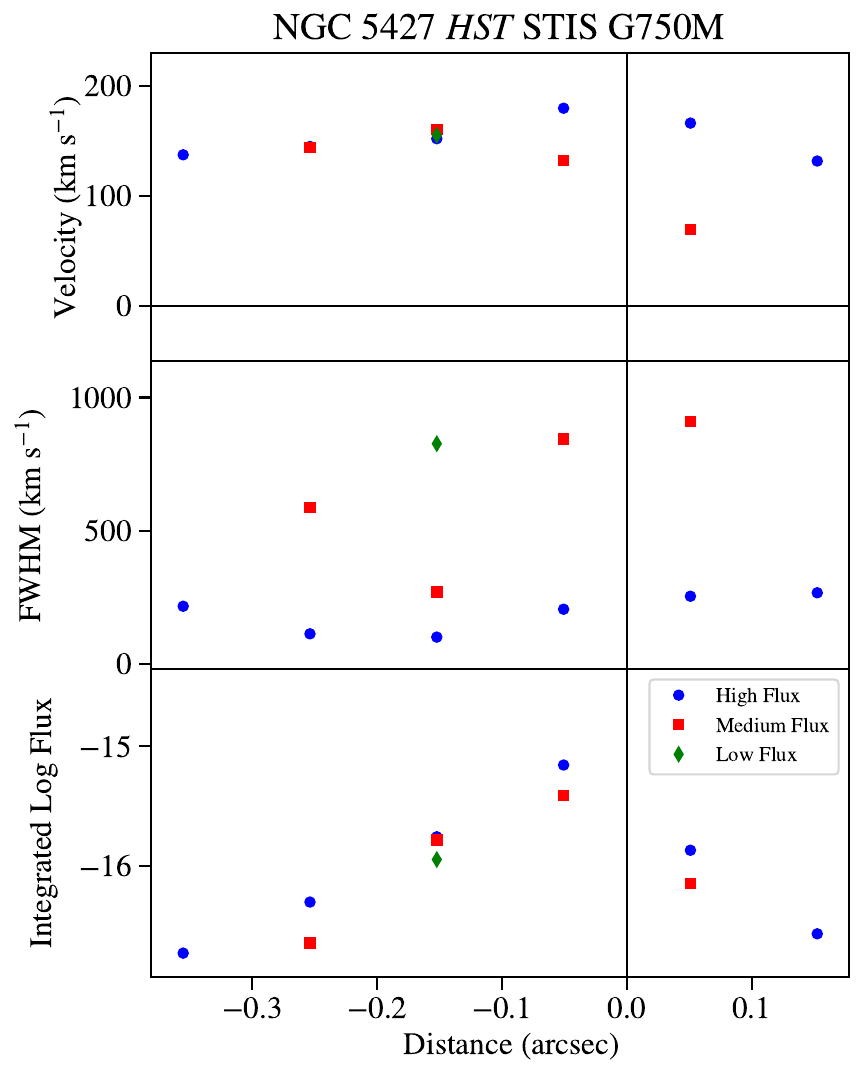}}
\caption{Kinematic Plots for the full sample of AGN, following the same structure detailed in Figure \ref{fig:Kinematics_Test}.}
\label{app:Kinematics_App}
\end{figure*}
\addtocounter{figure}{-1}
\begin{figure*}[ht!]
\centering
\subfigure{
\includegraphics[width=0.305\textwidth]{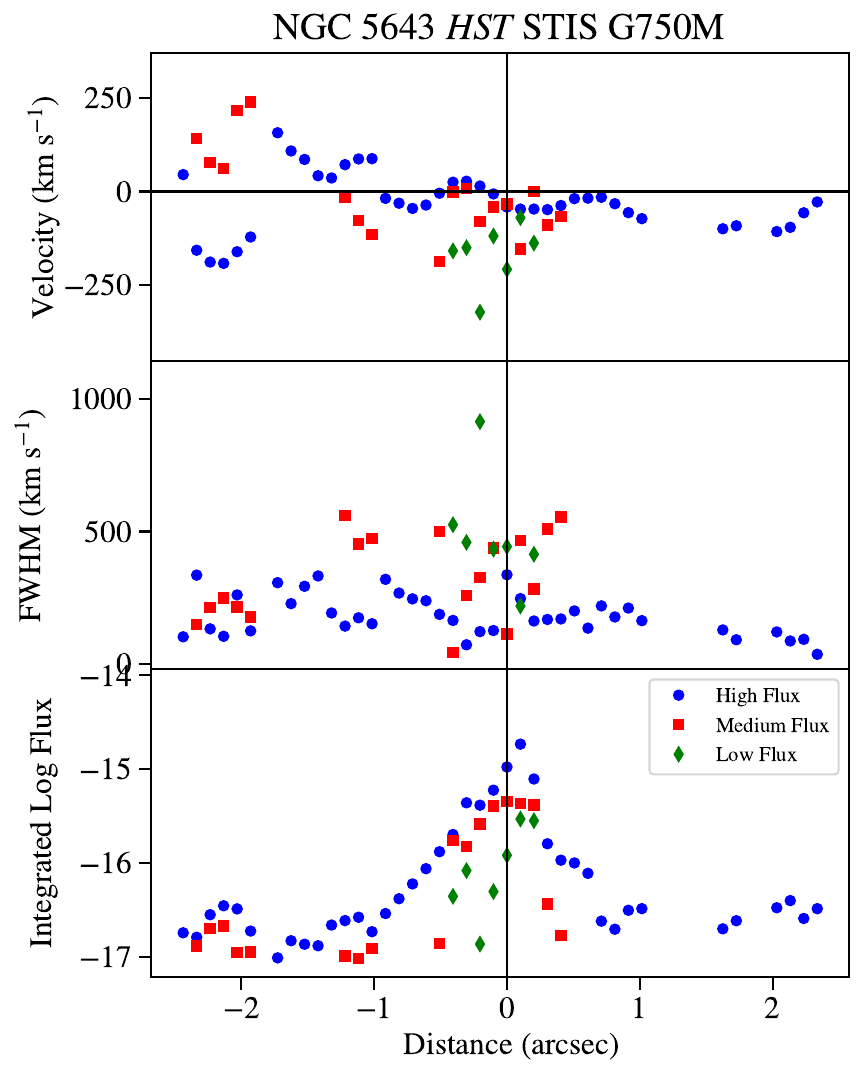}}
\subfigure{
\includegraphics[width=0.305\textwidth]{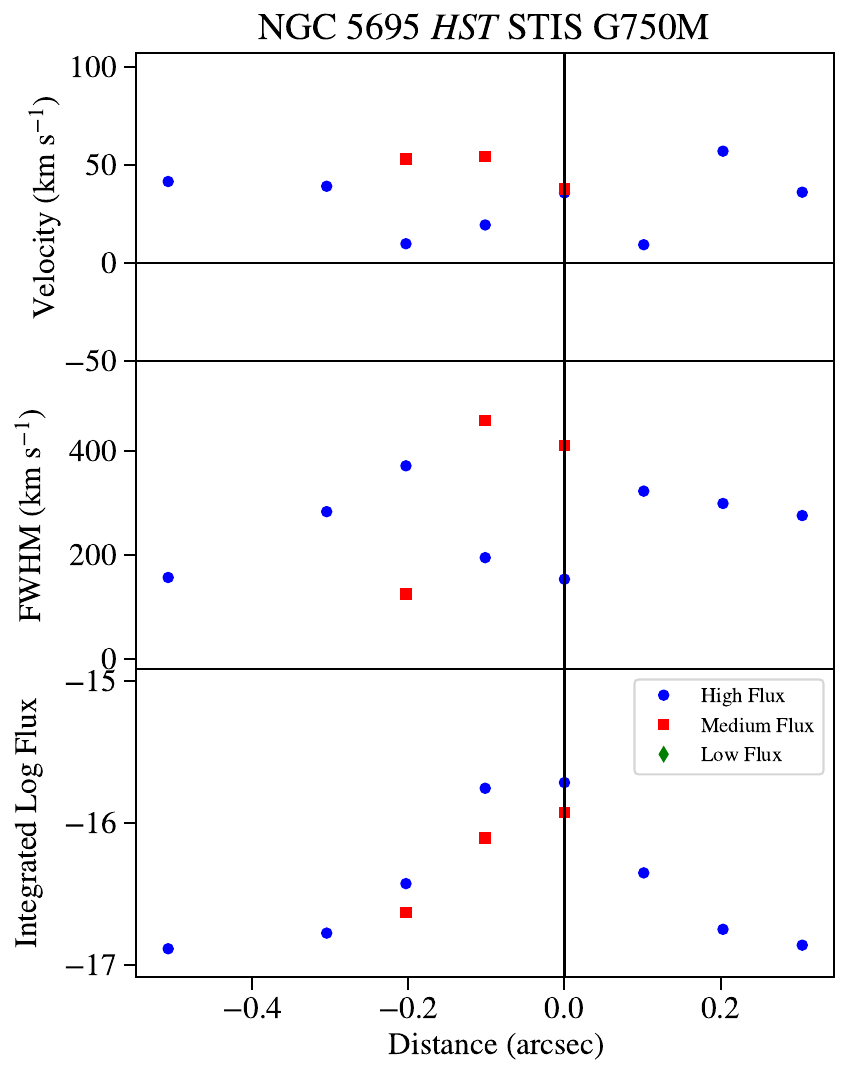}}
\subfigure{
\includegraphics[width=0.305\textwidth]{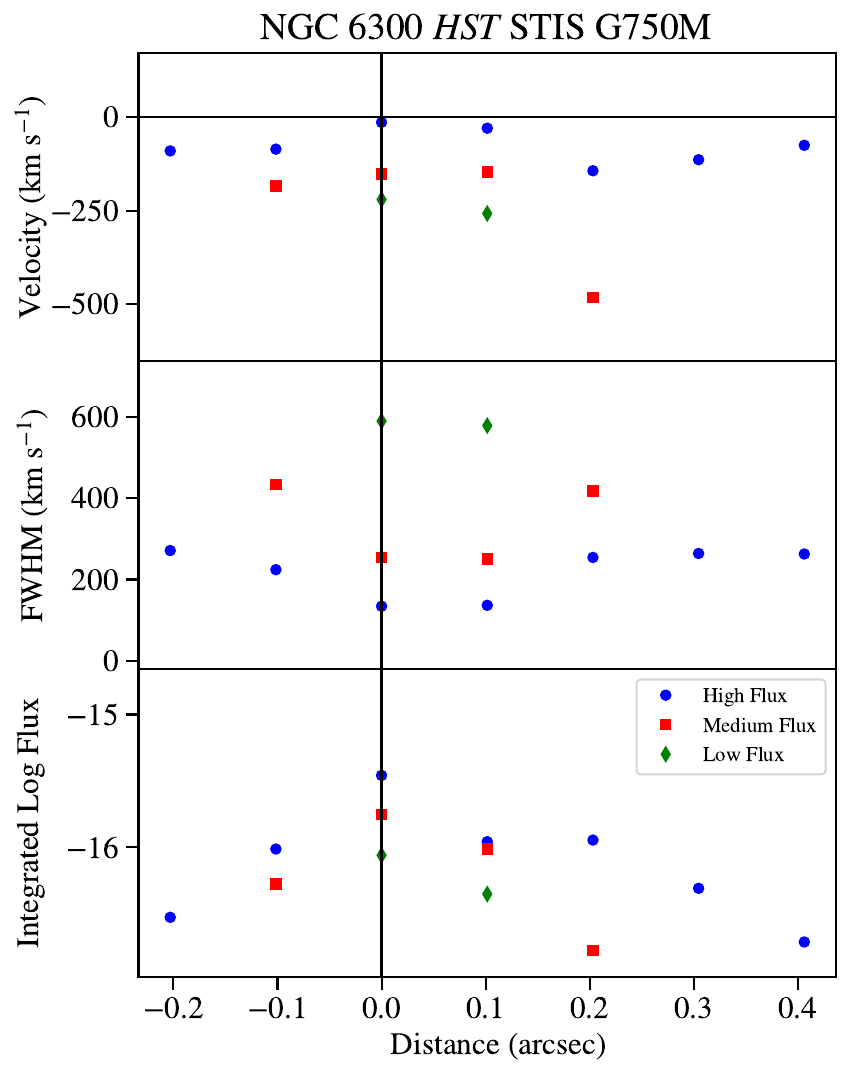}}
\subfigure{
\includegraphics[width=0.305\textwidth]{Plots/ngc7682_kinematics.pdf}}
\subfigure{
\includegraphics[width=0.305\textwidth]{Plots/ic3639_kinematics.pdf}}
\subfigure{
\includegraphics[width=0.305\textwidth]{Plots/um146_kinematics.pdf}}
\caption{\textit{continued.}}
\end{figure*}

\end{document}